\documentclass[12pt]{report}	

\usepackage{utdiss2}  		

\usepackage{amsmath,amsthm,amsfonts,amscd,amssymb} 
\usepackage{eucal} 	 	
\usepackage{verbatim}      	
\usepackage{makeidx}       	
\usepackage{graphicx} 
\usepackage{color}
\usepackage{subfigure}
\usepackage{citesort}         	%
\usepackage{url}		

\author{Jung Ho Ryu}  	

\address{Maryland, USA}  

\title{Congestion Control and Routing over Challenged Networks}

\supervisor
	{Sanjay Shakkottai}

\committeemembers
	[Gustavo de Veciana]
	[Sriram Vishwanath]
	[Christine Julien]
	{John Hasenbein}

\previousdegrees{B.S.E.E.; M.S.E.}
\oneandonehalfspacequote

\DeclareMathOperator*{\argmax}{arg\,max}
\DeclareMathOperator*{\argmin}{arg\,min}
\newcommand{\FR}{\mbox{$\mathbb F_{q}$}}
\newtheorem{assumption}{Assumption}
\newtheorem{defn}{Definition}

\newtheorem{theorem}{Theorem}

\newtheorem{lemma}{Lemma}
\def\done{\hspace*{\fill} \rule{1.8mm}{2.5mm}}

\newcommand{\LL}{\mathcal{L}}

\newcommand{\C}{\mathcal{C}}
\newcommand{\I}{\mathcal{I}}
\newcommand{\HH}{\mathcal{H}}

\newcommand{\NN}{\mathcal{N}}
\newcommand{\GG}{\mathcal{G}}
\newcommand{\tT}{\widetilde{T}}
\newcommand{\F}{\mathcal{F}}

\newcommand{\MM}{\mathcal{M}}
\newcommand{\defeq}{\mbox{$\stackrel{\small \Delta}{=}$}}

\theoremstyle{remark}

\newcommand{\latexe}{{\LaTeX\kern.125em2%
                      \lower.5ex\hbox{$\varepsilon$}}}

\chardef\bslash=`\\	

\makeatletter		
\def\square{\RIfM@\bgroup\else$\bgroup\aftergroup$\fi
  \vcenter{\hrule\hbox{\vrule\@height.6em\kern.6em\vrule}%
                                              \hrule}\egroup}
\makeatother		

\makeindex    

\begin{document}

\copyrightpage          

%
%
%
\commcertpage           

\titlepage              

\utabstract
\index{Abstract}%
\indent
This dissertation is a study on the design and analysis of novel, optimal routing and rate control algorithms in wireless, mobile communication networks.  Congestion control and routing algorithms upto now have been designed and optimized for wired or wireless mesh networks.  In those networks, optimal algorithms (optimal in the sense that either the throughput is maximized or delay is minimized, or the network operation cost is minimized) can be engineered based on the classic time scale decomposition assumption that the dynamics of the network are either fast enough so that these algorithms essentially see the average or slow enough that any changes can be tracked to allow the algorithms to adapt over time.  However, as technological advancements enable integration of ever more mobile nodes into communication networks, any rate control or routing algorithms based, for example, on averaging out the capacity of the wireless mobile link or tracking the instantaneous capacity will perform poorly.  The common element in our solution to engineering efficient routing and rate control algorithms for mobile wireless networks is to make the wireless mobile links seem as if they are wired or wireless links to all but few nodes that directly see the mobile links (either the mobiles or nodes that can transmit to or receive from the mobiles) through an appropriate use of queuing structures at these selected nodes.  This approach allows us to design end-to-end rate control or routing algorithms for wireless mobile networks so that neither averaging nor instantaneous tracking is necessary, as we have done in the following three networks.

A network where we can easily demonstrate the poor performance of a rate control algorithm based on either averaging or tracking is a simple wireless downlink network where a mobile node moves but stays within the coverage cell of a single base station.  In such a scenario, the time scale of the variations of the quality of the wireless channel between the mobile user and the base station can be such that the TCP-like congestion control algorithm at the source can not track the variation and is therefore unable to adjust the instantaneous coding rate at which the data stream can be encoded, i.e., the channel variation time scale is matched to the TCP round trip time scale.  On the other hand, setting the coding rate for the average case will still result in low throughput due to the high sensitivity of the TCP rate control algorithm to packet loss and the fact that below average channel conditions occur frequently.  In this dissertation, we will propose modifications to the TCP congestion control algorithm for this simple wireless mobile downlink network that will improve the throughput without the need for any tracking of the wireless channel.

Intermittently connected network (ICN) is another network where the classic assumption of time scale decomposition is no longer relevant.  An intermittently connected network is composed of multiple clusters of nodes that are geographically separated.  Each cluster is connected wirelessly internally, but inter-cluster communication between two nodes in different clusters must rely on mobile carrier nodes to transport data between clusters.  For instance, a mobile would make contact with a cluster and pick up data from that cluster, then move to a different cluster and drop off data into the second cluster.  On contact, a large amount of data can be transferred between a cluster and a mobile, but the time duration between successive mobile-cluster contacts can be relatively long.  In this network, an inter-cluster rate controller based on instantaneously tracking the mobile-cluster contacts can lead to under utilization of the network resources; if it is based on using long term average achievable rate of the mobile-cluster contacts, this can lead to large buffer requirements within the clusters.  We will design and analyze throughput optimal routing and rate control algorithm for ICNs with minimum delay based on a back-pressure algorithm that is neither based on averaging out or tracking the contacts.

The last type of network we study is networks with stationary nodes that are far apart from each other that rely on mobile nodes to communicate with each other.  Each mobile transport node can be on one of several fixed routes, and these mobiles drop off or pick up data to and from the stationaries that are on that route.  Each route has an associated cost that much be paid by the mobiles to be on (a longer route would have larger cost since it would require the mobile to expend more fuel) and stationaries pay different costs to have a packet picked up by the mobiles on different routes.  The challenge in this type of network is to design a distributed route selection algorithm for the mobiles and for the stationaries to stabilize the network and minimize the total network operation cost.  The sum cost minimization algorithm based on average source rates and mobility movement pattern would require global knowledge of the rates and movement pattern available at all stationaries and mobiles, rendering such algorithm centralized and weak in the presence of network disruptions.  Algorithms based on instantaneous contact, on the contrary, would make them impractical as the mobile-stationary contacts are extremely short and infrequent.


\tableofcontents   

\listoftables      
\listoffigures     

%
%
\chapter{Introduction}

Mobile communication networks have one essential problem that differentiates them from more traditional networks like the Internet or WiFi mesh networks.  Communication algorithms for those networks are engineered with the assumption that the network dynamics are either slow enough to be tracked or so fast that the algorithms would essentially see the time average.  For example, the round-trip time between any two computers in the Internet is now less than 50msecs.  This enables an end-to-end congestion controller like the Transport Control Protocol (TCP) to  detect network congestion quickly and adjust the transmission rate in response.  In this case, the TCP algorithm tracks the congestion inside the network and adjust the transmission rate accordingly.  On the other hand, in the based networks, the wireless channel between a base station and a mobile user fluctuates so rapidly over one data frame transmission time so that it can not be tracked; however, the average bit error rate (BER) in each frame is relatively static over multiple frames, and this allows channel coding algorithm with fixed coding rate to be used.  

The classic time-scale separation assumption of trackability or averaging no longer holds in mobile networks.  Consider a communication network used by soldiers deployed in remote terrains.  These soldiers have organized themselves into multiple, geographically separated clusters, and are equipped with wireless communication devices so that they may communicate with others in the same cluster.  However, because of the geographical separation and limited range of the wireless transceivers, they must rely on mobile data transporters to carry data between clusters.  When a mobile data transport comes into contact with a cluster, it can pick up a large amount of data per contact with that cluster; it can then move to the destination cluster, and drop off that data into that cluster.  If a rate control algorithm at the inter-cluster traffic source (inter-cluster traffic has the source and the destination in different clusters) adjusts the source rate by tracking the mobile-cluster contacts and in essence trying to track the instantaneous available inter-cluster communication data rate, then even though there is a temporary large increase in the available inter-cluster rate when the contact is made, there might not be enough rate available within the cluster through which the inter-cluster flow must travel; in effect, only small portion of the large instantaneous inter-cluster rate can be used.  

Alternatively, if the rate control algorithm tries to adjust the rates based on the ``average'' inter-cluster rates, then either there will be a large queue build up at every node (as we will demonstrate in the later chapter) or the algorithm runs into the problem of finding out what the ``average'' rate is, which makes it weak in the presence of changes and failures in the network.

In this dissertation, we design and analyze novel routing and rate control algorithms in mobile communication networks where tracking or averaging the network dynamics is impossible or leads to inefficient performance.  Towards that end, we examine three types of networks; we briefly introduce each type and highlight our contributions below.  

\section{Problem Statements and Contributions}

\subsection{TCP for Wireless Downlink Networks}

It is well-known that TCP connections perform poorly over wireless links due to channel fading. To combat this, techniques have been proposed where channel quality feedback is sent to the source, and the source utilizes coding techniques to adapt to the channel state. However, the round-trip timescales quite often are mismatched to the channel-change timescale, thus rendering these techniques to be ineffective in this
regime.  (By the time the feedback reaches the source, the channel state has changed.)

In this dissertation, we propose a source coding technique that when combined with a queuing strategy at the wireless router, eliminates the need for channel quality feedback to the source. We show that in a multi-path environment (e.g., the mobile is multi-homed to different wireless networks), the proposed scheme enables statistical multiplexing of resources, and thus increases TCP throughput dramatically.

\subsection{Time-Scale Decoupled Routing and Rate Control in Intermittently Connected Networks}

The second type of network we study in this dissertation is an intermittently connected network (ICN) composed of multiple clusters of wireless nodes. Within each cluster, nodes can communicate directly using the wireless links; however, these clusters are far away from each other such that direct communication between the clusters is impossible except through mobile contact nodes. These mobile contact nodes are
data carriers that shuffle between clusters and transport data from the source to the destination clusters.  Our dissertation here focuses on a queue-based cross-layer technique known as back-pressure algorithm. The algorithm is known to be throughput optimal, as well as resilient to disruptions in the network, making it an ideal candidate communication protocol for our intermittently connected network.

We design a back-pressure routing/rate control algorithm for ICNs. Though it is throughput optimal, the original back-pressure algorithm has several drawbacks when used in ICNs, including long end-to-end delays, large number of potential queues needed, and loss in throughput due to intermittency. We present a modified back-pressure algorithm that addresses these issues. 

\subsection{Efficient Data Transport with Mobile Carriers}

For the third type of network, we consider a network of stationary nodes that rely on mobile nodes to transport data between them. We assume the mobile nodes can control their mobility pattern to respond to data traffic loads, as well as satisfy some other secondary objectives, such as surveillance requirements. We study this problem in the framework of cost minimization, and we derive a dual iterative algorithm that results in optimal mobility pattern for minimizing network wide cost. 

\section{Organization}
Each subsequent chapter focuses one of the three network types above.  In Chapter 2, we study the problem of TCP congestion controller in cellular networks.  We briefly discuss the TCP background and present our modifications that will improve the TCP throughput using multi-homing.  We then present our simulation results.

Chapter 3 is on rate control and routing in intermittently connectedly network using a back-pressure algorithm.  We describe the back-pressure algorithm and highlight the benefits and shortcomings of the algorithm in ICN.  We then present our solutions and present our experimental results obtained from our test bed.

In Chapter 4, we present our mobility control algorithm that minimizes the network wide cost.  We present our network model and state our cost minimization optimization problem and iterative solution algorithm, followed by our experimental results.

We end this dissertation with a conclusion and a discussion on possible future research topics.

\chapter{TCP for Wireless Downlink Networks}
\section{Introduction}
\label{sec:intro_tcpnc}
The Transport Control Protocol (TCP)  is the most widely used congestion control protocol in the Internet.  When a router in the Internet is used beyond its capacity, its buffer will overflow and start to drop packets.  A source using TCP will interpret dropped packets as a signal of congestion, and it will promptly reduce its transmission rate to relieve the congestion.  

TCP was designed and optimized with the assumption that the networks that it was supposed to operate over have highly reliable node-to-node links such that dropped packets due to poor link quality are highly unlikely. Hence, a dropped packet meant only one thing -- congestion.  

\begin{figure}[h!]
\begin{center}
\subfigure[Average window size of plain TCP connection made over one wireless link with packet drop probability $d$ due to fading.  As $d$ increases, the average window size and the throughput drops rapidly.]{\includegraphics[width=0.48\textwidth]{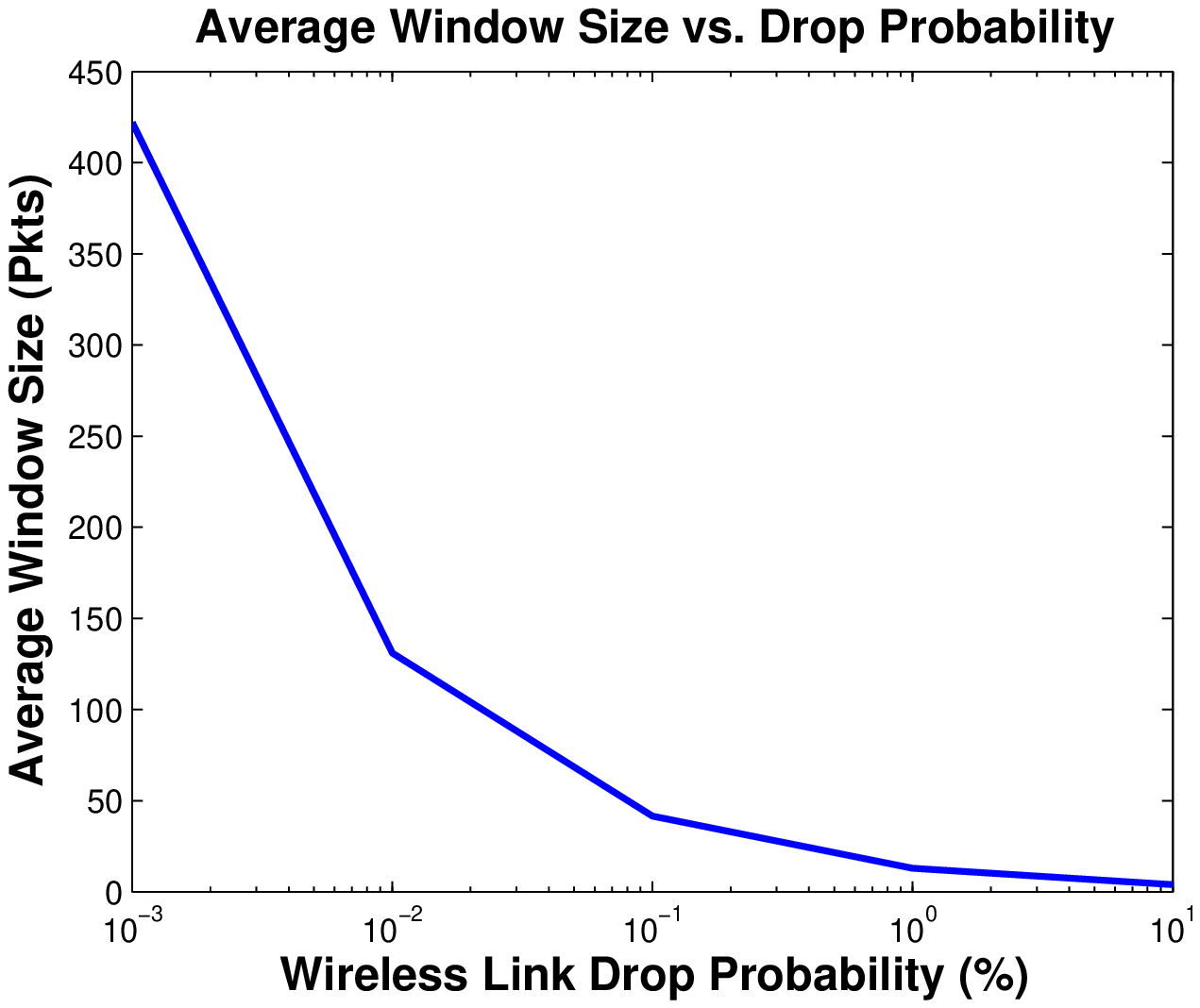}\label{f:avg_win1}}
\subfigure[Average window size of TCP using FEC at fixed coding rate as $p$ varies.  Even if the coding rate is sufficient for the average channel drop probability, if the drop probability changes every RTT, then the throughput will still be low.  The drop probability is equal to 0.05 with probability $p$, and is equal to 0.15 with probability $1-p$.]{\includegraphics[width=0.48\textwidth]{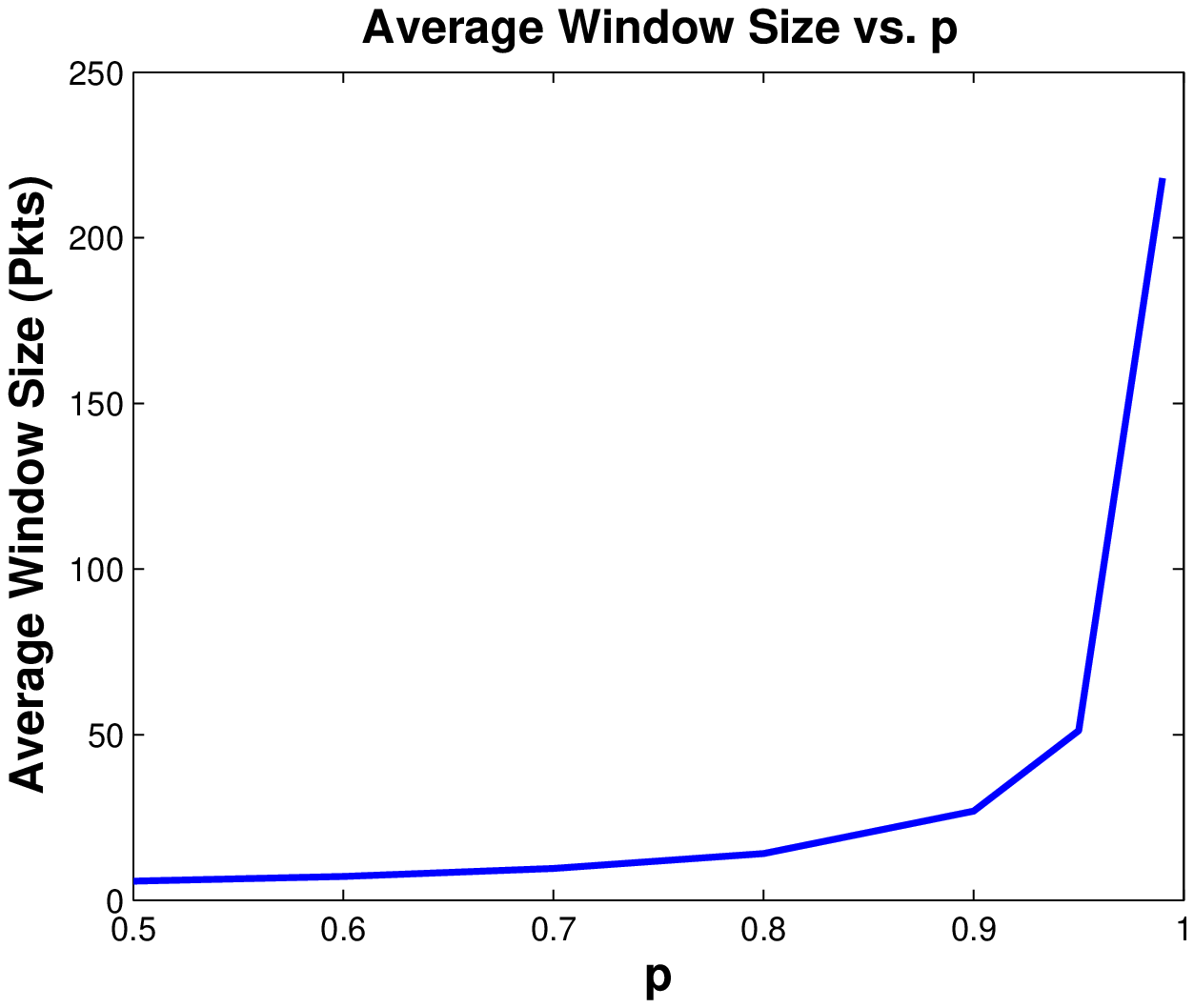}\label{f:avg_win2}}
\caption{Performance of TCP over one-hop wireless link}
\end{center}
\end{figure}

However, in wireless networks, TCP has no way of distinguishing congestion drops and drops due to poor quality wireless channel.  A typical wireless link is designed with average BER (Bit Error Rate) on the order of $10^{-5}$, which results in an average packet error/drop probability (PER) of 5-10\% assuming 1KB packet.  In addition, the average BER (and PER) of a wireless link can fluctuate over time, and the rate of fluctuation poses a significant problem for TCP.  If plain TCP is used over the wireless links without any modifications, this considerably reduces TCP's average congestion window size and prevents it from enlarging the window size to any significant portion of the ideal size, the bandwidth-delay product, resulting in a low utilization rate \cite{tianxuans05,lakmad97,balapadkatz97}.  Take for example, a plain TCP (Reno) connection made over one-hop wireless link.  Suppose the bandwidth-delay product is infinite for this connection.  However, the packet drop probability due to bad wireless channel is $d$.  In Figure \ref{f:avg_win1}, we plot the average window size of this TCP connection as $d$ varies from 0.001\% to 10\%.  As the figure shows, the average window size (and the throughput) decreases substantially as $d$ increases.  A similar performance graph is shown in Figure 5 of \cite{tianxuans05}.

\begin{figure}[t!]
\begin{center}
\includegraphics[scale=0.7]{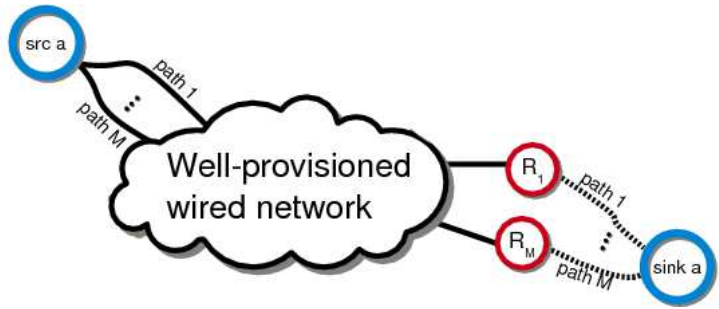}
\caption{A multipath TCP-RLC connection made over a well-provisioned wired network and $M$ wireless routers.  We exploit multi-path wireless channel diversity to increase throughput.  The dashed lines represent wireless links.}\label{f:net_fig}
\end{center}
\end{figure}

In this chapter, we address the problem of low TCP throughput in the simple topology of TCP senders connected via wireline network to intermediate
wireless routers and TCP receivers connected by a wireless channel to multiple intermediate routers (see Figure \ref{f:net_fig}).  An example scenario would be a cellular access network (such as UMTS/WiMax) where the cellular base station is connected to the wired backbone, and only the link between the base station and the mobile user is wireless. Although multi-homing is not currently implemented in current cellular networks, with the introduction of femto-cells, it is conceivable that in a campus scenario with a number of femto-cells, the mobile user may be able to receive downlink data simultaneously on multiple links from multiple femto-cells; this motivates the multi-path model in Figure \ref{f:net_fig}.

\subsection{Shortcomings of Existing Solutions}
To combat the adverse nature of the wireless network, multiple
solutions have been proposed, all involving a separation
of time-scales between the rate of channel variation and the
TCP congestion window evolution. One can break the TCP
connection between a wired server and a mobile into two
components: wired and wireless \cite{BPSK97}. However, this approach
needs a proxy at the wireless base-station, and breaks TCP
end-to-end semantics.

By contrast, one could protect TCP (without proxying at
the wireless router) from channel-level variations by suitable
physical layer schemes. Of these, the commonly deployed
solution in UMTS/WiMax systems involves channel coding,
adaptive modulation and/or automatic repeat request (ARQ
or hybrid ARQ) deployed in a lower layer protocol to deal
with packet drops resulting from channel variations that are
at a much faster rate than the end-to-end TCP round-trip time
(RTT). However, these schemes could lead to variations in the
rate provided to the TCP connections, and can lead to suboptimal
TCP performance  \cite{tcp_2000}. Papers such as \cite{ghsrzatocr06} improve
TCP performance over downlink wireless networks through
dynamically adjusting PHY layer parameters optimized for
TCP. However, such strategies require measurements both at
the transport layer like TCP sending rate as well as physical
layer information like channel quality at the cost of increased complexity at the cellular base station.

The alternate solution (see \cite{bengoetow02,LT_TCP,sun_tcp}) is to code the
data stream at a specific forward error coding rate at the
application layer so that the decoded TCP data stream can
withstand drops due to bad wireless channels. In \cite{LT_TCP}, the
authors use Reed-Solomon coding at a fixed rate to encode a
stream of TCP packets in order to deal with random losses.
In \cite{sun_tcp}, the authors use network coding combined with an
ACK scheme found in \cite {arq_nc} and TCP-Vegas like throughput
measurements to adapt TCP over wireless links. However,
such an approach requires the variation in channel drop rate
to be quasi-static relative to the time-scale of feeding back
this channel drop rate information to the source so that the
coding rate can be adjusted.

\subsection{Motivation}
\subsubsection{Channel Variation and RTT Have Same Time-Scale}
In many realistic settings, the packet drop rate of the wireless channel can change at the time scale of round-trip time of the TCP connection. For example, consider a mobile user traveling at 2-5km/h using the current UMTS network (carrier frequency $\approx$ 2GHz in the U.S.). This user's wireless channel coherence time is roughly around 20-50ms \cite{rapp}, a number well within the range of RTT for the Internet.  (Coherence time is roughly a measure of how long a wireless channel stays constant, and therefore a rough measure of how fast the packet drop rate changes.)  

In such scenarios, multiple ARQ requests and link-level
ACKs and NACKs are unhelpful -- they cause retransmission
delays and timeouts that may adversely affect the RTT estimation
and the retransmission time-out (RTO) mechanism,
and therefore throughput. Moreover, forward error correction
coding at a fixed rate (at the TCP source) is not helpful
since the drop rate at the wireless link is not quasi-static
relative to the feedback time scale. If the drop rate changes
every RTT, the information about the drop rate will not reach
the TCP sender in time to be useful since by the time the
information reaches the sender, the drop rate would have
changed. Thus, this mobile user’s wireless downlink channel
would be useless to track from the perspective of improving
his TCP throughput; the channel quality feedback reaches the
source too late, and is useless by the time the source gets it.

Take for example a TCP connection made over one wireless link.  The packet drop probability due to bad channel conditions changes every RTT time period.  Assume that the drop probability $d \in \{0.05, 0.15\}$, with the probability that $d=0.05$ is $p$ and the probability that $d=0.15$ is $1-p$.  Suppose that the TCP flow uses forward error correction coding at the rate of 10\%, i.e. for every 10 data packets, 1 coded packet is generated.  Using this FEC coding rate, the TCP data packets can reliably be delivered to the destination \textit{if the drop probability $d$ is 0.05}.  However, because the drop probability can be bad ($d=0.15$) often enough, the TCP throughput will be significantly small.  In Figure \ref{f:avg_win2}, we plot the average window size of the TCP connection as we vary $p$.  (We assumed that the bandwidth-delay product is infinite.)  As the graph shows, even though we have used FEC coding rate that is sufficient for the \textit{average} drop probability, the throughput is still low because the drop probability changes every RTT.

\textbf{In summary, coding at fixed rate will not work when the wireless channel variation and TCP RTT have the same time-scale.}
\subsubsection{Multiple Path Statistical Multiplexing}
There has been
much research into multipath TCP connections. The obvious
advantage of multipath TCP is that it can balance the load on
the multiple paths such that paths experiencing temporarily
high capacity can carry more packets than paths experiencing
low capacity. Further, multiple TCP connections can be useful
for load balancing among multiple wireless interfaces -- for
instance, in a situation where a mobile node is connected to a
3G network base-stations as well as a femto-cell base-station.
In this scenario, one would want to get statistical multiplexing
gain among the two wireless interfaces, as it is likely that
the wireless fading state between the two interfaces will
differ (e.g., when the 3G interface has a “bad” channel, the
femto-cell interface could have a “good” channel). However
to exploit this, a naive implementation would require that
packets stored at the 3G base-station be transferred to the
other base-station (femto) through a wired back-haul. Clearly,
time-scales of RTT over the back-haul and channel variation
would render this impractical.

A second issue one faces when running a TCP connection
over multiple paths is the problem of out-of-order packet
delivery, which can cause congestion window collapse even if
the network has plenty of capacity \cite{ye_m_tcp}. \cite{horizon} gets around this
problem by delaying and reordering received packets before
they are passed up to the TCP layer on the receive side. \cite{DSACK}
uses duplicate selective ACKs (DSACK) and dynamically
changes the duplicate ACK threshold to address the out-of-order
problem.

By using random linear coding, our proposed TCP modifications
can be naturally extended to multiple paths, and we
will show that coding + TCP enables the network to behave
as though packets are “virtually shared” among the different
base-stations without the need for a back-haul between
the various base-stations. This in-turn leads to multiplexing
gains. Further, coding + TCP can easily deal with out-of-order
delivery of packets.

%

\subsection{Other Related Work}
\noindent \textbf{Other coding approaches:} Recently, inspired by \cite{nc,xor_katabi} and others,
network coding schemes have been used in the context of
wireless networks in order to improve throughput. \cite{kumar_tcp_nc} and
\cite{piggycode} use network coding at intermediate nodes and exploit
the shared wireless spectrum to improve TCP throughput. In
our approach, we use random linear coding (RLC) \cite{lt_codes} at the
end nodes to improve TCP throughput, and the intermediate
router does not perform any coding operations. The concept
of using random linear coding for TCP over wired networks
has appeared recently \cite{jungle}; however, our work here is for
hybrid network with the goal of improving TCP throughput
over time-varying wireless channels.

\noindent \textbf{TCP window statistics under AQM:} There is a considerable body of literature~\cite{holmis01,peerapolmak} on modeling the TCP window process in the presence of active queue management (AQM) systems, especially random early detection (RED) \cite{flojac95}. \cite{peerapolmak} presents a weak limit of the window size process by proving a weak convergence of triangular arrays. \cite{bacmcdrey02} presents a fluid limit of the TCP window process, as the number of concurrent flows sharing a link goes to infinity, and the authors show that the deterministic limiting system provides a good approximation for the average queue size and total throughput.  None of the previous works mentioned above treats the situation when the loss rate can not be tracked due mismatch between the channel change time-scale and the RTT time-scale.  

This chapter shows throughput gains that can be achieved by multi-path diversity when TCP is combined with an ACK scheme similar to the one in \cite{sun_tcp} and priority queuing strategy found in \cite{cvq}, plus RLC \cite{lt_codes}.

\subsection{Main Contributions}
In this work, we employ (i) random linear coding, (ii) priority-based queuing at wireless routers and (iii) multi-path routing to demonstrate that throughput can be increased significantly for TCP over downlink wireless networks even when channel variations are on the same time-scale as RTT. {\em Our theoretical result shows that we can obtain full statistical multiplexing gain from multi-path TCP.}

Specifically, our analysis shows that we can achieve TCP throughput of $\Theta\left(\mathbb{E}\left[P\right]C\right)$ in multiple path (multi-homing) case with our modifications, in the absence of channel quality feedback from the destination to the sender.  Here, $\mathbb{E}\left[P\right]$ is the mean probability of successful packet transmission of the time-varying wireless channel; $C$ is the capacity of the wireless router.  

Further, our modifications to TCP, which we call TCP-RLC, present an orderwise gain over the performance of plain TCP, which is $\Theta(1)$, in the presence of random packet loss for wired-wireless hybrid networks, where the random packet loss rates change at the RTT time scale and cannot be tracked.

\section{Analytical Model}\label{sec:sysmodel_tcpnc}
We consider slotted time.  Each time slot is equal to round-trip time between the senders and the receivers.  The TCP-RLC source $i$ maintains a congestion window of size $W_i(t)$ for the $t$-th RTT interval. The congestion window is in units of packets; packets are assumed to be of fixed size.  In each RTT slot, source $i$ wants $W_i(t)$ data packets to be transferred to sink $i$.  We model the additive increase, multiplicative decrease (AIMD) evolution of the TCP congestion window as follows:
\begin{eqnarray}
W_i(t+1) &=& \mathbf 1_{\text{success}}\left(W_i(t) + 1\right) + \mathbf 1_{\text{drop}}\left\lceil W_i(t)/2\right\rceil\label{eq:winevolsingle} \\
&=& W_i(t) + 1 - \mathbf 1_{\text{drop}}\left\lfloor 1 + W_i(t)/2\right\rfloor \nonumber
\end{eqnarray}
where the random variable $\mathbf 1_{\text{success}} \defeq 1$ when all data packets transmitted in the $t$-th RTT interval for destination $i$ have been successfully received at the destination; $\mathbf 1_{\text{drop}} \defeq 1 - \mathbf 1_{\text{success}}$ takes the value $1$ when either (i) the receiver cannot recover one or more data packets corrupted by the packet drop process or (ii) a packet is marked by the router due to the presence of an \emph {active queue manager} (AQM).  AQM marks a connection with window size $W$ according to the probability given by $f(W)$, and thus the congestion window is deliberately halved. 

In our analytical model, we assume that the routers can measure the window size $W$ of a TCP flow, and based on the window size, each router can mark the flow with probability $f(W)$.  If a flow is marked by a router, the TCP source halves the TCP window size $W(t)$ and reduces the transmission rate by factor of two.  

We let $f(W)$ be the probability that a flow is marked when the window size is $W$, and we let $f_{\text{chan}}(W)$ denote the probability that the destination is unable to reconstruct the $W$ data packets due the channel packet drop/corruption processes over all paths, which would halve the TCP congestion window.  

The combined effects of the AQM with the marking function $f(w)$ and the packet drops from the wireless channels can be encapsulated into $f_{\text{eff}}(w)$ where
\begin{eqnarray}
f_{\text{eff}}(w) &=& 1-(1-f_{\text{eff}}(w)) \nonumber \\
 &=& 1 - (1-f(w))(1-f_{\text{chan}}(w)) \nonumber \\
&=& f(w) + f_{\text{chan}}(w) - f(w)f_{\text{chan}}(w).
\end{eqnarray}
Thus, the TCP congestion window will decrease by half or increase by one with probabilities $f_{\text{eff}}(w)$ and $1-f_{\text{eff}}(w)$, respectively, if the window size is $w$.

In this chapter, we are interested in finding the average throughput under the optimal AQM, i.e. $\mathbb{E}\left[W^*(t)\right]$, where $W^*(t)$ is the congestion window under the AQM that maximizes $\mathbb{E}\left[W^*(t)\right]$.  Note that the optimal AQM may be no AQM at all, but throughput under the optimal AQM has to be greater than that under some arbitrary AQM.  The presence of AQM greatly simplifies our analysis.  We later back our claims with simulation results that used no AQM. 

In practical scenarios, the evolution of the congestion window size is limited by the sender/receiver buffer size;  we will ignore this to simplify our analysis.  We will also neglect TCP timeouts for the same reason.

\subsection{Random Linear Coding}
In each time slot, the source $i$ takes $W_i(t)$ data packets and generates $r W_i(t)$ ($r>0$) coded packets as follows: let each data packet $x_{ik}, k = 1, 2, \ldots, W_i(t)$ be represented as an element of some finite field $\FR$; choose elements $\alpha_{ikj} \in \FR$ uniformly at random and generate a coded packet
\begin{equation*}
y_{ij} = \sum_{k = 1}^{W_i(t)} \alpha_{ikj} x_{ik} 
\end{equation*}
for $j=1, 2, \ldots, rW_i(t)$. The receiver can decode, with very high probability, any dropped data packets if sufficient number of linearly independent coded and data packets are received, as the field size from which the coding coefficients are drawn increases.  Hence, in the rest of the work, we will make the following assumption as a simplification. 
\begin{assumption}\label{ass:bigfield}
Suppose $W$ data packets are used to generate coded packets via RLC.  If $G$ coded packets are received by the TCP destination, then upto $G$ missing data packets out of the $W$ data packets can be recovered.
\end{assumption}

Thus, if the number of missing data packets from $W$ exceeds $G$ in an RTT slot, the congestion window will halve (and $\mathbf 1_{\text{drop}}=1$).  That is, if $G$ coded packets received and as long as no more than $G$ data packets of the original $W$ data packets are lost, then the receiver can recover all of the original data.  For detailed exposition on RLC and justification of assumption \ref{ass:bigfield}, see \cite{lun_thesis}.

\subsection{Network Topology}\label{sec:singleflow_model}
The network topology we consider in this chapter is a TCP-RLC connection made over $M\geq 1$ paths going through router $R_1$,..., $R_M$, each with capacity $C$ (black connection in Figure \ref{f:net_fig}).  Only the link between $R_i$, $i=1,...,M$ and the destination is wireless; the links between the routers and the source are wired.  The well-provisioned wired network is assumed with the paths having the same RTT, and the wireless routers do not store packets from one RTT slot to another in the analytical mode.  In practice, the paths will have different, but similar, RTT values.

We assume that the wired section of the network has greater capacity than the wireless section.  This is a reasonable assumption, since all of the currently existing downlink networks have wired back-plane network with excess capacity, and the wireless downlink channel is greatly limited due to tight spectrum and power constraints.

\subsection{Wireless Downlink Channel}
We model packet drops in the wireless channel between a wireless router and the TCP-RLC destination as a simple i.i.d. packet drop process whose parameter remains constant for each \emph{RTT-interval}. This is similar to the block-noise model common in wireless communication literature.  

Within each RTT-interval $t$, the probability that a packet transmitted over the air by the wireless router $i\in \{1,...,M\}$ for the destination is successfully received is given by $p_i(t)$.  The $j$-th packet transmitted over the air by the wireless router $i$ for the destination is corrupted (dropped) according to a Bernoulli error process $H_j^i(t)$ defined as
\begin{equation*}
H_j^i(t) = \left \{
\begin{array}{ll}
1 & \textnormal{w.p. $p_i(t)$ if } j^{\textnormal{th}} \textnormal{ packet is received correctly} \\
0 & \textnormal{w.p. $1-p_i(t)$ if } j^{\textnormal{th}} \textnormal{ packet dropped}
\end{array}
\right. \label{eq:H_def}
\end{equation*}
with parameter $p_i(t)\in \{ p_1=p_{\min},p_2,\ldots,p_{\Pi}=p_{\max}\}$ acting upon each packet over the air independently of other packets in the same RTT-interval.  We assume that $p_1>0$ and $p_1 < p_2 < \hdots < p_{\Pi}$.  We assume that $p_i(t)$ changes over time with $\mathbb{P}(p_i(t) = p_k) = \tilde{p}_k$. Thus, the channel packet-delivery-probability parameter itself changes with time (this corresponds to changing fading state over time), and at any time the actual packet delivery probability depends on the instantaneous value of this (random, time-varying) parameter. 

Lastly, we assume that $r> 2(1-p_{\min})/p_{\min}$ for multiple paths topology.


\subsection{Priority Transmission}\label{subsec:priority_tx}

In the multiple path network topology, the source maintains a path-level congestion window $w_l(t)$ for each path $l$; this is in addition to $W(t)$.  The idea in our multi-path algorithm is to have a rate control on each path.  The transmission rate of the high priority packets on path $l$ is controlled by $w_l(t)$. $w_l(t)$ will be the number of high priority (data) packets in transit towards the destination in time slot $t$; in addition, there will be $r w_l(t)$ low priority (coded) packets that will be in transit as well.  If $w_l(t)$ packets (either high or low priority) are successfully received by the destination, the rate controller on path $l$ at the source will increase $w_l$ by one so that in time slot $t+1$, $w_l(t+1) = w_l(t) + 1$.  If not, then $w_l$ will be halved, so that $w_l(t+1) = w_l(t)/2$.  

The main idea in our algorithm is to separate the problem of wireless link reliability on each path from the TCP rate control algorithm that operates over all paths ($W(t)$).  On each path, the only sign of congestion is if the total number of packets (high + low priority) received by the destination on path $l$ is less than the number of high priority packets it should have received on that path; that is, the router on path $l$ did not have enough capacity to either 1) send all high priority packets it received or 2) send enough low priority coded packets to compensate for any high priority packets that were dropped by the wireless link.  Unlike in the of the traditional TCP over wired links, gaps in the sequence of the received packet numbers are not a valid way of measuring congestion over wireless links. The mechanism that acts on $w_l(t)$ we described in the previous paragraph is a way to ``measure'' congestion on each path; $w_l(t)$ is the number of packets that path $l$ can guarantee to deliver in time slot $t$ to the destination through the wireless router $l$.

Thus, in time slot $t$, path $l$ can have any $w_l(t)$ of the $(1+r)W(t)$ data and coded packets delivered to the receiver.  The evolution of $w_l(t)$ is similar to $W(t)$:
\begin{eqnarray}
w_l(t+1) &=& \mathbf 1_{\text{success}}\left(w_l(t) + 1\right) + \mathbf 1_{\text{fail}}\left\lceil w_l(t)/2\right\rceil\label{eq:winevolmult}
\end{eqnarray}
where $1_{\text{success}}=1$ if $w_l(t)$ packets are successfully received in RTT slot $t$; $1_{\text{fail}}=1-1_{\text{success}}$.

Before the packets are sent out on a path, they are marked either high or low priority.  The number of high priority packets sent out in any given RTT slot is equal to $w_l(t)$.  For each high priority packet sent out, the path transmits $r$ low priority packets.  

Note that as long as $W(t)$ data packets are successfully received (or recovered), $W(t)$ will increase according to eq. (\ref{eq:winevolsingle}).  Since the available rates on the $M$ paths ($w_1(t)$, $w_2(t)$, ..., $w_M(t)$) evolve independently, we need a way to measure the total available instantaneous rate through all paths combined, and the main purpose of having $W(t)$ is to measure this total, combined rate through all paths.  The algorithm at the source is designed such that in time slot $t$, the main rate controller will take $W(t)$ data packets, generate $rW(t)$ coded packets, and leave both types of packets in some memory location.  Path-wise rate controller on path $l$ will take $w_l(t)$ data packets, mark them high priority, and send them out along with $rw_l(t)$ coded packets, which it will mark low priority.  If in time slot $t$, $W(t)$ data packets are successfully reconstructed at the destination, the main controller will deduce that the total available rate through all paths is equal to or greater than $W(t)$, and hence $W(t+1) = W(t) +1$.  Else, it will deduce the total available rate is less, and $W(t+1) = W(t)/2$.

\subsection{Wireless Router}
We assume that the low priority packets are transmitted by the router only when there are no high priority packets that can be transmitted.  For example, in the case $M=1$, the remainder of the nominal channel capacity $C$, $C-w_1(t)$, is used to transmit low priority packets in $t$-th RTT slot, where the evolution of $w_1(t)$ is defined in eq. (\ref{eq:winevolmult}).  

We also make the assumption that the wireless router maintains a pair of queues for each TCP-RLC connection made through that router.  While such a number that scales with the number of flows would be prohibitive for routers in the Internet core, we argue that it is reasonable for wireless downlink routers, as these routers serve relatively small number of mobile users in the same cell.  In addition, per flow queue maintenance is already done in cellular architectures for reasons of scheduling, etc. (see \cite{bender}).

For each TCP-RLC connection, one queue (FIFO) is used to handle high priority packets; the other queue (LIFO) is used for low priority packets.  (LIFO queue is used for low priority packets because the low priority packets are transmitted only when the router has excess capacity.  The router will run out of capacity often enough so that if FIFO is used for low priority packets, then after long enough time, it will be backed up with old low priority packets and any newly arriving low priority packets will be dropped due to (low priority) buffer overflow.)  In our analysis, we assumed that packets are not stored in the router's queues from one RTT time slot to another.  We relax this in our simulations.  

\section{Multiple Path Analysis} \label{sec:singleflow}
We first analyze the evolution of path-level congestion window.  Given that the path-level window size is $w_l(t)$ for path $l$ in time slot $t$, $w_l(t+1) = \lceil w_l(t)/2 \rceil$ if out of the $C$ packets transmitted by the router, fewer than $w_l(t)$ packets with the same block number are received; let $\chi_{\text{fail}}$ be this event.  

Using Chernoff's bound, we can show for any $\epsilon > 0$,
\begin{eqnarray*}
\mathbb{P}\left[\chi_{\text{fail}} | p_l(t) = p_1,~w_l(t) < (1-\epsilon)p_1 C \right] \leq \exp(-\epsilon^2 p_1 C/2 )
\end{eqnarray*}
Similarly, we can show that 
\begin{eqnarray*}
\mathbb{P}\left[\chi_{\text{fail}} | p_l(t) = p_1,~w_l(t) > (1+\epsilon)p_1 C \right]\geq 1 - \exp(-\epsilon^2 p_1 C/2 ).
\end{eqnarray*}
In addition, it is straight forward to show that for any $i=2,...,\Pi$,
\begin{eqnarray*}
\mathbb{P}\left[\chi_{\text{fail}} | p_l(t) = p_i,~w_l(t) < (1-\epsilon)p_1 C \right] \leq \mathbb{P}\left[\chi_{\text{fail}} | p_l(t) = p_1,~w_l(t) < (1-\epsilon)p_1 C \right]
\end{eqnarray*}
since $p_1 < p_i$.  Thus, 
\begin{eqnarray}\label{eq:wl_up}
\mathbb{P}\left[\chi_{\text{fail}} | w_l(t) < (1-\epsilon)p_1 C \right] & \leq & \sum_{i=1}^{\Pi} \tilde p_i \exp(-\epsilon^2 p_1 C/2 )\nonumber \\
& = & \exp(-\epsilon^2 p_1 C/2 )
\end{eqnarray}
and 
\begin{eqnarray}\label{eq:wl_down}
\mathbb{P}\left[\chi_{\text{fail}} | w_l(t) > (1+\epsilon)p_1 C \right] & \geq& \mathbb{P}\left[\chi_{\text{fail}} | p_l(t) = p_1,~w_l(t) > (1+\epsilon)p_1 C \right]\nonumber\\
&=&  \tilde p_1 (1-\exp(-\epsilon^2 p_1 C/2 )).
\end{eqnarray}

From eqs. (\ref{eq:wl_up}), (\ref{eq:wl_down}) and (\ref{eq:winevolmult}), we see that $0.5(1-\epsilon)p_1 C \leq w_l(t) \leq (1-\epsilon)p_1C$ w.p. at least $1-\exp(-\epsilon^2 p_1 C)$.  Since $(1+r)w_l(t)$ high and low priority packets are sent on path $l$ and $r>2(1-p_1)/ p_1$, the wireless router on path $l$ sends $C$ in each time slot w.p. at least $1-\exp(-\epsilon^2 p_1C)$.  

In a network with $M$ paths, the probability that $0.5(1-\epsilon)p_1 C \leq w_l(t) \leq (1-\epsilon)p_1C$, $\forall l=1,...,M$ is at least $(1-\exp(-\epsilon^2 p_1 C))^M \geq 1 - M\exp(-\epsilon^2 p_1 C)$ where $C>>M$.  Thus, we make the following reasonable assumption:

\begin{assumption}\label{as:path_assmpt}
On each path $l=1,...,M$, the wireless router transmits $C$ (high + low priority) packets to the TCP-RLC destination in each time slot.
\end{assumption}

From assumption \ref{as:path_assmpt}, we have that $W(t)$ is an irreducible, aperiodic Markov chain; $W(t+1)$ depends only on $W(t)$ and $p_l(t)$, $l=1,...,M$.  Let $\pi_{W}$ denote its stationary distribution.

Let $\chi_{\textnormal{failure}}' = \left\{\sum_{k=1}^{M}\sum_{j=1}^{C} H_j^{k}(t) < W(t)\right\}$.  Fix $\rho > 1$ and let $\Upsilon$ denote the event $\left\{\frac{1}{M}\sum_{i=1}^M p_i(t) \geq \frac{\rho W(t)}{MC}\right\}$.  Then 
\begin{eqnarray*}
f_{\textnormal{chan}}(W(t)) &=& \mathbb{P}\left[\Upsilon\right] \mathbb{P}\left[\chi_{\textnormal{failure}}' | \Upsilon\right]+\mathbb{P}\left[\neg \Upsilon\right] \mathbb{P}\left[\chi_{\textnormal{failure}}' |\neg \Upsilon\right].
\end{eqnarray*}

\begin{lemma}\label{lem:mp_f_chan_up_b}
$f_{\textnormal{chan}}(w)$ can be bounded as below 
\begin{eqnarray*}
\label{eq:multi_ub}f_{\textnormal{chan}}(w) \leq \left \{
\begin{array}{ll}
\exp\left(-MCL_1\right) \text{ if } 0 \leq \frac{\rho^2w}{MC} \leq p_1& \\
\exp\left(-MCL_2\right) + \exp\left(-M l'\left(1 - \frac{\rho w}{MC} \right)\right)\text{ if } p_1 < \frac{\rho^2w}{MC} < \mathbb{E}\left[P\right]
\end{array}
\right. \label{eq:f_chan_up_b}
\end{eqnarray*}
where $L_1$ and $L_2$ are non-zero constants, and 
\begin{eqnarray}l'(a) = \max_{-\infty < \Theta < \infty }\left\{\Theta a - \log M'(\Theta)\right\}\label{eq:lp_def_2}\end{eqnarray} 
and $M'(\Theta) = \mathbb{E}\left[e^{\Theta (1-P)}\right]$.
\end{lemma}

\noindent \emph{Proof: }
Let $\widehat{p}(t) = \frac{1}{M}\sum_{i=1}^{M} p_i(t)$.  If $p_1 \leq \frac{\rho^2 W(t)}{MC} < \mathbb{E}\left[P\right]$, by Chernoff's bound  
\begin{eqnarray*}
\mathbb{P}\left[\neg \Upsilon \right]&=& \mathbb{P}\left[\frac{1}{M}\sum_{i=1}^{M} \left(1-p(i,t)\right) > 1 - \frac{\rho W(t)}{MC} \right]\nonumber\\
 &\leq& \exp\left(-M l'\left(1 - \frac{\rho W(t)}{MC} \right)\right)\label{eq:b11}.
\end{eqnarray*} 

Then, 
\begin{eqnarray}
\mathbb{P}\left[\chi_{\textnormal{failure}}' | \Upsilon \right] & \leq & \mathbb{P}\left[\chi_{\textnormal{failure}}' |\widehat{p}(t) = \frac{\rho W(t)}{MC}\right] \label{eq:b12} \\
& \leq & e^{\left(-MC D\left(1- \frac{ W(t)}{MC}||1- \frac{\rho W(t)}{MC}\right)\right)} \label{eq:b13}
\end{eqnarray}
where $D(x||y)$ is the Kullback-Leibler distance between $x$ and $y$\footnote{$D(x||y) = x\log(x/y) + (1-x)\log((1-x)/(1-y))$.}; eq. (\ref{eq:b12}) follows because lowering $\widehat{p}(t)$ will increase the probability that the transmission will not be successful, and eq. (\ref{eq:b13}) by Chernoff's bound. 

Let $L_1 = \inf_{W:p_1 \leq \frac{\rho^2 W}{MC}\leq \mathbb{E}\left[P\right]} D\left(1 - \frac{W}{MC}||1 - \frac{\rho W}{MC}\right)$.  Thus, if $p_1 \leq \frac{\rho^2 W(t)}{MC} < \mathbb{E}\left[P\right]$, we have 
\begin{eqnarray*}
f_{\textnormal{chan}}(W) \leq \exp\left(-Ml'\left(1 - \frac{\rho W}{MC}\right)\right) + \exp\left(-MCL_1\right)\label{eq:b20}
\end{eqnarray*} 
since $\mathbb{P}\left[\Upsilon\right],\mathbb{P}\left[\chi_{\textnormal{failure}}' |\neg \Upsilon\right]\leq 1$.

If $\frac{\rho^2 W(t)}{MC} < p_1$, we have  $\mathbb{P}\left[\Upsilon\right]=1$ because $\hat p(t) \geq p_1$.  In addition, 
\begin{eqnarray}
\mathbb{P}\left[\chi_{\text{failure}}' | \Upsilon \right] &\leq& \mathbb{P}\left[\chi_{\text{failure}}' | \widehat{p}(t)  = p_1 \right]\label{eq:b15}\\
&\leq&e^{\left(-MC D \left(1-\frac{W(t)}{MC}|| 1 - p_1\right)\right)}. \label{eq:b16}
\end{eqnarray} 
Inequality (\ref{eq:b15}) follows from the fact that decreasing the probability that $H_j^i=1$ will increase the probability that not enough packets will be received by the receiver to decode $w$ packets.  Inequality (\ref{eq:b16}) follows from Chernoff bound.  Let $L_2 = \inf_{w: \frac{\rho^2 w}{MC}  \in [0, p_1)} D\left( 1- \frac{w}{MC}|| 1-p_1\right)$.  Then 
\begin{eqnarray*} 
\mathbb{P}\left[\chi_{\text{failure}}'| \Upsilon \right] \leq \exp\left(-MCL_2\right),\label{eq:b5}
\end{eqnarray*} 
and 
\begin{eqnarray*}
f_{\textnormal{chan}}(W) \leq \exp\left(-MCL_2\right) \label{eq:b18}
\end{eqnarray*} 
in the region $\frac{\rho^2 W(t)}{MC} < p_1$ since $\mathbb{P}\left[\neg \Upsilon\right]=0$
\done

\begin{theorem}\label{th:mp_thruput}
Fix $\rho>1$ and $M$.  Then, 
\begin{eqnarray*}
\mathbb{E}\left[W^*(t)\right]&\geq& \max\left\{ 0.75p_1MC/\rho^2-1, \min\left\{\frac{\mathbb{E}\left[P\right]MC}{\rho^2}\left[1 - 3\delta_1 - 2\left(e^{-1}+\delta_2\right)\right],\right.\right.\\
&&\left.\left. \beta \left[1 - 3\lfloor \beta \rfloor^{-1} - 2\left(e^{-1}+\delta_2\right)\right] \right\}\right\}
\end{eqnarray*}
where $\delta_1,\delta_2 \rightarrow 0$ as $C\rightarrow\infty$ and where $\beta \in [p_1MC/\rho^2,\mathbb{E}\left[P\right]MC/\rho^2]$ is the solution to the equality 
\begin{eqnarray}
1/\beta = \exp\left(-Ml'(1-\rho \beta/MC)\right) \label{eq:beta_eq}
\end{eqnarray}
(for $M$, $C$ large enough, we can show a solution exists) and where $l'(\cdot)$ is defined in equation (\ref{eq:lp_def_2}).\label{lem:mp_down}
\end{theorem}

\noindent\emph{Proof: }If $\exists \beta \in [p_1MC/\rho^2, \mathbb{E}\left[P\right]MC/\rho^2]$ such that $\beta$ is the solution to eq. (\ref{eq:beta_eq}), then this case corresponds to the situation when there are enough paths to start gaining path diversity, but not enough paths to gain complete path diversity.  (We assume $C$ is large enough so that $l'(1-\mathbb{E}\left[P\right]/\rho)<CL$.)  

Let 
\begin{eqnarray}f(w) = \left\{
\begin{array}{ll}
\frac{2(\beta)^{-1} - f_{\textnormal{chan}}(w)}{1-f_{\textnormal{chan}}(w)}&\mbox{if $w < \lfloor \beta\rfloor $}  \\
1&\mbox{if $w \geq \lfloor \beta\rfloor $}
\end{array}
\right.\label{eq:AQM_eq}\end{eqnarray} 
so that $f_{\text{eff}}(w) = 2\beta^{-1}$ if $w <\lfloor\beta\rfloor$ and 1 if $w \geq \lfloor \beta \rfloor $.  Under this AQM marking scheme, 
\begin{eqnarray*}
\mathbb{E}\left[W(t+1)\right] &=& \mathbb{E}\left[W(t) + 1 \right]  - \mathbb{E}\left[1_{\text{drop}}\lfloor 1 + W(t)/2 \rfloor \right]\nonumber\\
1 &=&  \mathbb{E}\left[f_{\text{eff}}(W) \lfloor 1 + W(t)/2 \rfloor \right]\nonumber\\
1 &\leq&  \mathbb{E}\left[f_{\text{eff}}(W) (1 + W(t)/2) \right]\nonumber\\
1 &\leq&  \mathbb{E}\left[f_{\text{eff}}(W)\right] + \mathbb{E}\left[f_{\text{eff}}(W)W/2\right].\label{eq:multi_marking_bound}
\end{eqnarray*}

Thus,
\begin{eqnarray*}\label{eq:yy}
\mathbb{E}\left[f_{\text{eff}}(W)\right] = 2\beta^{-1} + (1-2\beta^{-1})\pi_{W}\left(\lfloor \beta \rfloor \right)
\end{eqnarray*} 
and 
\begin{eqnarray}\label{eq:zz}
\mathbb{E}\left[f_{\text{eff}}(W)\frac{W}{2}\right] =\beta^{-1}\mathbb{E}\left[W\right] + (1-2\beta^{-1})\frac{\lfloor \beta\rfloor }{2}\pi_{W}\left(\lfloor \beta \rfloor \right).
\end{eqnarray}

Let $S_k(t)=1$ if $W(t)=k$ and 0 if $W(t)\neq k$.  For $k>\lfloor \lfloor \beta \rfloor / 2 \rfloor $, we have 
\begin{eqnarray*}
\mathbb{P}\left[S_k(t)\right] &=& \mathbb{P}\left[S_{k-1}(t-1)\right]\mathbb{P}\left[W(t)=k|W(t-1)=k-1\right] \\
&=&  (1-2\beta^{-1})\mathbb{P}\left[S_{k-1}(t-1)\right].
\end{eqnarray*} 

Thus, 
\begin{eqnarray*}
\mathbb{P}\left[S_{\lfloor \beta \rfloor}\left(t+\lceil \lfloor \beta \rfloor/2\rceil\right)\right]&=& \mathbb{P}\left[S_{\lfloor \lfloor \beta \rfloor /2 \rfloor - 1}(t)\right]\prod_{k=\lfloor \lfloor \beta \rfloor /2 \rfloor }^{\lfloor \beta \rfloor} \mathbb{P} \left[W=k|W=k-1\right]\\
&=&  (1-2(\beta)^{-1})^{\lceil \lfloor \beta\rfloor/2\rceil }\mathbb{P}\left[S_{\lfloor \lfloor \beta \rfloor /2 \rfloor - 1}(t)\right].
\end{eqnarray*}

Let $T_{ \lfloor \lfloor \beta \rfloor /2 \rfloor - 1}$ be the time takes $W(t)$ to return to $\lfloor \lfloor \beta \rfloor /2 \rfloor - 1$.  It is known that 
\begin{eqnarray*}
\mathbb{P}\left[S_{\lfloor \lfloor \beta \rfloor /2 \rfloor - 1}(t)\right] = \mathbb{E}_{\lfloor \lfloor \beta \rfloor /2 \rfloor - 1}\left[T_{\lfloor \lfloor \beta \rfloor /2 \rfloor - 1}\right]^{-1}.
\end{eqnarray*}
	
If starting from state $\lfloor \lfloor \beta \rfloor /2 \rfloor - 1$, the window size increases by $k\geq 0$ before halving, the total number of steps before $W(t)$ returns to $\lfloor \lfloor \beta \rfloor /2 \rfloor - 1$ is at least $k+\lfloor \lfloor \beta \rfloor /2 \rfloor - 1 - \lceil ( \lfloor \lfloor \beta \rfloor /2 \rfloor - 1 + k)/2\rceil$.  Minimizing over $k$, the least number of steps before $W(t)$ enters state $\lfloor \lfloor \beta \rfloor /2 \rfloor - 1$ again is at least $\lfloor(\lfloor \lfloor \beta \rfloor /2 \rfloor - 1)/2\rfloor$.  Thus, $\mathbb{E}_{\lfloor \lfloor \beta \rfloor /2 \rfloor - 1}\left[T_{\lfloor \lfloor \beta \rfloor /2 \rfloor - 1}\right] > \lfloor(\lfloor \lfloor \beta \rfloor /2 \rfloor - 1)/2\rfloor$ and 
\begin{eqnarray*}
\mathbb{P}\left[S_{\lfloor \lfloor \beta \rfloor /2 \rfloor - 1}(t)\right] \leq \left(\lfloor(\lfloor \lfloor \beta \rfloor /2 \rfloor - 1)/2\rfloor\right)^{-1}.
\end{eqnarray*} 

So we have 
\begin{eqnarray}\pi_W\left(\lfloor \beta \rfloor \right)\leq \frac{(1-2\beta^{-1})^{\lceil \lfloor \beta \rfloor/2\rceil }}{\left(\lfloor(\lfloor \lfloor \beta \rfloor /2 \rfloor - 1)/2\rfloor\right) }\label{eq:pi_w_up}
\end{eqnarray} 

Combining above with equation (\ref{eq:yy}), we get 
\begin{eqnarray}
\mathbb{E}\left[f_{\text{eff}}(W)\right] &\leq& \beta^{-1} +\frac{(1-2\beta^{-1})^{\lceil \lfloor \beta\rfloor/2\rceil +1}}{\left(\lfloor(\lfloor \lfloor \beta\rfloor /2 \rfloor - 1)/2\rfloor\right) }\notag\\
&\leq& \beta^{-1} +\frac{1}{\left(\lfloor(\lfloor \lfloor \beta\rfloor /2 \rfloor - 1)/2\rfloor\right)}\notag\\
&\leq& 3\left(\lfloor \beta \rfloor\right)^{-1} .\notag
\end{eqnarray}  

Combining equation (\ref{eq:zz}) with equation (\ref{eq:pi_w_up}), we get 
\begin{eqnarray*}
\mathbb{E}\left[f_{\text{eff}}(W)\frac{W}{2}\right] &=&\beta^{-1}\mathbb{E}\left[W\right] + (1-2\beta^{-1})\frac{\lfloor \beta \rfloor }{2}\pi_{W}\left(\lfloor \beta \rfloor \right)\\
&\leq&\beta^{-1}\mathbb{E}\left[W\right] + \frac{\lfloor \beta \rfloor }{2}\frac{(1-2\beta^{-1})^{\lceil \lfloor \beta \rfloor/2\rceil +1}}{\left(\lfloor(\lfloor \lfloor \beta \rfloor /2 \rfloor - 1)/2\rfloor\right) }
\end{eqnarray*}

Since $(1-2\beta^{-1})^{\lceil \lfloor \beta \rfloor/2\rceil +1}$ converges to $e^{-1}$, we have 
\begin{eqnarray*}
\mathbb{E}\left[f_{\text{eff}}(W)\frac{W}{2}\right]\leq \beta^{-1}\mathbb{E}\left[W\right] + 2(e^{-1} + \delta)
\end{eqnarray*}
where $\delta$ is some small, positive constant.  Combining the bounds on $\mathbb{E}\left[f_{\text{eff}}(W)\frac{W}{2}\right]$ and $\mathbb{E}\left[f_{\text{eff}}(W)\right]$ and substituting into inequality (\ref{eq:multi_marking_bound}), we get 
\begin{eqnarray*}
1 \leq 3(\lfloor \beta \rfloor)^{-1}  + \beta^{-1}\mathbb{E}\left[W\right] + 2(e^{-1}+\delta).
\end{eqnarray*}
Solving for $\mathbb{E}[W]$ and using $\mathbb{E}\left[W^*(t)\right]\geq \mathbb{E}\left[W(t)\right]$, we get 
\begin{eqnarray*}
\beta\left[1 - 3\lfloor \beta\rfloor^{-1} - 2\left(e^{-1}+\delta_2\right)\right] \leq \mathbb{E}[W^*].
\end{eqnarray*}

If $\nexists \beta \in [p_1MC/\rho^2, \mathbb{E}\left[P\right]MC/\rho^2]$ such that $1/\beta = \exp\left(-Ml'(1-\rho \beta/MC)\right)$, then either $1/\beta < \exp\left(-Ml'(1-\rho \beta/MC)\right)$ for $\beta = p_1MC/\rho^2$, or $1/\beta > \exp\left(-Ml'(1-\rho \beta/MC)\right)$ for $\beta = \mathbb{E}\left[P\right]MC/\rho^2$.

If $1/\beta < \exp\left(-Ml'(1-\rho \beta/MC)\right)$ for $\beta = p_1MC/\rho^2$, then there are not enough paths to gain path diversity.  We can use no AQM (i.e., $f(W) = 0, \forall W$), and we will be guaranteed throughput at least $0.75p_1C/\rho^2-1$ as $C\rightarrow \infty$.

If $1/\beta > \exp\left(-Ml'(1-\rho \beta/MC)\right)$ for $\beta = \mathbb{E}\left[P\right]MC/\rho^2$, then there are sufficient number of paths to gain all path diversity. 

Let $\beta = \left(\frac{\mathbb{E}\left[P\right]}{\rho^2}MC\right)$ and let 
\begin{eqnarray*}
f(w) = \left\{
\begin{array}{ll}
\frac{2\beta^{-1} - f_{\textnormal{chan}}(w)}{1-f_{\textnormal{chan}}(w)}&\mbox{if $w < \beta$}  \\
1&\mbox{if $w \geq \beta$,}
\end{array}
\right.\end{eqnarray*}
so that $f_{\text{eff}}(w) = \beta^{-1}$ if $w<\beta$ and 1 else.  We can follow the same analysis as in the case when the solution to eq. (\ref{eq:beta_eq}) exists (just let $\beta$ be equal to $\mathbb{E}\left[P\right]MC/\rho^2$ and the result will follow through) and arrive at the conclusion that 
\begin{eqnarray*}
\mathbb{E}\left[W^*(t)\right]\geq \frac{\mathbb{E}\left[P\right]MC}{\rho^2}\left[1 - 3\lfloor \beta\rfloor^{-1} - 2\left(e^{-1}+\delta_2\right)\right].
\end{eqnarray*}
\done

Due to the fact that $w_l(t) \leq p_1 C$ and $r > 2(1-p_1)/p_1$, the capacity of the wired section only need to be twice that of the wireless section; i.e. the wired capacity needs to be slightly greater than $2 C$.

\section{Simulation Results}\label{sec:sim_results}
Before we present our simulation results, we note that implementation of TCP-RLC uses novel techniques not found in original version of TCP.  We provide brief description of some of the novel techniques in the appendix; we refer any readers interested in the implementation details to the appendix.

We simulate single flow with multiple paths, where we exploit path diversity.  In our simulations, the random coefficients used to encode each packet are drawn from a field of size 8191; thus, for coding block size of smaller than $\sim500$, the probability of two coded packets in the same coding block not being ``independent'' is negligible.  We fixed the size of all packets to 256 bytes.  Each flow had two buffers allocated at the router whose capacities were equal to the transmission rate times the RTT.  We have used bimodal channel profile, $\left\{(p_{\min},\widetilde{p}_{\min}),(p_{\max},\widetilde{p}_{\max})\right\}$.  $p_{\max}$ is set to 1 in all our simulations.  We varied $p_1$ from 0.1 to 0.5 in increments of 0.1 and set $\widetilde{p}_1 = 0.1$ and $\widetilde{p}_2 =0.9$.  Thus, $\mathbb{E}\left[P\right]$ varied from 0.91 to 0.95. This corresponds to the scenario where the downlink channel can be controlled to provide good capacity most of the time (90\% of the time), but bad channel conditions can occur frequently enough to destroy TCP throughput (10\% of the time).  Note that using fixed coding rate adjusted for the average channel quality would not work for this scenario; coding is not needed most of the time and becomes useless when needed because the coding rate is insufficient against poor quality channel conditions that can occur frequently enough.  We varied the number of paths from 1 to 8.  We set the total capacity to 1Mbps, so that per path capacity is 1Mbps/$M$, where $M$ is the number of paths in our simulation.  In our simulations, all paths have the same bimodal channel profile and have the same capacity.    
(Our simulation scenario corresponds to the situation when the quality of the wireless channels can be controlled (for example, by increasing transmission power temporarily) upto certain degree to be good enough most of the time, but there are occasions when the channels are so degraded that nothing can be done.)

Each time the wireless channel changed, it stayed constant for some random time according to the uniform distribution with parameters 100ms to 200ms; on average, the channels stayed constant for 150ms.  RTT for each path was drawn randomly from uniform distribution [100ms, 200ms].  Time-out clock is set to expire after 3 $\times$measured RTT.  RTT was measured using IIR filtering: measured RTT = 0.9 $\times$ old measured RTT + 0.1 $\times$ new measured RTT. The redundancy factor $r$ was such that $(1+r)/p_{\min}\approx 2$, with $r$ being an integer.  We assume perfect uplink channel from the mobiles to the wireless routers for the end-to-end TCP ACK's.

We used no AQM; the presence of AQM were assumed mainly to simplify our analysis.  Theorem \ref{th:mp_thruput} says that the TCP throughput achieved by using the AQM function in Eq. (\ref{eq:AQM_eq}) provides a lower bound on the optimal TCP throughput.  When the AQM function in Eq. (\ref{eq:AQM_eq}) is used, the packet drop probability is deliberately increased (compared to when no AQM is used).  Though we have no mathematical proof that TCP throughput achieved by not deliberately increasing the drop probability is larger than that achieved by deliberately doing so, we believe that using the AQM function in Eq. (\ref{eq:AQM_eq}) and deliberately increasing the drop probability only decreases the throughput.

\begin{figure*}[t!]
\begin{center}
\includegraphics[width=1\textwidth]{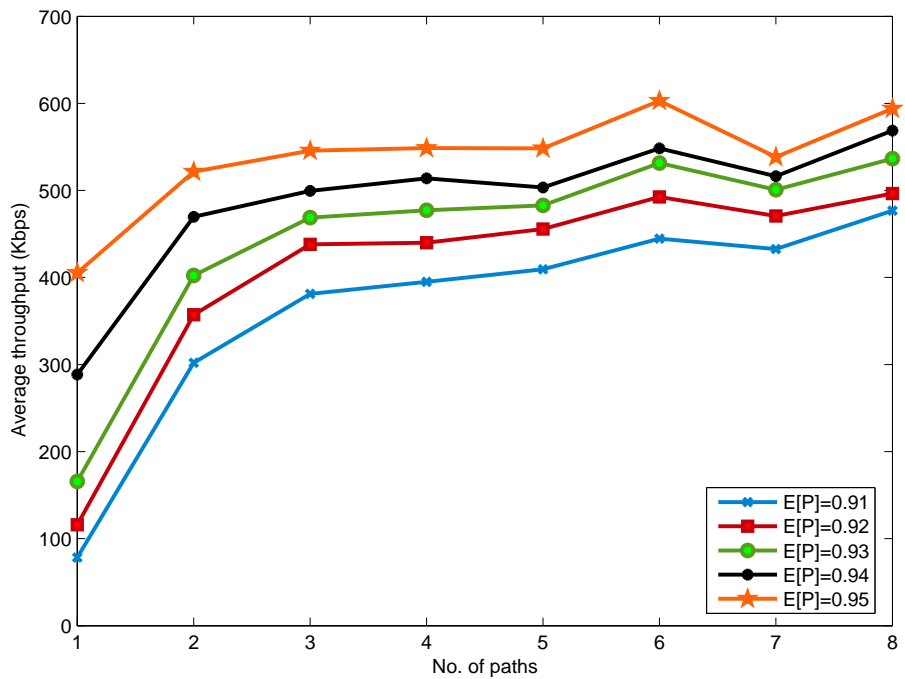}
\caption{Simulation results: Average throughput vs. no. of paths for multiple path TCP-RLC connection.  Note that the throughput of $M$ separate traditional TCP connections (whether over wired or wireless downlink networks) does not increase with the number of paths.  In addition, TCP throughput over wireless downlink network with fixed FEC coding rate is $\Theta(1)$.  Thus, the throughput of the traditional TCP over wireless downlink network where the channel fluctuation changes at the time scale of the RTT is low and does not increase with the number of paths.}\label{f:mp_thru}
\end{center}
\end{figure*}

The average throughputs we obtained are shown in Figure \ref{f:mp_thru} as a function of the number of paths.  As the number of paths increases, the average throughput increases towards $\mathbb{E}\left[P\right]CM$ ($CM$ is fixed) in a concave manner, indicating that going from one path to two paths gives much gain in throughput, especially when $p_{\min}$ is small.  The throughput should reach 700Kbps (for $\mathbb{E}\left[P\right]=0.91$) to 730Kbps (for $\mathbb{E}\left[P\right]=0.95$).   

%



\section{Appendix}

\subsection{TCP-RLC Protocol}
In this subsection, we give a brief description of NS-2 simulation implementation details.  We break the description of TCP-RLC into three subsections.
\subsubsection{Source Architecture}
\paragraph{ACK and pseudo ACK}
TCP-RLC uses two types of ACK's: ACK, as used in the plain TCP and pseudo ACK, which we describe here.  

Plain ACK's cumulatively acknowledge the reception of all packets with
frame numbers smaller than or equal to the ACK.

In our context, a lost data packet can be ``made up'' by a future
coded packet, and we would like the sliding congestion window to slide
forward and have delay-bandwidth product worth of packets in transit.
Thus, we use a strategy similar to that in \cite{sun_tcp}, where
degrees of freedom are ACKed.  In our context, we refer to this as a
pseudo ACK, which simply ACKs any out of order data packet or coded
packet that helps in decoding the smallest-index missing packet (e.g.,
if the sink has received packets 1, 2, 3, and 7, the smallest-index
missing packet is 4).  Note that with regular TCP, out-of-order
packets would trigger duplicates ACK's that would lead to a loss of
throughput.

To summarize: ACK -- cumulatively acknowledges in-order packet
arrivals; pseudo ACK -- acknowledges out-of-order data packet/coded
packet arrivals that help in decoding the missing packets.

\paragraph{Sliding congestion window}
\begin{figure}
\begin{center}
\includegraphics[scale=0.8]{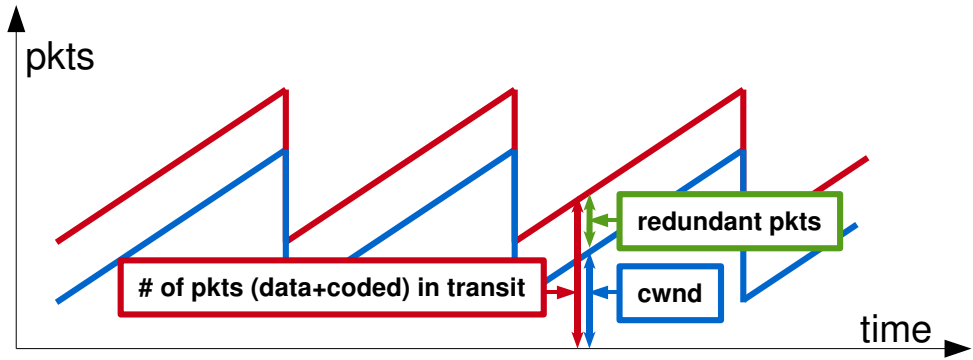}
\caption{Evolution of congestion window: the total number of packets in transit is composed of cwnd high priority packets (data packets) + redundant packets (RLC encoded packets).  The number of redundant packet is some multiple of cwnd, and has to be greater than $p_1$.  Note that without priority transmission at the wireless router, this will reduce the throughput.}\label{f:cwnd}
\end{center}
\end{figure}
The source maintains a congestion window $W$, which is the same as the size of the coding block for TCP-RLC.  All packets transmitted are marked either high or low priority; the source allows only $W$ high priority packets to be in transit.   Each time a high priority packet is transmitted, it transmits $r$ low priority packets as well, where $r\geq1/p_{\min} - 1$.

The source maintains variables, last-ACK and SN.  All packets with frame numbers lower than last-ACK are assumed to have been successfully received.  SN is the frame number of the starting packet in the coding block currently being transmitted.  If pseudo ACK or ACK arrives ``acknowledging'' the reception of SN, a new coding block is encoded and readied for transmission, with the coding block size being $W+1$ packets.  This is because acknowledgment of packet number SN implies a RTT has been elapsed, which is enough time for $W$ packets to have been successfully received and decoded by the sink.

\paragraph{Multiple paths}
When multiple paths are used, the marking of packets is done by the
paths independently.  Each path is maintained by a path controller and
the controller maintains a congestion window, $cwnd_i$; another
top-level controller maintains $cwnd$, which is used as the size of
the coding block.  After the packets are encoded, they are passed to path controllers (thus, coded
packets are ``mixed'' across paths, which in-turn leads to statistical
multiplexing across paths).  Each path marks packets either high or
low priority.  In each RTT slot, the number of high priority packets
in transit is equal to $cwnd_i$. The number of low priority packets
(coded packets) per path is equal to $r \times cwnd_i.$ Each packet
going out on a path contains the block number and block size, which is
also equal to $cwnd_i$.  In each RTT slot, if the sink on path $i$
receives $cwnd_i$ packets (either high or low priority), $cwnd_i$
increases by one; else $cwnd_i$ reduces by half.  Note that single
path is just a special case of multiple paths; and the variables
$cwnd$ and $cwnd_1$ are the same. (In the single-path case, the role
played by the separate path controllers is subsumed into the
source controller.)

There are two levels of ACK's; one level for source controller (source
level ACK, consisting of ACK and p-ACK) and the other for path
controllers (path level ACK). Note that: {\em (i)} path level ACK is a
new ACK introduced for multi-path TCP -- there is no equivalent in the
single-path case, and {\em (ii)} the three types of ACKs described here
is abstracted into a single indicator function for success/drop in the
analysis (see eq. (\ref{eq:winevolsingle}) in section \ref{sec:sysmodel_tcpnc}). Source controller level ACK's
affect $cwnd$ and moves the coding window; path controller level ACK's
affect $cwnd_i$'s and moves the block numbers.

\subsubsection{Destination Architecture}
Upon reception of a packet, the destination examines if it is the next
expected (i.e., smallest-index missing) packet.  If it is, the
destination sends an ACK cumulatively acknowledging all packets upto
and including the just received packet.  If not, the destination sees
if the packet is an innovative packet that can be used to decode the
next expected.  (A packet is innovative if it is linearly independent
of all packets received so far, i.e. it can help in decoding the next
expected packet.  For a complete definition of innovative packets, see
\cite{lun_thesis} and \cite{holun08}.) In case the packet is helpful, the
destination sends a pseudo ACK with next expected packet number +
total number of innovative packets accumulated that can help in
decoding the next expected packet.  If the packet does not help in
decoding the next expected packet, the destination sends a duplicate
ACK.

\subsubsection{Wireless Router Architecture}
\begin{figure}[h!]
\begin{center}
\subfigure[]{\includegraphics[width=0.60\textwidth]{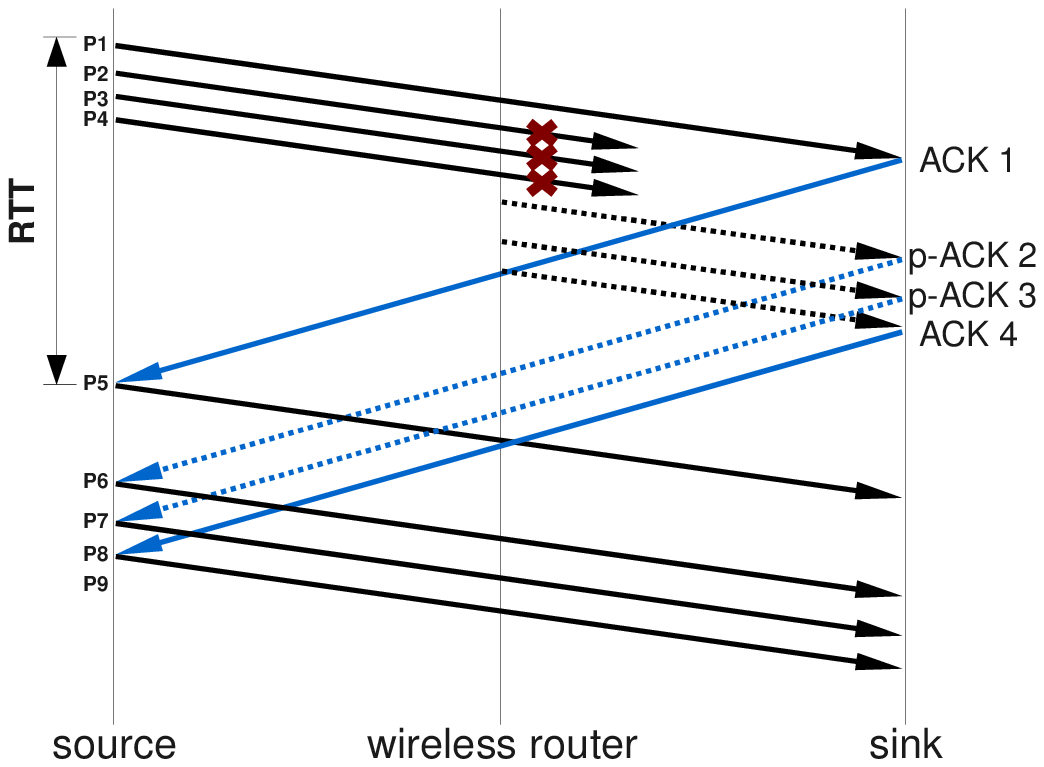}\label{f:cwnd_fig_1}}\\
\subfigure[]{\includegraphics[width=0.60\textwidth]{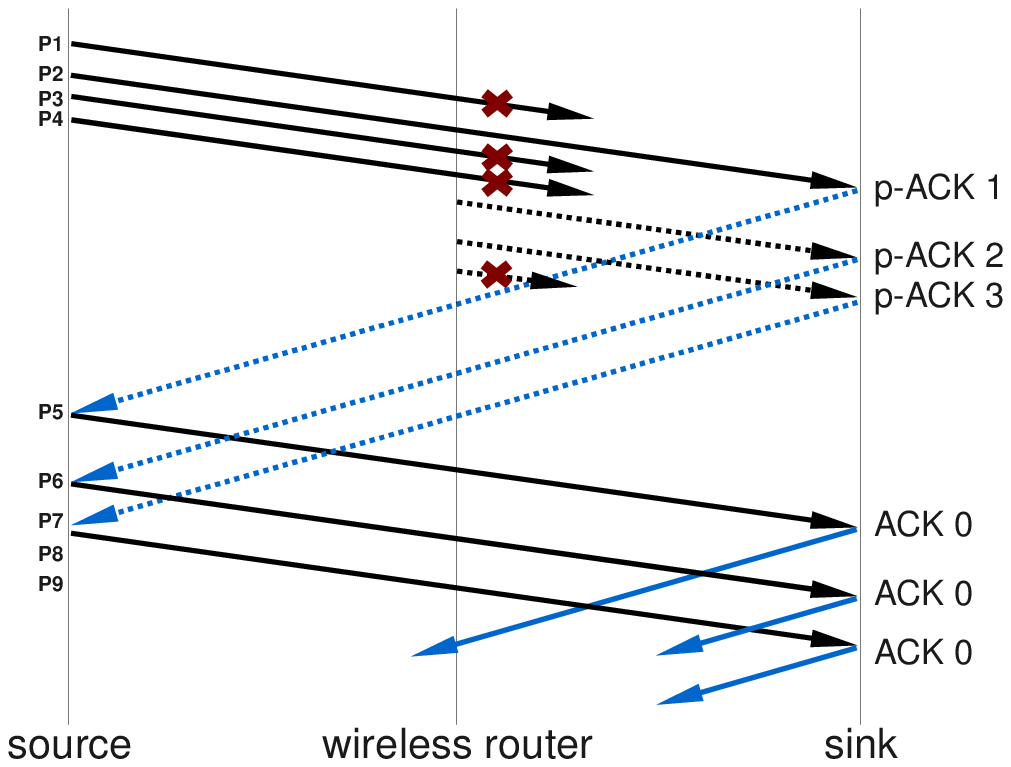}\label{f:cwnd_fig_2}}\\
\subfigure[]{\includegraphics[width=0.60\textwidth]{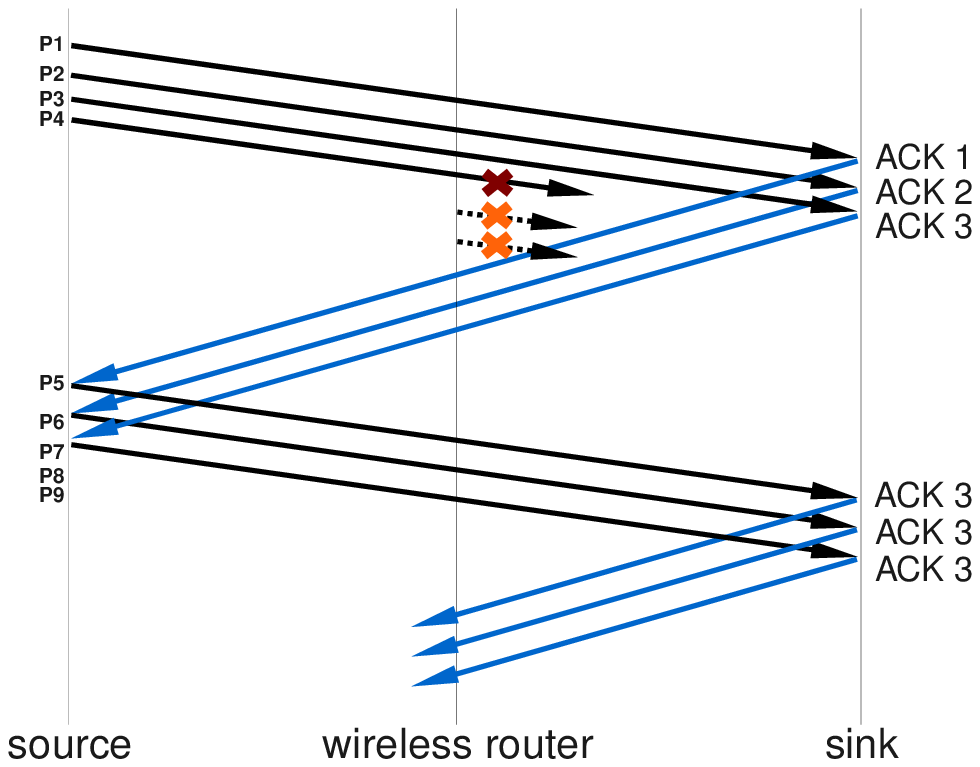}\label{f:cwnd_fig_3}}\\
\subfigure{\includegraphics[scale=0.5]{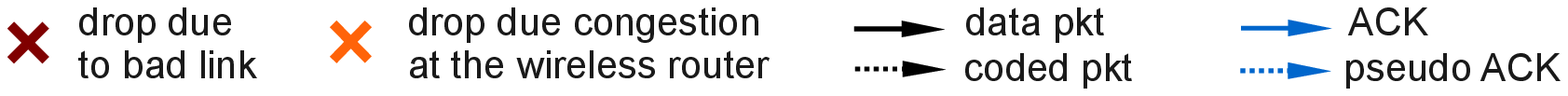}}
\caption{Illustrative example}
\end{center}
\end{figure}


As mentioned, the wireless router maintains two buffers for each flow.  A FIFO buffer is maintained for high priority packets; a LIFO buffer is maintained for low priority packets.  The FIFO buffer does not need to be large, but large enough to handle packet processing execution.  The LIFO buffer needs to be large enough to handle RTT worth of packets. In our implementation, we do not have an explicit AQM mechanism at the router (other than tail-drop); however, even without AQM we observe large performance gains through simulations.

\subsubsection{Illustrative Example}
We illustrate the additive increase, multiplicative decrease component of TCP-RLC and the pseudo ACK's using three examples.  The examples are for when a TCP connection uses a single path, and the data and coded packets are marked high and low priority, respectively.  Although the TCP source uses retransmission time-outs for when there is no response from the sink for long period of time, we do not show this in our examples.
\begin{enumerate}
 \item Additive increase: When there is no congestion at the wireless router and the sink receives enough data and coded packets in a coding block, the sink will be able to recover any missing data packets.  In Figure \ref{f:cwnd_fig_1}, the four packets P1-P4 are encoded together.  While P1 is successfully received by the sink, P2-P4 are dropped/lost due to bad wireless channel.  However, the sink has received enough coded packets to recover packet P2-P4.  As the sink receives these coded packets, it sends out a pseudo ACK for each one.  This enables the source to move the congestion window forward, keeping the ``pipe'' between the source and the sink full.  When the sink recover P2-P4, it sends out an ACK acknowledging the successful reception of packets upto P4.  Note that the congestion window is increased by one packet, and thus new packets P5-P9 are encoded together.
 \item Multiplicative decrease (bad wireless channel): When too many packets that belong in the same coding block are dropped due to bad wireless channel, duplicate ACK's will be triggered when packets that do not help in decoding the next expected packet.  In Figure \ref{f:cwnd_fig_2}, data packet P1, P3 and P4 are dropped and not enough coded packets arrive due to bad wireless channel.  When packets P5-P7 arrive at the sink, duplicate ACK's are sent out and the source will cut the congestion window.  Note that when p-ACK 1 is received by the source, a new coding block P5-P9 is encoded and passed to path controllers.  However, when the duplicate ACK's are received, the source decreases the congestion window by half to two packets, and restarts transmission starting from P1.  
 \item Multiplicative decrease (congestion at wireless router):  Where there is a congestion at the wireless router, the router will be busy transmitting data (high priority) packets, and no or very few coded (low priority) packets will be transmitted.  In Figure \ref{f:cwnd_fig_3}, packet P1-P3 are successfully received, but no coded packets are received to help the sink recover the packet P4, which has been lost due to then bad wireless channel condition.  Thus, the sink will send out duplicate ACK's and the source will cut the congestion window and restart transmission starting from P4.
\end{enumerate}

\chapter{Time-Scale Decoupled Routing and Rate Control in Intermittently Connected Networks}

\section{Introduction}
There recently has been much interest in intermittently connected networks (ICNs). Practical use of such networks include military scenarios in which geographically separated clusters of soldiers are deployed in a battlefield.  Each cluster is connected wirelessly internally, but the clusters of soldiers rely on unmanned aerial vehicles to transport battlefield information between clusters.  In such a case,  building a communication network can take too much time as these combat units must be deployed rapidly, and wireless connections are often susceptible to enemy jamming or the communication/RF ranges might not be large enough.

Another scenario of interest is a sensor network composed of multiple clusters, with each cluster containing low power sensor nodes.  The data collected by the sensor nodes must either be transmitted to a data fusion node or to another cluster.  To do this, the sensor nodes rely on mobile nodes that provide inter-cluster connectivity; thus resulting in a network with multiple time-scales and intermittent connectivity.

\begin{figure}
  \centering
    \includegraphics[width=0.75\textwidth]{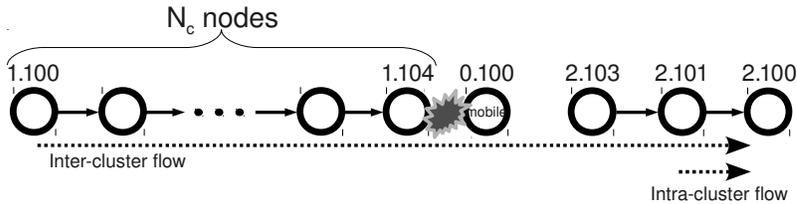}   
  \caption{Our ICN consists of two clusters and one mobile node connecting these clusters.  We have one intra-cluster flow from 2.101 to 2.100, and one inter-cluster from flow 1.100 to 2.100, which must rely on the mobile to transport data from the left to the right cluster.  }\label{fig:test_network}
\end{figure}

{In this dissertation, we consider a network of clusters of nodes connected via ``mobile'' nodes (see Figure~\ref{fig:test_network} for an example).  Internally, each cluster has many nodes connected via a (multi-hop) wireless network.  Each cluster has at least one \textit{gateway node}. (We will call the other nodes in the cluster \textit{internal nodes}.) These gateways are the designated representatives of the clusters, and they are the only ones able to communicate with the mobiles -- traffic from one cluster to another cluster (inter-cluster traffic) must be funneled through the gateways, both in the source cluster and in the destination cluster. The mobiles and gateways exchange packets (pick-ups and drop-offs) on contact. Each contact is made over a high capacity link and is long enough for a large quantity of data to be exchanged. The mobiles then move between clusters, and on contact with a gateway in the destination cluster, packet drop-offs are made.}

A key challenge in the network above is the fact that \textit{intermittently connected networks have several time-scales of link variability}.  For instance, wireless communication between soldiers within the same cluster is likely to occur at a time-scale several orders of magnitude faster than communication across clusters (which needs to use the mobile carriers). In this context, there are essentially two time-scales: \textit{(a)} within a cluster, where wireless links are formed in an order of tens of milliseconds, and \textit{(b)} across clusters where the  time-scale could be tens of seconds, to minutes. To communicate from one node to another node in the same cluster poses no significant  problem -- one can use existing protocols such as TCP. However, for two nodes in two different clusters to communicate, they must use the mobile nodes, as these mobiles move between clusters to physically transport data. Hence, the mobile communication time scale is many, many times greater than the electronic communication time scale. Any communication protocol that relies on fast feed-back (in the order of milliseconds to tens/hundreds of milliseconds) incurs severe performance degradation. The mobiles may be able to transport a large quantity of data (of the order of mega or giga bytes) in one ``move,'' but to move from one cluster in one part of a network to another part still takes time (of the order of seconds to minutes).

The design and development of communication protocols for intermittently connected networks, therefore, must start with an algorithm with as few assumptions about the underlying network structure as possible.  The back-pressure (BP) routing algorithm \cite{TasEph92} was introduced nearly two decades ago by Tassiulas and Ephremides with only modest assumptions about the stability of links, their ``anytime'' availability, or feasibility of fast feed-back mechanism; yet remarkably, it is throughput optimal {(throughput performance achieved using any other routing   algorithm can be obtained using the back-pressure algorithm \cite{TasEph92})} as well as resilient to changes in the network.  The BP routing algorithm is a dynamic routing and scheduling algorithm for queuing networks based on congestion gradients.  The congestion gradients are computed using the differences in queue lengths at neighboring nodes (the routing part).  Then, the back-pressure algorithm activates the links so as to maximize the sum link weights of the activated links, where the link weights are set to the congestion gradient (the scheduling part).  Over the years, there has been continued effort to further develop back-pressure type algorithms to include congestion control and to deal with state-space explosion and delay characteristics \cite{NeelyThesis,Sto05,ErySri06,NeeModLi05,AkyAndGupHobSanSto08,YinSriTow07,YinShaRed09,MoeSriKriGna10,WarJanHaRhe09,AthBuiJi10}.

However, the traditional back-pressure algorithm is impractical in intermittently connected networks, even though it is throughput optimal.   This is because the delay performance and the buffer requirement of the back-pressure algorithm in the heterogeneous connectivity setting of the ICNs increase with the product of the network size and the time scale of the intermittent connections -- i.e., the larger the network or more intermittent and sporadic the connections, the larger the delay and buffer requirement.  However, we believe that the back-pressure algorithm is a reasonable starting point for developing rate control/routing protocols for intermittently connected networks.  In this dissertation, we design, implement and evaluate the performance of two-scale back-pressure algorithms specially tailored for ICNs.

\section{Related Works}
The back-pressure algorithm {and distributed contention resolution mechanism in wireless networks, in one form or another, have been studied and} implemented in \cite{WarJanHaRhe09,RadGka08,MoeSriKriGna10,SriMoeKri09,LLYCPC09,AziStaThi09,RajShaShi09,JanPsoGov10}. \cite{WarJanHaRhe09} improves TCP performance over a wireless ad-hoc network by utilizing the back-pressure scheduling algorithm with a backlog-based contention resolution algorithm.  \cite{RadGka08} improves multi-path TCP performance by
taking advantage of the dynamic and resilient route discovery algorithmic nature of BP.  The authors in \cite{MoeSriKriGna10} have implemented and studied back-pressure routing over a wireless sensor network. They have used the utility-based framework of the traditional BP algorithm, and have developed implementations with good routing performance for data gathering (rate control is not studied in \cite{MoeSriKriGna10}). Their chief objective is to deal with the poor delay performance of BP.  \cite{SriMoeKri09} is an implementational study of how the performance of BP is affected by network conditions, such as the number of active flows, and under what scenarios backlog-based contention resolution algorithm is not necessary.  \cite{LLYCPC09} studies utility maximization with queue-length based throughput optimal CSMA for single-hop flows (with no routing or intermittent connectivity).  {More recent works on contention resolution mechanism are \cite{AziStaThi09,RajShaShi09,JanPsoGov10}.    In \cite{AziStaThi09}, a queue-based contention resolution scheme is proposed; however, the proposed algorithm only uses the local estimates of the neighbors' queue lengths to change the contention window parameter of IEEE 802.11, unlike the original back-pressure algorithm which requires explicit neighborhood queue length feedback.  The authors of \cite{RajShaShi09} proposed another form of queue-based contention resolution algorithm, which they conjecture does not require any neighborhood queue length feedback or message passing; however, the algorithm in \cite{RajShaShi09} requires larger buffers, resulting in long delays.  Channel access mechanism not based on queue lengths is proposed in \cite{JanPsoGov10}.  Here,  the authors propose a distributed time-sharing algorithm that allocates time slots based on the number of flows in the wireless network.}  Lastly, \cite{AkyAndGupHobSanSto08} is not an implementation, but discusses a lot of issues related to BP routing with rate control implementation.  Our study differs from all of them in that we focus on the multiple time-scales issue in an intermittently connected network (thus, queues throughout the network get ``poisoned'' with the traditional BP), and study modifications that loosely decouple the time-scales for efficient rate control.

A cluster-based back-pressure algorithm has been first studied in \cite{YinSriTow07} to reduce the number of queues in the context of
traditional networks. However, the algorithm as proposed in \cite{YinSriTow07} in general does not separate between the fast intra-cluster time scale and the slow inter-cluster timescales due to intermittently connected mobile carriers, thus leading to potentially large queue lengths at all nodes along a path. (We will demonstrate this in Section \ref{sec:motivation}.)  Our proposed algorithms in this dissertation explicitly decouples the two time scales by separating the network into two layers, with each layer operating its own back-pressure algorithm, and by allowing the two layers to interact in a controlled way at nodes that participate in inter-cluster traffic.  This explicit separation is the key property that leads to much smaller buffer usage and end-to-end delays and more efficient network resource utilization.

Initially, the approach taken in intermittently connected networks and DTNs for routing was based on packet replications.  The simplest way to make sure packets are delivered is to flood the ``mobile'' portion of the network so that the likelihood of a packet reaching the destination increases as more and more replicas are made \cite{VahBec00}.  A more refined approach is to control the number of replicas of a packet so that there is a balance between increasing the likelihood and still leaving some capacity for new packets to be injected into the network
\cite{BalLevVen07,SpyPsoRag05,BurGalJenLev06,VahBec00,LinDorSch03}.  Another refined approach is to learn the
intermittently connected topology and use this knowledge to route/replicate through the ``best'' contacts and encounters and avoid congestion \cite{JaiFalPat04,JonLiWar05,SonKotJaiHe04,ThoNelBak10,LeeYiJeo10}.  

\cite{OttKutDwe06,WhiCon09,HarAlm06} study networks that are closer to ours.  In \cite{WhiCon09}, distant groups of nodes are connected via mobiles, much like our network but with general random mobility.  At the intra-group level, a MANET routing protocol is used for route discovery, and at the inter-group level, the Spray-and-Wait algorithm \cite{SpyPsoRag05} is used among mobiles to decrease forwarding time and increase delivery probability.  \cite{OttKutDwe06} augments AODV with DTN routing to discover routes and whether those routes support DTN routing and to what extent they support end-to-end IP routing and hop-by-hop DTN routing.  \cite{HarAlm06} studies how two properties of the mobile nodes, namely whether a mobile is dedicated to serve a specific region (ownership) and whether the mobile movement can be scheduled and controlled by regions (scheduling time), affect performance metrics such as delay and efficiency.

Because replication-based algorithms inject multiple copies of a packet, they suffer from throughput drops.  However, all the aforementioned replication-based algorithms are valuable as they provide insight into engineering an efficient and robust {ICN} protocol.  There is (to the best of our knowledge) no literature on rate control over ICNs. {We demonstrate} in this dissertation that it is possible to obtain utility maximizing rate allocation, even though there is the ``mobile-gateway'' time scale that operates much slower than the wireless communication time scale, and all inter-cluster packets have to pass through the two different time scales.

 \begin{figure}
  \centering
    \subfigure[Inter-cluster and intra-cluster rates]{\includegraphics[scale=0.6]{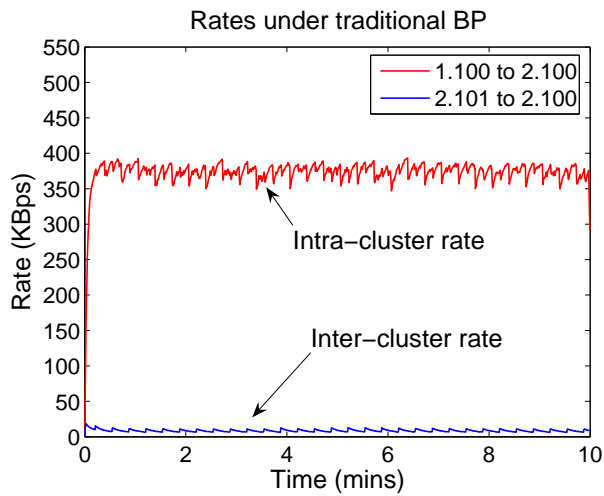} \label{fig:bp_rate}}
    \subfigure[End-to-end delay as a function of $N_c$]{\includegraphics[scale=0.6]{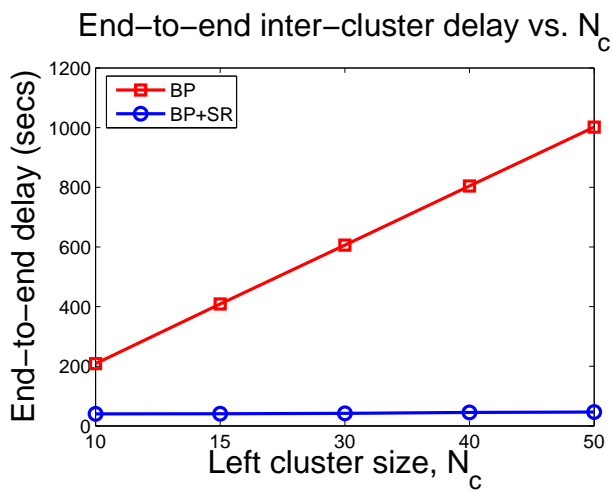} \label{f:N_del_res}}    
  \caption{The inter-cluster rate suffers serious throughput degradation under the traditional BP with rate controller because the inter-cluster source mistakenly sees the intermittent link as a low-capacity (with low delay) link and not as a high-capacity (with high delay) link (see Figure \ref{fig:bp_rate}).  Under the traditional BP, even if the inter-cluster rate is fixed to the correct value, the inter-cluster end-to-end delay grow to be extremely large as the source cluster size increases (see Figure \ref{f:N_del_res}).}  
\end{figure}

\section{Motivation: Difficulties with Traditional Back-Pressure}\label{sec:motivation}
Consider a simple, intermittently connected line network as shown in Figure \ref{fig:test_network}.  We have two clusters geographically separated.  We have two gateways (1.104 and 2.103) representing the left and right clusters, respectively.  In the left cluster, we have $N_c$ nodes, and in the right cluster, we have three nodes.  Between these two clusters, we have a ``mobile'' contact node 0.100
that moves from one cluster to the other every ten seconds.  On contact, the mobile and the gateways (the designated nodes in the cluster that can communicate with the mobiles) can exchange a large quantity of packets.  Finally, there are two flows; an inter-cluster flow originating from 1.100 and an intra-cluster flow originating from 2.101; both flows are destined for 2.100.

In this network, routing is straightforward.  But the question here is:  \textit{What is the rate at which these flows can transmit data?}  An
even more basic question is: \textit{Can these flows attain high and sustainable throughput}\footnote{By ``high'' we mean close to the
maximum throughput possible, and by ``sustainable'' we mean stochastically stable.}, provided that the link capacity between the mobiles and gateways is high enough (albeit with extreme delays)?  How close can we get to the maximum throughput allowed by the network?  Can we obtain utility-maximizing rate allocation over an ICN?  What will be the delay performance in such networks?
  
The answer to this is clearly negative, if TCP is used for rate control, and we shall see that even with traditional back-pressure algorithms \cite{TasEph92,ErySri06}  that have a theoretical guarantee that the above is possible, in a practical setting, the answer still seems to be negative! 

To put the above statement in context, we know that the back-pressure (BP) routing/rate control algorithm is throughput optimal \cite{TasEph92}, meaning that if any routing/rate control algorithm can give us certain throughput performance, so can the back-pressure algorithm.  \textit{Contra-positively, if the back-pressure algorithm {cannot} give a certain throughput performance, no other algorithm can do so}.  Further, a rate controller based on utility maximization can be added to this framework \cite{Sto05,ErySri06,NeeModLi05} that is theoretically utility
maximizing, and it chooses rates that (averaged over a long time-scale) lead to {high and sustainable} throughput corresponding to
the rates determined via an optimization problem \cite{Sto05,ErySri06,NeeModLi05}.

We first consider the performance of a traditional BP based routing/rate control algorithm.  In Figure \ref{fig:bp_rate}, we plot the
rate trace of the two (inter- and intra-cluster) flows (with $N_c=2$).  ({Figure \ref{fig:bp_rate} is obtained experimentally.} Each source
uses the BP congestion algorithm \cite{NeelyThesis,Sto05,ErySri06,NeeModLi05} which we will describe later).  In the figure, we can see that the inter-cluster traffic performs very poorly, even though the mobile-gateway contact has enough capacity.  The reason is simple -- the BP congestion control uses the local queue length as a congestion signal, and between two successive contacts that can be seconds or minutes apart, there is a large queue build-up to the point that the inter-cluster source mistakenly believes that the network has low-capacity (and low-delay) links.  Because of this, the inter-cluster source is not able to fully utilize the contacts (see Figure \ref{fig:bp_rate}).

Importantly, this rate achieved by the inter-cluster traffic is much lower than that predicted by the theory (the theory predicts that
intra-cluster rate is $\approx$ 200KBps, and the inter-cluster rate $\approx$ 100KBps). This is because the theoretical results hold only
when the utilities of users are scaled down by a large constant -- this is to (intuitively) enable {\bf all} queues in the network to build up
to a large enough value in order to ``dilute'' the effects of the ``burstiness'' of the intermittently connected link.  However, scaling down the utilities by a large constant will result in long queues over the entire network, even at nodes that do not participate in inter-cluster traffic.

Furthermore, even if the inter-cluster source is aware of the presence of these intermittent mobile-gateway links and therefore can transmit at the correct rate (i.e., a genie computes the rate and tells this to the source), the problem can manifest itself in another way. 

Consider the same network as in Figure \ref{fig:test_network}, but we just have the inter-cluster flow from 1.100 to 2.100; however, we vary the left cluster size $N_c$ from 10 to 50.  {See Figure \ref{f:N_del_res}.  (Figure \ref{f:N_del_res} is obtained from simulations.)} We run the traditional BP algorithm (with no rate controller, as the genie has solved this problem) and fix the source rate at 200KBps. In this case, there is large backlog that builds up not only at node 1.104 (the intermittently connected gateway), but also at every node in the left cluster.  

\subsection{Main Contributions}

In this chapter, we design, implement and empirically study the performance of a modified back-pressure algorithm that has been
coupled with a utility based rate controller for an intermittently connected network.  Our contributions are:

\begin{enumerate}
\item We present a modified back-pressure routing algorithm that can separate the two time scales of ICNs.  Separating the time scales improves the end-to-end delay performance and provides a throughput arbitrarily close to the theory.  A key advantage of this modified BP algorithm is that it maintains large queues only at nodes which are intermittently connected; at all other nodes, the queue sizes remain small. 
\item On top of our modified back-pressure routing algorithm, we implement a rate control on our testbed. {The essential components of our testbed are built with modified MadWifi and Click \cite{click}.  The nodes in our testbed are organized into multiple clusters, with intermittent connectivity emulated using an Ethernet switch.}  
\item Using this testbed, we first show that the traditional back-pressure algorithms coupled with a utility function-based rate controller is not suitable for intermittently connected networks, and leads to a large divergence between theoretically predicted rates and actual measurements (shown for an intermittently connected line network). We then present measurement results using our time scale decoupled algorithm on a line network and on a larger sized network. 
\item Finally, we present a practical implementation of shadow queues developed by the authors in \cite{BuiSriSto09}. Using our implementation, we demonstrate there is a nice trade-off between shorter end-to-end inter-cluster delay and the network capacity  utilization.
\end{enumerate}

\section{Network Model}\label{sec:bp_alg}
The time is slotted, with $t$ denoting the $t^{\text{th}}$ time slot.  The intermittently connected network consists of multiple clusters.  Each cluster $\C_i$ is represented by a graph $\GG_{\C_i} = (\NN_{\C_i}, \LL_{\C_i}),$ where $\NN_{\C_i}$ is the set of nodes in $\C_i$ and $\LL_{\C_i}$ is the set of links. Let $\C(n)$ denote the cluster to which node $n$ belongs.  The clusters are geographically separated, and two nodes in distinct clusters cannot communicate with each other directly (they must rely on the mobiles to transport data).

The clusters are connected by a set $\MM$ of mobile carrier nodes that move around to carry packets from one cluster to another. For each cluster, the set of nodes that can communicate with the mobiles is fixed. These nodes are named as gateways; those nodes that cannot communicate with the mobiles directly are called internal nodes.  Let $\I_{\C_i}$ and $\HH_{\C_i}$ denote the set of internal nodes and gateways in $\C_i$, respectively.

We use $g(i,j)$ to denote gateway $j$ in cluster $\C_i$. To simplify the notations, we assume that gateways $g(\cdot,j)$ have access to mobile $m(j)$ only. The mobiles change gateways every $T$ time slots, which is called a super time slot. We let $\tau$ denote the $\tau^{\text{th}}$ super time slot.  Here, $T$ is a very large number.  A time slot is the time scale of one intra-cluster packet transmission, and $T$ is the time scale of the mobility (thus, a time slot is roughly a few milliseconds long, and $T$ is roughly $> 10^{3}$ to reflect the mobility time scale which is seconds or minutes long).  In our model, the size of the clusters is $<< T$.  

We assume that the mobility of the mobiles follows a Markov process.  Given that mobile $m(j)$ is at gateway $g(i_1,j)$ at the beginning of super time slot $\tau,$ the probability that it moves to gateway $g(i_2,j)$ at the beginning of the super time slot $\tau+1$ is 
\begin{eqnarray*}
  &\mathbb{P}\left(m(j)-g(i_2,j) \text{ in } \tau+1|m(j)-g(i_1,j)\text{ in } \tau\right)\\
  &=\mathbf{P}_{m(j)}(i_1, i_2) 
\end{eqnarray*}
where ``$-$'' means that the mobile and the gateway are in contact.  Let $\mathbf{P}_{m(j)}$ denote the transition probability matrix of mobile $m(j)$ and let $\pi_{m(j)}$ be the corresponding stationary distribution. The Markov chains are assumed to be aperiodic and irreducible. The assumption that $g(\cdot, j)$ only have access to $m(j)$ is not necessary; we make this assumption to simplify our notations.

\subsection{Traffic Model}
\label{sec:traffic_model}
A traffic flow is defined by its source and destination. We assume that the sources and destinations are all internal nodes. If the source and the destination lie in the same cluster, then the traffic flow is an \textit{intra-cluster} traffic, and the intra-cluster traffic can be routed only within the cluster. If the source and the destination lie in different clusters,
the traffic flow is an \textit{inter-cluster} traffic. We let $[s,d]$ denote the flow from $s$ to $d,$ $\F$ denote the set of all flows, and $\F_{\text{inter}}$ and $\F_{\text{intra}}$ be the sets of all inter- and intra-cluster flows, respectively. We let $x_{s}^{d}$ be the number of packets source $s$ generates per time slot {for destination $d$}, $\textbf{x} = \{x_s^d:[s,d]\in \F \},$ and $\textbf{x}_{\text{intra}(\C)}$ be the set of intra-cluster traffic rates in cluster $\C$.

Note that all inter-cluster traffic flows must be forwarded to the gateways in source clusters, then carried over to the gateways in destination clusters via the mobile nodes before reaching their destinations.

\subsection{Communication Model}
Let $\mu_{(n_1,n_2)}[t]$ denote the transmission rate (packets/time slot) of link $(n_1, n_2)$ at time $t,$ and $\vec{\mu}_{\C_i}[t] = \{\mu_{(n_1,n_2)}[t],~(n_1,n_2)\in \LL_{\C_i} \}$.  Let $\Gamma_{\C_i}$ be the convex hull of the set of all feasible transmission rates in cluster $\C_i$.  We note that in general, $\vec{\mu}_{\C_i}[t]$ and $\Gamma_{\C_i}$ depend on the interference model used for cluster $\C_i$.

We assume that a mobile and a gateway can send $R$ packets to each other per contact. We assume that the transmissions between mobiles and gateways do not cause interference to other transmissions.

{For an example, see the network depicted in Figure \ref{f:sim_net}.  Here, we have three clusters, with two gateways in each and each cluster connected in a grid.   We have pairs of nodes in different clusters communicating using the two mobiles (inter-cluster flows), which shuffle between these clusters transporting data from one cluster to another.  In addition, we have intra-cluster flows in each cluster.}

\section{Two-Scale BP with Queue Reduction: BP+SR}\label{sec:bp_sr}
In this section, we introduce our two-scale back-pressure routing algorithm (which we refer to as BP+SR (Source Routing)) that separates the times scales of inter-cluster and intra-cluster connections, while at the same time, reducing the number of queues that need to be maintained.  We build on this two-scale back-pressure algorithm in section \ref{sec:bp_with_rc} to implement a utility-maximizing rate controller.

\subsection{Queuing Architecture}
In our algorithm, the network maintains two types of queues.  The first type, referred to as type-I, will be denoted by $q$, and the second type, type-II, will be denoted by $u$. 

Any internal node $n_1$ maintains a type-II queue $u_{n_1}^{g}$ for each gateway $g$ in the same cluster and a type-I queue $q_{n_1}^{n_2}$ for each node $n_2$ in the same cluster. A gateway $g_1$ maintains a type-II queue $u_{g_1}^{g_2}$ for each of other gateways $g_2$ in the network {(even for gateways in the same cluster)}.

For each node $n$ in the same cluster, gateway $g$ maintains both a type-I queue and a type-II queue for node $n$.  A mobile $m$ maintains a separate type-II queue $u_m^g$ for each gateway $g$ in the network. We use $q_a^b[t]$ ($u_a^b[\tau]$) to denote the length of the type-I (type-II) queue maintained by node $a$ for node $b$ at the beginning of the time slot $t$ (super time slot $\tau$).  Note that $u_n^n = 0$ and $q_n^n = 0$ at all times $\forall~n$.

\begin{figure*}[ht!]
  \centering
    \includegraphics[scale=0.5]{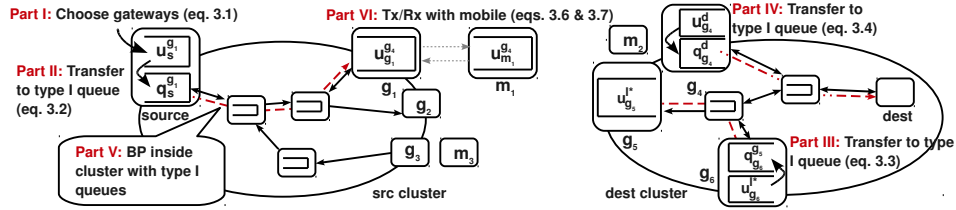}     
    \caption{In this figure, we break down the BP+SR algorithm into six parts.  Part I is the source routing.  Parts V and VI are the intra- and inter-cluster back-pressure routing algorithms, respectively.  The two back-pressure algorithms interact via packet transfers between type I and II queues.  Part III is for the load balancing over different gateways in the same cluster using the intra-cluster back-pressure, i.e., the gateways use the intra-cluster back-pressure as the back haul link for transferring packets between them. }\label{f:rout_big} 
\end{figure*}

\subsection{BP+SR Algorithm}
We now present our two-scale BP routing algorithm, BP+SR (Source Routing).  We give a high-level highlight using Figure \ref{f:rout_big} before discussing the algorithm in details.  First, an inter-cluster traffic source takes a group of packets and chooses the optimal gateways (in both source and destination clusters) through which these packets should be routed (Part I).  The optimal gateways chosen can change over time.  These packets are routed from the source to the chosen optimal source gateway, and from the chosen optimal destination gateway to the destination using the intra-cluster back-pressure inside the respective clusters (Part V); they are routed from the source gateway to the destination gateway using the inter-cluster back-pressure (Parts III and VI).  The interaction between the two back-pressure routing algorithms  happens through packets transfers between the two types of queues (Parts II, III, and IV).  Part VII is presented for mathematical convenience and is not in the actual protocol.

\noindent{\bf Part I: Selecting source and destination gateways}

At the beginning of super time slot $\tau$, the inter-cluster traffic source $s$ picks the source and destination gateways $g_s^*[\tau]$ and $g_{d}^*[\tau]$ such that
\begin{eqnarray}\label{eq:src_routing}
\left(g_s^*[\tau], g_{d}^*[\tau]\right) \in \arg\min_{\substack{g_s \in \HH_{\C(s)}\\g_d\in \HH_{\C(d)}}}  \left( u_{s}^{g_s}[\tau] + u_{g_s}^{g_d}[\tau] + u_{g_d}^{d}[\tau]\right).
\end{eqnarray}
This route selection is done at the beginning of each super time slot (i.e., every $T$ time slots).  {The source $s$ makes the routing decision (Eq. \ref{eq:src_routing}) independently of other inter- and intra-cluster sources.  The source and destination gateways are chosen such that the pair minimizes the total queue lengths considering the intra-cluster path from the source to its gateway, the inter-cluster path between the two gateways, and the intra-cluster path between the destination gateway and the destination node.}

\noindent{\bf Part II: Traffic control at the source nodes}
\begin{list}{\labelitemi}{\leftmargin=1em}
\item For an \textbf{inter-cluster} flow $[s,d],$ the source node $s$ deposits newly arrived packets into queue $u_{s}^{g_s^*[\tau]}$ during time slot $t\in [\tau T, (\tau+1)T).$   The identities of the source gateway $g_s^{*}[\tau]$ and the destination gateway $g_d^{*}[\tau]$ are recorded in the headers of the packets.

\item For an \textbf{intra-cluster} flow $[s, d],$ the source node $s$ deposits the new arrived packets into queue $q_{{s}}^{{d}}.$

\item Define $\theta_{s}^{g_s}[\tau] = u_s^{g_s}[\tau]/K_s,$ where $K_s = T/|{\cal C}(s)|.$  ($|\C|$ is the number of nodes in cluster $\C$.)  Consider the queues associated with gateway $g_s \in \HH_{\C(s)}.$ If \begin{eqnarray}\label{eq:src_cond}
\theta_{s}^{g_s}[\tau] > q_{s}^{g_s}[t]
\end{eqnarray} at time $t\in [\tau T, (\tau+1)T),$ $\eta$ packets are transferred from queue $u_{s}^{g_s}$ to queue $q_{s}^{g_s}$ at the beginning of time slot $t$.  ($\eta$ is some positive value greater than the largest transmission rate out of any node inside a cluster.)  

When a packet arrives at the source gateway $g_s^*$, the source gateway would set the next destination of that packet to $g_d^*$, which it would find in the packet header.  (See Eq. (\ref{eq:src_routing}).)  The gateway would then insert that packet into the queue $u_{g_s^*}^{g_d^*}$.

\end{list}

To achieve throughput optimality, an inter-cluster traffic source must do source routing as in Eq. (\ref{eq:src_routing}), and when it does source routing, the length of the queue $u_{s}^{g_s}$ will become of order $\Theta(T)$, and there must be a way to release the packets stored in the queue $u_s^{g_s}$ into the cluster so they can reach the source gateway.  The purpose of Eq. (\ref{eq:src_cond}) is exactly that -- to release/transfer the packets from $u_s^{g_s}$ (type-I queue) to $q_s^{g_s}$ (type-II queue) at a controlled and acceptable rate to the cluster.  The factor $|\C(s)|$ is needed so as to prevent the inter-cluster end-to-end delay from scaling with the cluster size.

\noindent{\bf Part III: Traffic control at the gateway nodes}

The gateway $g_1$ computes $l_{g_1,g_2}[\tau] \in \argmax_{l\in \left(\bigcup_{\C}\HH_{\C}\right) }(u_{g_1}^{l}[\tau] - u_{g_2}^{l}[\tau] )$ at the beginning of each super time slot for each gateway $g_2$ in the same cluster.  Define $\theta_{g_1}^{g_2} [\tau] = \left\{u_{g_1}^{l_{g_1,g_2}[\tau]}[\tau] - u_{g_2}^{l_{g_1,g_2}[\tau]}[\tau]\right\}/K_{g_1}$ where $K_{g_1} = T/|\C(g_1)|$.  At each time slot $t\in [\tau T, (\tau+1)T)$, $g_1$ transfers $\eta$ packets from $u_{g_1}^{l_{g_1,g_2}[\tau]}$ to $q_{g_1}^{g_2}$ if
\begin{eqnarray}\label{eq:gg_cond}
\theta_{g_1}^{g_2}[\tau] > q_{g_1}^{g_2}[t].
\end{eqnarray}
The next destination of the transferred packets is temporarily set to $g_2$; when $g_2$ receives those packets, they are inserted into $u_{g_2}^{g_d^*}$, where $g_d^*$ (from Eq. (\ref{eq:src_routing})) can be found in the packet headers.

Part III is used by the gateways (that are in the same cluster) to balance the load amongst themselves using the intra-cluster resources.  That is, the gateways can us any available bandwidth in the cluster to shift load from one gateway to another.

\noindent{\bf Part IV: Traffic control at the destination gateways}

When the packets arrive at their destination gateways, they are deposited into queue $u_{g_d}^{d}$.  Let $\theta_{g_d}^{d}[\tau] = u_{g_d}^{d}[\tau]/K_{g_d}$, where $K_{g_d} = T/|\C(g_d)|$.  In each time slot $t\in [\tau T, (\tau+1)T)$, $\eta$ packets are transferred from $u_{g_d}^{d}$ to $q_{g_d}^{d}$ if
\begin{eqnarray}\label{eq:gd_cond}
\theta_{g_d}^{d}[\tau] > q_{g_d}^{d}[t].
\end{eqnarray}

\noindent{\bf Part V: Routing and scheduling within a cluster}

In each time slot $t$, each cluster $\mathcal{C}$ computes $\vec \mu_{\mathcal{C}}[t]$ such that
\begin{eqnarray}
\vec{\mu}_{\mathcal{C}}[t] = \argmax_{\vec{\mu} \in \Gamma_{\mathcal{C}}} \left\{ \sum_{(m,n)\in \mathcal{L}_{\mathcal{C}}} \mu_{(m,n)} P_{(m,n)}[t]\right\}\label{eq:bp_max}
\end{eqnarray}
where $P_{(m,n)}[t] = q_m^{j_{(m,n)}[t]} [t] - q_n^{j_{(m,n)}[t]}[t]$ and $j_{(m,n)}[t] = \argmax_{j} \left\{ q_m^{j} [t] - q_n^{j}[t]\right\}$.  After the computation, node $m$ transmits $\mu_{(m,n)}[t]$ packets out of queue $j_{(m,n)}[t]$ to node $n$ in time slot $t$.  $\Gamma_{\mathcal{C}}$ is the set of all feasible rate in the cluster $\mathcal{C}$.

\noindent{\bf Part VI: Routing between gateways and mobiles}

At the beginning of each super time slot $\tau$, the mobile $m$ and the gateway $g$ that are in contact compute the following:
\begin{eqnarray}
j_{(m,g)}[\tau] = \argmax_{j \text{ gateway }} \left\{u_m^{j}[\tau] - u_g^{j}[\tau]\right\},\label{eq:gmb_1}\\
j_{(g,m)}[\tau] = \argmax_{j \text{ gateway}} \left\{u_g^{j}[\tau] - u_m^{j}[\tau]\right\}.\label{eq:gmb_2}
\end{eqnarray}
Afterwards, $m$ transmits $R$ packets from the queue $u_m^{j_{(m,g)}[\tau]}$ to $g$, and the gateway $g$ transmits $R$ packets from the queue $u_g^{j_{(g,m)}[\tau]}$ to $m$, maximizing $R\left(P_{(m,g)}[\tau] + P_{(g,m)}[\tau]\right)$.


\noindent{\bf Part VII: Real and regulated queues}

For analytical purposes, we assume that each type-II queue at a gateway consists of two parts: a real type-II queue, denoted $\tilde{u}$, and a regulated type-II queue, denoted $u$. At gateway $g_1$, the amount of packets transferred from the real queue $\tilde{u}_{g_1}^{g_2}$ to the regulated queue $u_{g_1}^{g_2}$ in time slot $t$ is $y_{g_1}^{g_2}[t] = (1+\delta) \sum_{[s,d]\in \F_{\text{inter}},~s\in\I_{\C(g_1)}}  x_{s,g_1}^{g_2,d}[t]$ where $g_2$ is another gateway, and the amount of packets transferred from $\tilde{u}_{g_1}^{d}$ to $u_{g_1}^{d}$ is $y_{g_1}^{d} [t] = (1+\delta) \sum_{[s,d]\in \F_{\text{inter}},~d\in \I_{\C(g_1)}} x_{s,g_2}^{g_1,d}[t]$ where $d$ is an internal node in $\C(g_1)$ and $\delta>0$, and 
\begin{eqnarray*}
x_{s,g_s}^{g_d,d}[t] = \left\{\begin{array}{ll}
x_s^d & \text{ if } g_s = g_s^*[\tau]\hbox{ and }g_d = g_d^*[\tau] \\
0 & \text{ else }
\end{array}
\right.,
\end{eqnarray*}
is the amount of inter-cluster traffic $[s,d]$ that is assigned to the gateways $g_s^*[\tau]\in \HH_{\C(s)}$ and $g_d^*[\tau]\in \HH_{\C(d)}$ in time slot $t \in [\tau T , (\tau+1)T)$.  Though regulated queues are not used in practice, they are needed to prove the stability and throughput optimality of our algorithm.

\textit{Remark: }In our two-scale BP+SR algorithm with queue reduction, the source nodes need to know the queue-lengths of all gateways in the source cluster and destination clusters; this is needed to reduce the number of queues maintained at the other internal nodes in the same cluster.  We note that it would be difficult to obtain this information instantaneously as required in the algorithm. However, similar to the cluster-based back-pressure proposed in \cite{YinSriTow07}, we can use the delayed queue length information, and the algorithm is also throughput-optimal. (The analysis of our algorithm would, however, have to account for the presence of the two time-scales.) We skip this because similar analysis is provided in \cite{YinSriTow07}.


\subsection{Throughput Optimality}
We now prove that our BP+SR is throughput optimal.
\begin{theorem}\label{th:t_op}
  Fix any $\delta > 0.$ Given external arrival $\textbf{x}$ such that
  $(1+\delta+\epsilon)\textbf{x}$ is supportable for some $\epsilon>0$
  (i.e., there exists an algorithm that can stabilize the network with
  traffic load $(1+\delta+\epsilon)\textbf{x}$), all queues are
  bounded under the BP+SR algorithm.
\end{theorem}
\noindent{\textit{Sketch of the proof: }}We first bound the lengths of type-II queues when the routing algorithms (Eq. (\ref{eq:src_routing}), (\ref{eq:src_cond}), (\ref{eq:gg_cond}), and (\ref{eq:gd_cond})) are updated every $\tilde{T}$ super time slots; we will refer $\tilde{T}$ super time slots as a super-super time slot.  We assume that $\tilde{T}$ is large enough so that for any mobile $m$ and any gateway $g$ that $m$ comes into contact with, $m$ makes at least $(1+\epsilon)^{-1}(\pi_m)_g \tilde{T}$ contacts with $g$ over $\tilde{T}$ super time slots.  Let $\tilde{\tau}$ denote $\tilde{\tau}$-th super-super time slot.  Then, we will use this bound to obtain the upper bound when routing algorithms are updated every super time slot.

We define a Lyapunov function $V[\tilde{\tau}\tilde T] = \sum_{n} \sum_{j} (u_n^j[\tilde{\tau}\tilde T ])^2$, and let 
\begin{eqnarray*}
\Delta_{\tilde T} V[\tilde{\tau}\tilde T ]&=& V[(\tilde \tau +1)\tilde T] - V[\tilde \tau \tilde T].
\end{eqnarray*}
Using our BP+SR algorithms (\ref{eq:src_routing}), (\ref{eq:src_cond}), (\ref{eq:gg_cond}), (\ref{eq:gd_cond}), (\ref{eq:bp_max}), (\ref{eq:gmb_1}), and (\ref{eq:gmb_2}), and the fact that $(1+\delta+\epsilon)\textbf{x}$ is supportable, we can show that $\Delta_{\tilde T} V[\tilde{\tau}\tilde T ] < -\delta$ if $u_{n_1}^{n_2}[\tilde \tau \tilde T] > U_{\max}$ for some $n_1$, $n_2$ and for some $U_{\max}$, from which we can show that $V[\tilde \tau \tilde T] \leq (\bar K)^2$ and $u_{n_1}^{n_2} [\tilde \tau \tilde T] \leq \bar K$, $\forall$ $n_1$, $n_2$, where $\bar K$ is some positive constant.

The probability of the mobile not exhibiting the stationary distribution in $\tT$ super time slots is exponentially decreasing in $\tT$.  Thus, we can obtain an expected upper bound on the type-II queues when our algorithms (\ref{eq:src_routing}), (\ref{eq:src_cond}), (\ref{eq:gg_cond}), and (\ref{eq:gd_cond}) are updated every super time slot.  We can then use Theorem 1 of \cite{NeeModLi05} to bound the type-I queues since in eqs. (\ref{eq:src_cond}), (\ref{eq:gg_cond}) and (\ref{eq:gd_cond}), type-II queue sizes are used as a linear utility function.  Each super time slot is long enough so that within each cluster, any utility-based back-pressure rate control algorithm (that is updated every time slot) using the type-I queues converges. For the complete proof, see the Appendix of this chapter.
\done

\subsection{Buffer Usage \& Delay}\label{sec:buff_usage}
Consider a line network with two gateways and a mobile, as shown in Figure \ref{fig:test_network} (without the intra-cluster traffic).  Let there be an inter-cluster flow $[s,d]$ (with $s$ being 1.100 and $d$ being 2.100 in Fig. \ref{fig:test_network}).  Let $g_s$ and $g_d$ be the source and destination gateways, respectively ($g_s$ is 1.104 and $g_d$ is 2.103).  The source cluster size is $N_c$, so that the number of hops from $s$ to $g_s$ is $N_c -1$.  Assume that the mobile $m$ (0.100 in the figure) shuffles between $g_s$ and $g_d$ every $T$ time slots, so that two consecutive contacts at a gateway are $2T$ time slots apart{; the mobile makes contact with the source gateway in time slots $2T$, $4T$, $6T$, ...  The system starts at time slot 0.}  Let $s$ generate traffic for $d$ at the average rate of $1-\gamma$, $1>\gamma>0$ packet per time slot (in each time slot, $s$ generates one packet with probability $1-\gamma$ and generates nothing with probability $\gamma$).  Assume that each node $i$ can transmit one packet to $i-1$ and receive one packet from $i+1$ simultaneously per time slot and that all links are directed.  {We assume directed links only for the purpose of analysis in the subsection, and t}his restriction is placed to prevent packet looping in our analysis.  Assume that the mobile and the gateway can exchange $2T$ packets per contact.  We will use $i$ to denote the nodes in the source cluster, with $i=1$ being $g_s$, $i=2$ being the immediate neighbor of $g_s$, $i=3$ being the node two hops from $g_s$, and so on. 

{\begin{assumption}\label{assume:gen_assume} For the purpose of the following lemma, we assume that $T$ is large enough so that the number of packets generated by the source over $2T$ time slots is between $2T(1-\gamma-\epsilon)$ and $2T(1-\gamma+\epsilon)$, for a sufficiently small $\epsilon$.  In addition, we assume that packets are transmitted out of the queues at the beginning of a time slot, and inserted into the queues in the middle of a time slot, and that the source generates packets at the end of a time slot.  \end{assumption}

We will use $\tilde q$, $\bar q$, $\hat q$ to denote the queue size at the beginning, in the middle, and at the end of a time slot, respectively.  We will also use $G_n$ to denote the total number of packets the source generates from time slots $2nT$ to $2(n+1)T-1$.  By Assumption \ref{assume:gen_assume}, $2T(1-\gamma-\epsilon) \leq G_n \leq 2T(1-\gamma+\epsilon)$.}

{The assumption that the number of packets the source generates is between $2T(1-\gamma-\epsilon)$ and $2T(1-\gamma+\epsilon)$ over $2T$ time slots can be justified using the Chernoff bound: assume i.i.d. Bernoulli arrivals, and for any $\epsilon>0$, the probability that the source generates more than $2T(1-\gamma+\epsilon)$ packets in $2T$ time slots is less than or equal to
\begin{eqnarray*}
\exp(-2T\times D((1-\gamma+\epsilon) || (1-\gamma)),
\end{eqnarray*}
and the probability that the source generates fewer than $2T(1-\gamma-\epsilon)$ packets in $2T$ time slots is less than or equal to
\begin{eqnarray*}
\exp(-2T\times D((1-\gamma-\epsilon) || (1-\gamma)),
\end{eqnarray*}
where 
\begin{eqnarray*}
 D(x||y) = x \log \left(\frac{x}{y}\right) + (1-x) \log \left(\frac{1-x}{1-y}\right).
\end{eqnarray*}
When $T$ is sufficiently large, both probabilities are very small.}

\begin{lemma}
 Suppose all queues are initially empty.  {Under BP, there exists a time slot $t_d$ such that all packets the source generates after $t_d$ would experience a delay of at least $(N_c-1)(2T(1-\gamma-\epsilon)-1)$ before being picked up by the mobile, where $\epsilon>0$ so that Assumption \ref{assume:gen_assume} holds.  Under BP+SR, the delay any packet would experience before being picked up by the mobile is at most $N_c^2 + 3T$.}
\end{lemma}
\noindent\textit{Proof:} \underline{Under traditional BP:} Let $\{n_1, n_2, ...\}$ be an infinite sequence of positive integers such that $n_l \neq n_{l'}$ if $l\neq l'$ and at the beginning of time slots $2n_lT$ (just before the mobile picks up packets), 
\begin{eqnarray}
\tilde q_1^d[2n_lT] \geq 2T(1-\gamma).\label{eq:beg_q1d} 
\end{eqnarray}
Such a sequence exists because the mobile has to transport at least $2T(1-\gamma)$ packets infinitely often in order to support the source rate of $1-\gamma$ packets per time slot.  We set $t_d=2n_lT$.  


\noindent{\emph{Sketch of Proof:} }We first show that new packets generated in time slots $2n_lT$, $2n_lT+1$, ... , $2(n_l+1)T-1$ would experience delay of at least $(N_c-1)(2T(1-\gamma-\epsilon)-1)$ before reaching the source gateway because $\tilde q_i^d[2n_lT] \geq 2T(1-\gamma)-1$ for $i=2,...,N_c$.  To show this, we need to show that $\tilde q_i^d[2n_lT] \geq 2T(1-\gamma)-1$, $i=2,...,N_c$ if $\tilde q_1^d[2n_lT] \geq 2T(1-\gamma)$ (see Eq. (\ref{eq:q_i_d_ineq})).

We then show that any new packets generated in time slots $2(n_l+1)T$, $2(n_l+1)T+1$, ... , $2(n_l+2)T-1$ would also experience delay of at least $(N_c-1)(2T(1-\gamma-\epsilon)-1)$.  For this, we show that
\begin{eqnarray*}
 \sum_{i=2}^{N_c} \tilde q_i^d[2(n_l+1)T] \geq (N_c-1)(2T(1-\gamma-\epsilon)-1)
\end{eqnarray*}
which requires the fact that $\tilde q_i^d[2n_lT] \geq 2T(1-\gamma)-1$, $i=1,...,N_c$.  See Eq. (\ref{eq:eq18}).

We finally show that for $k>1$ such that $n_l + k < n_{l+1}$, any new packets generated in time slots $2(n_l+k)T$, $2(n_l+k)T+1$, ... , $2(n_l+k)T-1$ would also experience delay of at least $(N_c-1)(2T(1-\gamma-\epsilon)-1)$ by showing that 
\begin{eqnarray*}
 \sum_{i=2}^{N_c} \tilde q_i^d[2(n_l+k)T] \geq (N_c-1)(2T(1-\gamma-\epsilon)-1)
\end{eqnarray*}
which requires that $\sum_{i=2}^{N_c} \tilde q_i^d[2(n_l+k-1)T] \geq (N_c-1)(2T(1-\gamma-\epsilon))$.  See Eq. (\ref{eq:k2q}).

\noindent{\emph{Details of Proof:} }For any $i=1,...,N_C-1$ and at the beginning of any time slot $t$, if $\tilde q_i^d[t] = q+1$, where $q\geq 0$, then
\begin{eqnarray}
 \exists l_{i,t}>0 \text{ s.t. }\tilde q^d_{i+1}[t-l_{i,t}]\geq q+1. \label{eq:eq_l_it}
\end{eqnarray}

 We prove Eq. (\ref{eq:eq_l_it}) by contradiction.  If for all $l_{i,t}>0$ such that $\tilde q^d_{i+1}[t-l_{i,t}]< q+1$, then $\tilde q_i^d$ could not have increased to $q+1$ since node $i+1$ would transmit a packet to node $i$ in some time slot $\tau$ if and only if $\tilde q_{i+1}^d[\tau] - \tilde q_{i}^d[\tau] > 0$ and node $i+1$ is the only node that transmits to node $i$; since $\tilde q^d_{i+1}[t-l_{i,t}]< q+1$ for all $l_{i,t}$, the maximum that $\tilde q_{i}^d[t]$ can be is $q$, which is a contradiction.  There can be multiple values for $l_{i,t}$ for $\tilde q^d_{i+1}[t-l_{i,t}]\geq q+1$ to hold; we let $l_{i,t}$ take on the smallest value $>0$ so that $\tilde q^d_{i+1}[t-l_{i,t}]\geq q+1$.

We now show that if $\tilde q_i^d[t] = q+1$, then $\tilde q^d_{i+1}[t']\geq q$, $t\geq t' > t-l_{i,t}$ by contradiction.  Suppose there is $t'$ such that $t\geq t' > t-l_{i,t}$ and $\tilde q^d_{i+1}[t']< q$.  Since $\tilde q^d_{i+1}[t-l_{i,t}] \geq q+1$, this implies that $\tilde q^d_{i+1}[t''] = q$ in some time slot $t''$ between $t-l_{i,t}$ and $t'$, and transmitted a packet to node $i$ without having received a packet from node $i+2$ (or generated a packet); this is the only way node $i+1$ could decrease its queue length from $q$ to $q-1$.  Since a packet has been transmitted from node $i+1$ to node $i$ in time slot $t''$, $\tilde q^d_{i+1}[t''] - \tilde q^d_{i}[t''] > 0$, which implies that $\tilde q^d_{i}[t''] \leq q-1$.  Since $\tilde q^d_i$ has to increase to $q+1$ by time $t$, there is a time slot $t'''$ between $t''$ and $t$ so that $\tilde q^d_{i+1}[t'''] \geq q+1$ (otherwise, $\tilde q_i^d[t]$ could not reach $q+1$), which violates the assumption that the smallest value for $l_{i,t}$ is chosen so that $\tilde q^d_{i+1}[t-l_{i,t}]\geq q+1$ holds.  Thus, 
\begin{eqnarray}
 \tilde q^d_{i+1}[t']\geq q, t\geq t' > t-l_{i,t}.\label{eq:t_minus_l_it}
\end{eqnarray}

Note that in addition, if $\tilde q_i^d[t], \tilde q_{i+1}^d[t] \geq q$, then as long as 
\begin{eqnarray}
\tilde q_i^d[\tilde t] \geq q \text{ for }\tilde t \geq t, ~\tilde q_{i+1}^d[\tilde t] \geq q \label{eq:q_tilde_t}
\end{eqnarray}
since node $i+1$ will not transmit to node $i$ if $\tilde q_{i+1}^d[\tilde t] - \tilde q_{i}^d[\tilde t] \leq 0$.

By eqs. (\ref{eq:beg_q1d}) and (\ref{eq:eq_l_it}), there is $l_{2,2n_lT}$ such that 
\begin{eqnarray}
\tilde q_2^d[2n_lT - l_{2,2n_lT}] \geq 2T(1-\gamma). \label{eq:q_2_d}
\end{eqnarray}
By Eq. (\ref{eq:t_minus_l_it}), $\tilde q_2^d[2n_lT] \geq 2T(1-\gamma)-1$.

By eqs. (\ref{eq:q_2_d}) and (\ref{eq:eq_l_it}), there is $l_{3,2n_lT-l_{2,2n_lT}}$ such that $\tilde q_3^d[2n_lT - l_{2,2n_lT} - l_{3,2n_lT-l_{2,2n_lT}}] \geq 2T(1-\gamma)$.  By Eq. (\ref{eq:t_minus_l_it}), $\tilde q_3^d[2n_lT - l_{2,2n_lT}] \geq 2T(1-\gamma)-1$.  By eqs. (\ref{eq:q_2_d}) and (\ref{eq:q_tilde_t}), $q_3^d[2n_lT] \geq 2T(1-\gamma)-1$.

Similar to the way we showed that $\tilde q_3^d[2n_lT] \geq 2T(1-\gamma)-1$, we can show that $\tilde q_4^d[2n_lT]$, $\tilde q_5^d[2n_lT]$, ... , $\tilde q_{N_c}^d[2n_lT]$ are also $\geq 2T(1-\gamma)-1$.  Thus,
\begin{eqnarray}
 \tilde q_i^d[2n_lT] \geq 2T(1-\gamma) -1\label{eq:q_i_d_ineq}
\end{eqnarray}
for $i=2,...,N_c$ if $\tilde q_1^d[2n_lT] \geq 2T(1-\gamma)$.  Thus, any packet the source generates in time slots $2n_lT$, $2n_lT+1$, ... , $2(n_l+1)T-1$ would see a delay of at least $(N_c-1)(2T(1-\gamma)-1)$ before reaching the source gateway.

Now we show by contradiction that for any $l>0$, 
\begin{eqnarray}
 \tilde q_1^d[2(n_l +1)T] &\leq&  L_{(n_l+1)}  \label{eq:q_1_lim}
\end{eqnarray}
where 
\begin{eqnarray}
 L_{(n_l+1)} &=&  \frac{1}{N_c} \left( \sum_{i=2}^{N_c} \tilde q_i^d[2n_lT] + G_{n_l} + \max\{\tilde q_1^d[2n_lT] -2T, 0\}\right) \label{eq:L_def}
\end{eqnarray}
and $\max\{\tilde q_1^d[2n_lT] -2T, 0\}$ is the number of packets remaining in $q_1^d$ after the mobile picked up packets from the source gateway in time slot $2n_lT$.

Suppose $\tilde q_1^d[2(n_l +1)T] \geq L_{(n_l+1)}+1$ for some $l$.  Then, similar to the way we showed Eq. (\ref{eq:q_i_d_ineq}), we can show that $\tilde q_i^d[2(n_l +1)T] \geq L_{(n_l+1)}$ for $i=2, ... , N_c$.  Then, if $\tilde q_1^d[2(n_l+1)T] \geq L_{(n_l+1)}+1$ then $\sum_{i=1}^{N_c} \tilde q_i^d[2(n_l+1)T] \geq N_c L_{(n_l+1)} + 1$ at the very beginning of the time slot $2(n_1+1)T$, which is impossible because at the very end of time slot $2(n_l+1)T-1$, 
\begin{eqnarray}
 \sum_{i=1}^{N_c} \hat q_i^{d}[2(n_l+1)T - 1] &=&  \left( \sum_{i=2}^{N_c} \tilde q_i^d[2n_lT] + G_{n_l} + \max\{\tilde q_1^d[2n_lT] -2T, 0\}\right)\nonumber\\
&=& N_c L_{(n_l+1)}.\label{eq:sum_q}
\end{eqnarray}
Eq. (\ref{eq:sum_q}) should equal $\sum_{i=1}^{N_c} \tilde q_i^d[2(n_l+1)T]$ because by Assumption \ref{assume:gen_assume}, the number of packets in the queues at the very end of a time slot is the same as at the very beginning of the next time slot.  Hence, $\tilde q_1^d[2(n_l+1)T] \leq L_{(n_l+1)}$, and by subtracting $\tilde q_1^d[2(n_l+1)T]$ from LHS and $L_{(n_l+1)}$ from RHS of Eq. (\ref{eq:sum_q}), 
\begin{eqnarray}
 \sum_{i=2}^{N_c} \tilde q_i^{d}[2(n_l+1)T] &\geq& (N_c-1) L_{(n_l+1)}.\label{eq:sum_q2}
\end{eqnarray}
(Note that $\tilde q_i^d[2(n_1+1)T] = \hat q_i^d[2(n_1+1)T-1]$.)

By definition Eq. (\ref{eq:L_def}), 
\begin{eqnarray*}
 L_{(n_l+1)} &\geq& \frac{1}{N_c} \left(\sum_{i=2}^{N_c} \tilde q_i^d[2n_lT] + G_{n_l}\right)\\
&\geq& \frac{1}{N_c} \left((N_c-1)(2T(1-\gamma)-1) + 2T(1-\gamma - \epsilon)\right)\\
&\geq& 2T(1-\gamma-\epsilon)-1
\end{eqnarray*}
where the second Ineq. holds by Eq. (\ref{eq:q_i_d_ineq}) and by Assumption \ref{assume:gen_assume}.  Combining this with Eq. (\ref{eq:sum_q2}), 
\begin{eqnarray}
 \sum_{i=2}^{N_c} \tilde q_i^{d}[2(n_l+1)T] &\geq& (N_c-1)(2T(1-\gamma-\epsilon)-1).\label{eq:eq18}
\end{eqnarray}

Any packet that the source generates in time slots $2(n_l+1)T$, $2(n_l+1)T+1$, ... , $2(n_l+2)T-1$ would see a delay of at least $(N_c-1)(2T(1-\gamma-\epsilon)-1)$ before reaching the source gateway.  This is because the rate of the link from node 2 to node 1 is one packet per time slot and there are at least $(N_c -1)(2T(1-\gamma-\epsilon)-1)$ packets before the newly generated packets.

For all $k>1$ such that $n_l + k <n_{l+1}$, we can show that 
\begin{eqnarray*}
 \tilde q_1^d[2(n_l+k)T] &\leq& \frac{1}{N_c} \left(\sum_{i=2}^{N_c} \tilde q_i^d[2(n_l+k-1)T] +G_{n_l+k-1}\right. \\
& &\left.\max \left\{\tilde q_1^d[2(n_l+k-1)T] - 2T,0 \right\}\right) \\
&=& L_{n_l+k}
\end{eqnarray*}
from which we can show that 
\begin{eqnarray}
 \sum_{i=2}^{N_c} \tilde q_i^d [2(n_l+k)T] \geq (N_c-1)L_{n_l+k}. \label{eq:refNc}
\end{eqnarray}
(Replace $n_l+1$ and $n_l$ with $n_l+k$ and $n_l+k-1$, respectively, in eqs. (\ref{eq:q_1_lim}) through (\ref{eq:sum_q2}) and follow the same reasoning.)  From there, we can further show that 
\begin{eqnarray*}
 L_{n_l+k} &\geq& \frac{1}{N_c} \left(\sum_{i=2}^{N_c} \tilde q_i^d[2(n_l+k-1)T] +G_{n_l+k-1}\right)\\
&\geq&\frac{1}{N_c} \left((N_c-1)(2T(1-\gamma-\epsilon)-1) + 2T(1-\gamma-\epsilon)\right)\\
&\geq& 2T(1-\gamma-\epsilon)-1.
\end{eqnarray*}
For example, $L_{n_l+2} \geq 2T(1-\gamma-\epsilon)-1$ because of Eq. (\ref{eq:sum_q2}), which implies that
\begin{eqnarray}
 \sum_{i=2}^{N_c} \tilde q_i^d [2(n_l+2)T] \geq (N_c-1)(2T(1-\gamma-\epsilon)-1) \label{eq:k2q}
\end{eqnarray}
by plugging in the value for $L_{n_l+2}$ into Eq. (\ref{eq:refNc}).  Because of Eq. (\ref{eq:k2q}), $L_{n_l+3} \geq 2T(1-\gamma-\epsilon)-1$, which then implies $\sum_{i=2}^{N_c} \tilde q_i^d [2(n_l+3)T] \geq (N_c-1)(2T(1-\gamma-\epsilon)-1)$. Following the same reasoning, we can show that $\sum_{i=2}^{N_c} \tilde q_i^d [2(n_l+k)T] \geq (N_c-1)(2T(1-\gamma-\epsilon)-1)$ for $k$ such that $n_l + k < n_{l+1}$.

Thus, any packet the source generates in time slots $2(n_l+k)T$, $2(n_l+k)T+1$, ... , $2(n_l+k+1)T-1$ would see a delay of at least $(N_c-1)(2T(1-\gamma-\epsilon)-1)$ before reaching the source gateway.

For $q_1^d$ to reach at least $2T(1-\gamma)$ in time slot $2nT$, $q_2^d$ must have been at least $2T(1-\gamma)$ at some point in time before $2nT$, or otherwise $q_1^d$ could not reach $2T(1-\gamma)$.  Since no packet can be transmitted out of $q_2^d$ into $q_3^d$ (because the links are directed from node $i+1$ to node $i$), $q_2^d[2nT] \geq 2T(1-\gamma)-1$.  Likewise, since $q_2^d$ must reach at least $2T(1-\gamma)$, $q_3^d$ must have been at least $2T(1-\gamma)$ at some point in time before when $q_2^d$ reaches $2T(1-\gamma)$.  Thus, $q_3^d[2nT] \geq 2T(1-\gamma)-1$, and using the same reasoning, $q_i^d[2nT] \geq 2T(1-\gamma)-1$, $i=2, ..., N_c$.  

Since the net change in $q_2^d$ in each time slot is zero, $q_2^d[t] \geq 2T(1-\gamma)-1$ $\forall~t\geq 0$.    Because the link between nodes 3 and 4 is directed, $q_3^d[t] \geq 2T(1-\gamma)-1$ $\forall ~t\geq 0$.  Using the same logic, $q_i^d[t] \geq 2T(1-\gamma)-1$ $\forall~t\geq 0$ and $\forall~i\geq 2$.

\noindent{\underline{Under BP+SR:}} {Note that $q_1^{g_s} = q_{g_s}^{g_s} = 0$ $\forall ~t$.  Whenever $q_2^{g_s}[t] =1$ for some time slot $t$, in the next time slot, one packet will be transmitted out of $q_2^{g_s}$ to $q_1^{g_2}$.  Since at most one packet can be transmitted into $q_2^{g_2}$ each time slot, we have $q_2^{g_s}[t] < 2$ $\forall~t$.  Likewise, whenever $q_3^{g_s}[t] = 2$ for some time slot $t$, in the next time slot, one packet will be transmitted from $q_3^{g_s}$ to $q_2^{g_s}$, and since at most one packet can be transmitted to $q_3^{g_s}$, we have $q_3^{g_s}[t] < 3$ $\forall ~t$.  Continuing with this logic, we have $q_i^{g_s}[t] < i$ $\forall~t$.}

{In each time slot $t$, if $u_{s}^{g_s}[t] \geq K_s q_s^{g_s}[t]$ (with $K_s = T/N_c$), then $\eta=1$ packet will be transferred from $u_s^{g_s}$ to $q_s^{g_s}$.  Since $q_s^{g_s}[t] < N_c$ $\forall~t$, we have $u_{s}^{g_s}[t] < K_s N_c = T$ $\forall~t$.  }

{Note that the length of $u_{g_s}^{g_d}$ is upper bounded by $2T$, since only $2T$ packets can accumulate at $g_s$ before the mobile comes and picks up the $2T$ packets. Thus, there is at most $N_c^2 + 3T$ packets waiting to be transported by the mobile ahead of any newly generated packet.  Thus, because $q_1^{g_s}=0$ $\forall t$, one packet will be transmitted over the link between node 2 and node 1 if there are any packets in the cluster.}
\done

Considering a line network as we have done here is the same as fixing the route the inter-cluster flow takes to reach the gateways.  Though our analysis here is for a line network, we believe the claim holds in two dimensional clusters where the hop count to reach the gateways increases as the cluster size increases since the proof above essentially shows that what matters is the hop count to the gateway, not the size of the cluster.

{If there are multiple inter-cluster flows, the characteristic of the delay that one would observe under BP+SR and the traditional BP would remain unchanged.  That is, under BP+SR, the delay to reach the mobile transport would be linear in $N_c^2$ and $T$, whereas under the tradition BP, the delay would be linear in the product $N_c T$.  This is because under BP, each internal node would have to match the time scale of the intermittent connectivity by having large queues (on the order of $\theta(T)$), whereas under BP+SR, this is not necessary.}

Since we assume that the internal nodes all use FIFO queues, the end-to-end delay under BP scales linearly with $N_c T$, which is what we observed in Figure \ref{f:N_del_res}.  

{One last case we consider is when we have multiple mobiles with different shuttling times; for example, in the line network we have one mobile that shuttles between the two clusters every 100 time slots and another mobile that takes a million time slots to do the same.  In such cases, the average inter-cluster delay will be on the order of the longer shuttling time.}

\section{Two-Scale BP with Rate Control}\label{sec:bp_with_rc}
When the back-pressure rate control algorithm is implemented on ICN, the previously defined queuing architecture needs to modified slightly.  The main reason for this is that the back-pressure rate control algorithm uses local queue lengths to detect congestion.  The length of queue $u_s^{g_s}$ (see Eq. (\ref{eq:src_routing})) maintained at an inter-cluster source $s$ would only tell about any congestion between $s$ and $g_s$.  Thus, the length of $u_s^{g_s}$ is useless to measure the level of congestion between the source and the ultimate destination.  Thus, in our two-scale BP with rate control, we eliminate type-I queues for gateways, and instead have queues directly to the destinations.

\subsection{Queuing Architecture}
The queuing architecture for internal nodes is the same as under the traditional back-pressure algorithm, i.e., an internal node $i$ maintains a queue for each destination that it receives a packet for -- if $i$ receives a packet destined for $n$, it will create and maintain the queue $q_i^n$.  Each gateway node $g$ does the same, and in addition for each \textit{inter-cluster} destination $d$ in the network, it will create and maintain the queue $\hat q_{g}^{d}$. 

To understand how the queuing architecture works for our two-scale BP with rate control, consider a cluster $\C$ and a gateway $g$ in $\C$.  Suppose an inter-cluster flow $[s,d]$ originates from $\C$.  The gateway $g$ maintains $q_{g}^{d}$ and $\hat q_{g}^d$.  All internal nodes in $\C$ maintain a queue for $d$, and the internal nodes in the neighborhood of $g$ transmit packets for $d$ into $g$'s $q_{g}^d$.  Once the packets for $d$ arrives at $g$, they are immediately placed into $\hat q_{g}^d$ (bypassing $q_g^d$).  The gateway $g$ does not advertise the queue length of $\hat q_{g}^d$ internally; instead, it advertises $\hat q_{g}^d / T$ as the queue length of $q_g^d$.  (Note $q_g^d$ is empty at all times because any packet that comes into $q_g^d$ is immediately taken out and placed into $\hat q_g^d$.  Though it is empty, the length of $q_g^d$ is advertised to the internal nodes as $\hat q_{g}^d /T$.)  Each internal node $i$ in the cluster advertises $q_i^d$  without any scaling, and the back-pressure algorithm operates within the cluster based on the \textit{advertised} queue lengths of $q_n^d$ $\forall~n\in \C$. 

For each inter-cluster destination $d$, each mobile $m$ maintains $\hat q_m^d$.  When a gateway comes in contact with a mobile, they use $\hat q_m^d$ and $\hat  q_g^d$ to compute the back-pressure between them.  (In BP+SR, the type-II queues at the gateways are for destination gateways.  Here, $\hat q_g^d$ is a queue that the gateway $g$ maintains for the actual inter-cluster traffic destination $d$.)

Once the inter-cluster flow packet reaches a gateway $g_d$ in the destination cluster, it is placed into $\hat q_{g_d}^d$.  $g_d$ maintains \textit{two queues} $q_{g_d}^d$ and $\hat q_{g_d}^d$ for $d$ as mentioned before.  In each time slot $t$, $g_d$ transfers $\eta$ ($\eta << R$, where $R$ is the number of packets transferred between mobiles and gateways on contact) packets from $\hat{q}_{g_d}^d$ to $q_{g_d}^d$ if and only if
\begin{eqnarray}
\frac{\hat{q}_g^n[t]}{T} \geq q_g^n[t].
\end{eqnarray}
Once put into $q_{g_d}^d$, the packets are routed to the destination using back-pressure routing in the destination cluster.

To reduce the number of queues to be maintained by each node in our implementation, a gateway that receives packets destined for a different cluster (i.e., is a way-point gateway), does not send out the inter-cluster packets to the internal nodes within its cluster.  This way, an internal node only has to maintain a queue for every other node in the same cluster as itself and only for the other nodes in different clusters that are destinations of inter-cluster traffics originating from the same cluster.

\subsection{Impact of $T$ Estimation}
Our two-scale BP with rate control algorithm requires the knowledge of the time scale difference between the intra-cluster wireless packet transmission and the mobility.  But in fact, even a rough estimate (anything $\Theta(T)$) of the difference is good enough,
and the throughput optimality would still hold.  {The best scaling factor $T$ would be the ratio of the time duration it takes the mobiles to make two contacts to the intra-cluster time slot; however, this is difficult to measure precisely.  If too large of an estimate is used for $T$, it would result in longer queues at the gateways and a longer time before the inter-cluster rates converge {for two-scale BP with rate control; for BP+SR, the time it takes for the source routing to converge would be increased as well}.  If too small of an estimate is used, this would result in fluctuations in the instantaneous inter- and intra-cluster rates {for two-scale BP}.  This rate fluctuations can be seen in Figures \ref{fig:inter_less_rates}, \ref{fig:same_rate}, and \ref{fig:inter_more_rate}.}  {For BP+SR, too small of an estimate of T would result in an increase in delay; for example, if $T=1$ is used, this would result in end-to-end delay that scales with the cluster size, as depicted in Figure \ref{f:N_del_res}.}

{In networks with multiple mobiles with different values of $T$, the throughput optimality property of BP+SR and two-scale BP with rate control will still hold.  In such case, our algorithms (without any further modifications) can use the estimate of the largest $T$, and the delay performance result of BP+SR in Section \ref{sec:buff_usage} would have to be adjusted so that it scales with the largest $T$.  However, the fact that only the nodes involved in inter-cluster traffic (gateways, mobiles, and the inter-cluster source) have to maintain large queues still holds in networks with multiple mobiles with different $T$.}

\subsection{Rate Controlled BP}\label{sec:mod_BP}
Utility maximization has been addressed in a back-pressure framework \cite{NeelyThesis,Sto05,ErySri06,NeeModLi05,AkyAndGupHobSanSto08} to address rate control issues.

We use the formulation in \cite{AkyAndGupHobSanSto08} to describe the idea here.  Each flow $[s,d]$ (whether inter- or intra-cluster) has a utility function $U_{[s,d]}(x_s^d)$, which is a function of the rate $x_s^d$ it is served at.  We assume that all utility functions are strictly concave with continuous derivatives.  The utility maximization problem is the following:
\begin{eqnarray}\label{eq:util_max}
  \max_{\textbf{x} \in \Lambda } \sum_{[s,d]\in \F} U_{[s,d]}(x_s^d).
\end{eqnarray}
Let $x_s^d[t]$ denote the rate at which the flow $f$ is served in time slot $t$.  The rate control algorithm that maximizes
(\ref{eq:util_max}) is the following.  In each time slot $t$, the source $s$ injects $\kappa>0$ packets into the queue $q_{s}^{d}$ if and only if  
\begin{eqnarray}\label{eq:rate_control}
  U'_{[s,d]}(x_s^d[t]) - \beta q_{s}^{d}[t] > 0,
\end{eqnarray}
where $\beta>0$ is a control parameter and $U'_{[s,d]}$ is the first derivative of flow $[s,d]$'s utility function.  The parameter $\beta$ controls how close to the optimal rate allocation the system performs, but this comes at the price of longer queues. We also refer to \cite{AkyAndGupHobSanSto08} for a discussion on this implementation.

\begin{figure}
  \centering
    \includegraphics[scale=0.6]{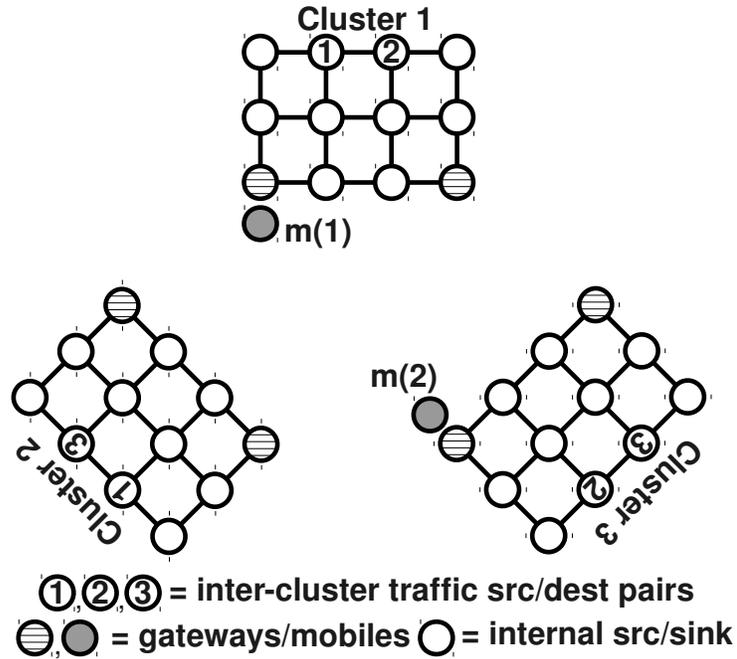} 
  \caption{$3\times 4$ ICN used in simulations for Section \ref{subsec:bpsr_sim}.  We also simulate with cluster size $6\times 4$.  The pair of nodes labeled 1 communicate with each other, likewise for pairs 2 and 3.  Under the traditional BP, the end-to-end inter-cluster delay increases (by approximately twice) as the cluster size increased from $3\times 4$ to $6\times 4$ (see Figure \ref{f:distro}); however, the delay is invariant to the increase in cluster size under BP+SR.  Internal nodes in the clusters maintained large queues to accommodate the inter-cluster flows under BP; under BP+SR, they did not.} \label{f:sim_net}
\end{figure}
\begin{figure*}[h!]
\centering
 \includegraphics[scale=0.6]{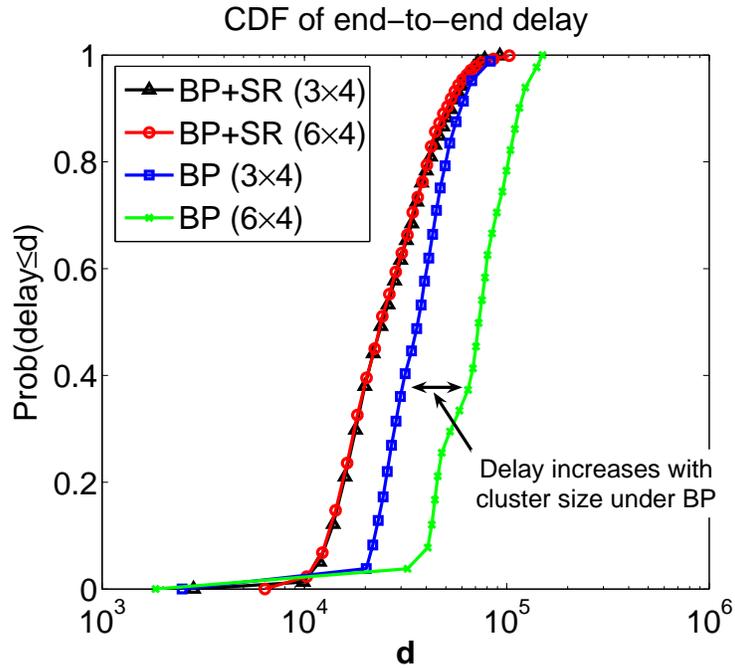}
\caption{CDF of end-to-end delay for inter-cluster traffic} \label{f:distro}
\end{figure*}

\begin{figure*}[h!]
\centering
 \includegraphics[scale=0.6]{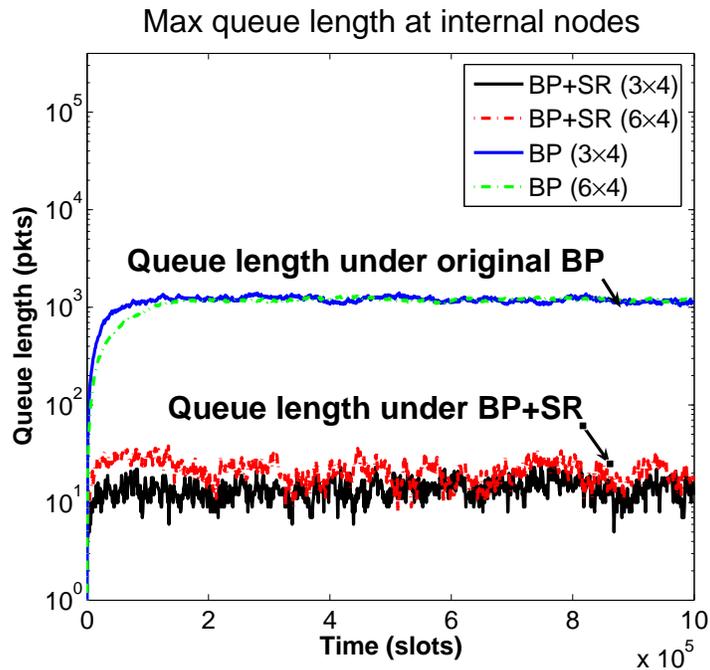}
\caption{Max (type I) queue length at internal node} \label{f:int}
\end{figure*}

\begin{figure*}[h!]
\centering
 \includegraphics[scale=0.7]{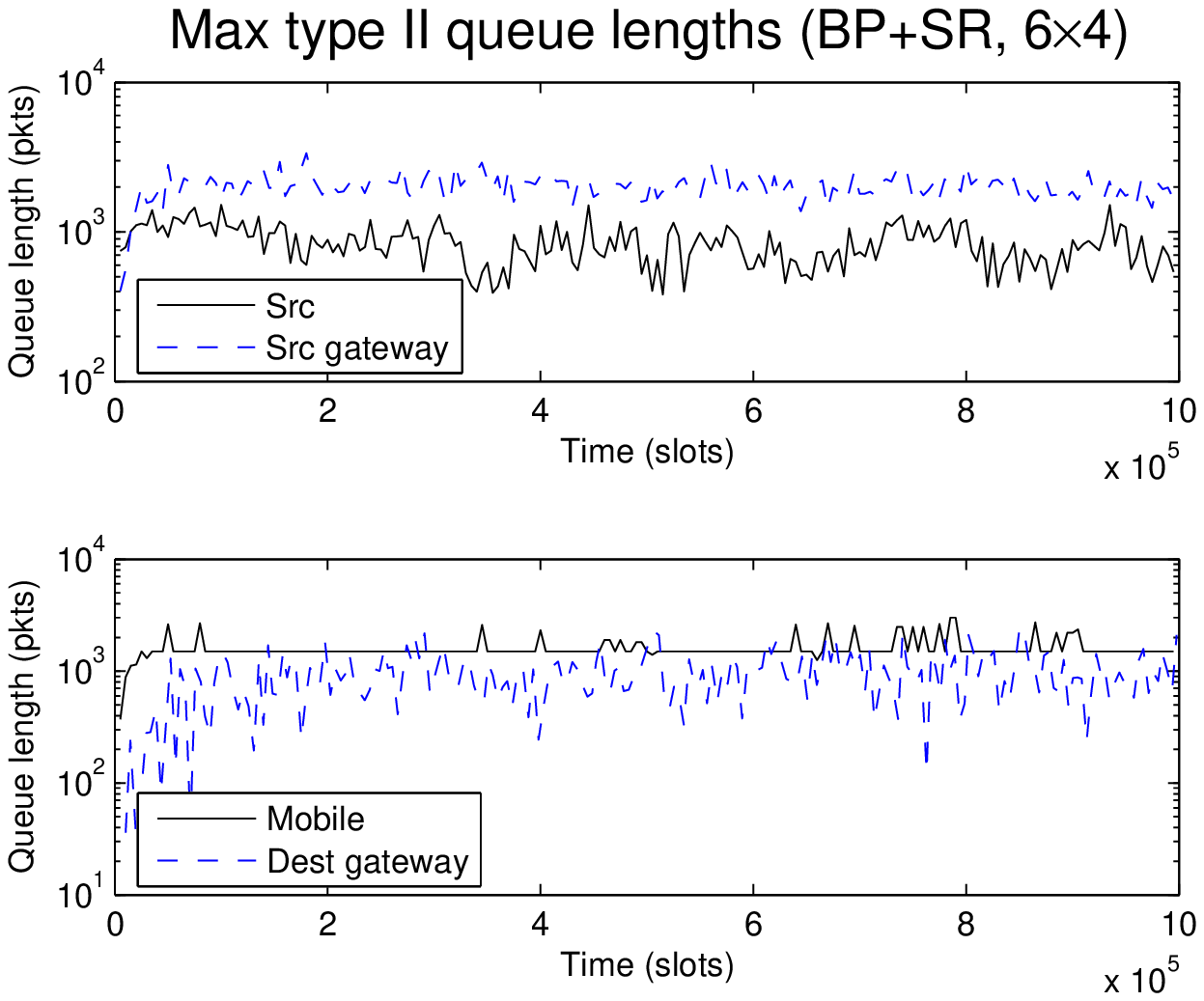}
\caption{Max (type II) queue length at inter-cluster traffic nodes} \label{f:type2q}
\end{figure*}

\subsection{Comparison of Two-Scale BP Algorithms}
{The two-scale BP algorithms we presented in Sections \ref{sec:bp_sr} and \ref{sec:bp_with_rc} both separate the time scale of the intermittent connectivity from the time scale of the internal connections.  Under BP+SR, the rates at which sources generate data are fixed; under two-scale BP with rate control, the source rates are dynamically adjusted to maximize the sum utility.  In addition, under BP+SR the internal nodes in the source cluster maintain queues for the gateways, instead of inter-cluster destinations, which allows the inter-cluster source to perform source routing.  This, however, {cannot} be done when rate control is implemented because the inter-cluster congestion signal {cannot} be passed back to the inter-cluster flow source; when the inter-cluster source maintains queues for gateways and not for inter-cluster destinations, the large gateway queues can signal congestion for gateways and not for the inter-cluster destination.}

\section{Experimental Results}\label{sec:imple}
We present two sets of experimental results; one for two-scale BP with queue reduction (BP+SR) and another with rate control.  {The gain of the BP+SR compared to traditional BP is proportional to the product of the network size and the time scale of the intermittent connections. In small size networks, if the time scale of intermittent connection is large, we will also see a significant gain of the BP+SR.} 

Because of the limited size of our wireless testbed, we instead simulate a large network for BP+SR.
\subsection{Two-Scale BP with Queue Reduction: BP+SR}\label{subsec:bpsr_sim}
We consider two networks.  The first network consists of three clusters, with each cluster composed of 12 nodes ($3\times 4$); the second network consists of three clusters, with each cluster composed of 24 nodes ($6\times 4$).  Figure \ref{f:sim_net} is the first network we simulate.  There are three pairs of inter-cluster traffic sources/sinks, labeled with 1, 2 and 3, i.e. the two nodes labeled with 1 communicate with each other and likewise for 2's and 3's.  The inter-cluster traffic sources generate data at a rate of 0.4 pkts/time slot.  Mobile $m(1)$ comes into contact with gateways $g(i,1)$, $i=1,2,3$ only; similarly for $m(2)$.  $T$ is set to 1000.  In the simulation, we have randomly generated intra-cluster traffic, such that the combined intra- and inter-cluster traffic utilizes all internal links at 90\% of their capacity.  Each internal link has a capacity of 1 pkt/time slot.  $\eta$ is set to 10.  {(If $\eta$ is too large, then the queue lengths would fluctuate by large amounts.  If $\eta$ is too small, then the transfer algorithms Eq. (\ref{eq:src_cond}), (\ref{eq:gg_cond}), and (\ref{eq:gd_cond}) would need to be executed often, increasing the processing burden.)}

The probability that mobile $m(1)$ goes to gateway $g((i+1~\text{mod}~3),1)$ when it is at gateway $g(i,1)$ is 0.8 (gateway $g(i,1)$ belongs to cluster $i$); the probabilities that it stays at $g(i,1)$ or goes to $g((i-1~\text{mod}~3),1)$ both equal to 0.1.  The probability that mobile $m(2)$ goes to $g((i-1~\text{mod}~3),2)$ is 0.8; the probabilities that it stay or goes to $g((i+1~\text{mod}~3),2)$ both equal to 0.1.

The number of packet transferred per contact between a mobile and a gateway is $1500$ pkts/contact ($R=1500$ pkts/contact).

We compare our BP+SR to the traditional BP, as the traditional BP (and its variants) is the only other throughput optimal routing algorithm.  Figure \ref{f:int} shows the evolution of the longest queue under BP and the evolution of the longest type-I queue under BP+SR.  We can see that the longest type I queue is substantially smaller than the longest queue under BP.  In our simulations, we found that under BP \textit{each node in the network} had six queues of order $\Theta(T)$, corresponding to the six different inter-cluster traffic destinations.  Under BP+SR, each inter-cluster source had two queues of order $\Theta(T)$, corresponding to the two gateways in each cluster.  The other nodes with $\Theta(T)$ queues were gateways and mobiles.

In Figure \ref{f:distro}, we show the CDF of the total end-to-end packet delays for inter-cluster traffic.  The average end-to-end delay under BP+SR is roughly 28,000 time slots in both $3\times 4$ and $6\times 4$ cases; the delay under BP is 38,000 time slots in $3\times 4$ case and 78,000 in $6\times 4$ case.  As discussed in Section \ref{sec:buff_usage}, the delay doubled under BP as the cluster size doubled.  The delays for intra-cluster flows stayed the same under BP+SR; BP+SR and BP both showed short delays for intra-cluster flows.

\subsection{Two-Scale BP with Rate Control: Implementation}\label{sec:IMP}
We implemented our two-scale BP on our 16-node testbed.  Our implementation consists of two parts.  The first part is the modification of the MadWifi wireless device driver {for Atheros 5212} to support differentiated levels of channel access on a frame-by-frame basis through varying MAC contention parameters such as AIFS and the contention window sizes.  The second part is our implementation of the modified back-pressure routing algorithm on the Click Modular Router \cite{click}, which utilizes the modified MadWifi to
approximately solve the MaxWeight optimization problem (\ref{eq:bp_max}) without global knowledge.  We describe each part
below.

\noindent \underline{MAC and PHY:} We modified the MadWifi driver so that it supports four hardware queues, with each queue having
different AIFS, $\text{CW}_{\max}$ and $\text{CW}_{\min}$ values shown in Table \ref{table:mac}.  (When two wireless transmissions contend for access to the same channel, the wireless transmission with smaller MAC parameter values will statistically have more access.)  Each hardware queue is given a priority number ranging from 0 to 3.  The modified device driver inspects the TOS field of the IP header of a packet, and injects it into the hardware queue with the same priority number as the TOS field.  If the {BP (Click-layer)} queue difference between the source hop and the destination hop is greater than or equal to 25, between 24 and 14, between 13 and 6, and lower than or equal to 5, we assign TOS levels 0, 1, 2 and 3, respectively (we represent this mapping by the threshold array $L = \{25, 13, 5\}$).  {Here, the threshold arrays are chosen experimentally.  If the thresholds were too close to or too far from the optimal, we did not observe desired behaviors, i.e., a pair of nodes with a large queue difference should have more access to the channel than another pair with a smaller difference.  In our implementation, only one transmission packet/frame is stored in the hardware queues at a node at any given time.  Our modification of the MadWifi device driver is very similar to the one in \cite{WarJanHaRhe09}, with only minor differences.}

\noindent \underline{Routing and Rate Control}: We have implemented the traditional as well as the modified back-pressure
algorithm (presented in Section \ref{sec:mod_BP}) in the Click Modular Router.  Each packet (with 1KB payload) that is sent out is
assigned a value between 0 to 3 that is written to the TOS field.  The assigned value depends on the queue length difference that the wireless transmission source has with the next-hop destination.  Each node broadcasts a beacon on its wireless card every
500msecs.  The beacon contains the information about the queues the node maintains.  The nodes also use the beacons to discover their neighbors.  All data packets received by a node are acknowledged, and the {BP} ACKs sent to the transmitting node also contain the queue information.  Thus, ACKs and beacons are used to calculate the queue differences.  { (IEEE 802.11 ACKs are transmitted, but they are not used by the back-pressure implementation layer.)} The transmitting node retransmits a data packet if an ACK for that packet is not received within 250msecs.  The hop-by-hop ACK guarantees that all packets are received correctly by the final destination, and the ACKs are also used to throttle the transmission rate.  We did not use any RTS-CTS in our implementation. 

A source node inspects the queue for its destination every 5msecs. It runs the back-pressure rate control algorithm in Eq. (\ref{eq:rate_control}) with $\kappa=3;$ i.e., it generates three packets if (\ref{eq:rate_control}) is positive, and generates no packet other-wise (let this variable be $\text{NumGen}$). Then it uses the following update algorithm to estimate the rate $x$ (in Bps):
\begin{eqnarray*} 
x = 0.999x + 0.001 \frac{\text{packet\_size $\times$ NumGen}}{5\text{msecs}}.
\end{eqnarray*}

\begin{table}[t!]	
  \begin{center}
    \begin{tabular}{ | c | c | c | c | c |}
      \hline
      Priority \# & 0 & 1 & 2 & 3 \\ 
      \hline
      AIFS & 1 & 3 & 5 & 7 \\
      \hline
      CWmin & 1 & 7 & 31 & 255 \\ \hline
      CWmax & 7 & 63 & 255 & 1023 \\
      \hline
    \end{tabular}
  \end{center}
  \caption{MAC scheduling parameters of the four MadWifi hardware queues}\label{table:mac}
\end{table}

\begin{figure*}[ht!]
  \begin{center}
    \subfigure[$K_1=K_2=200$. The short flow rate should be two times the long flow rate.]{\includegraphics[width=0.45\textwidth]{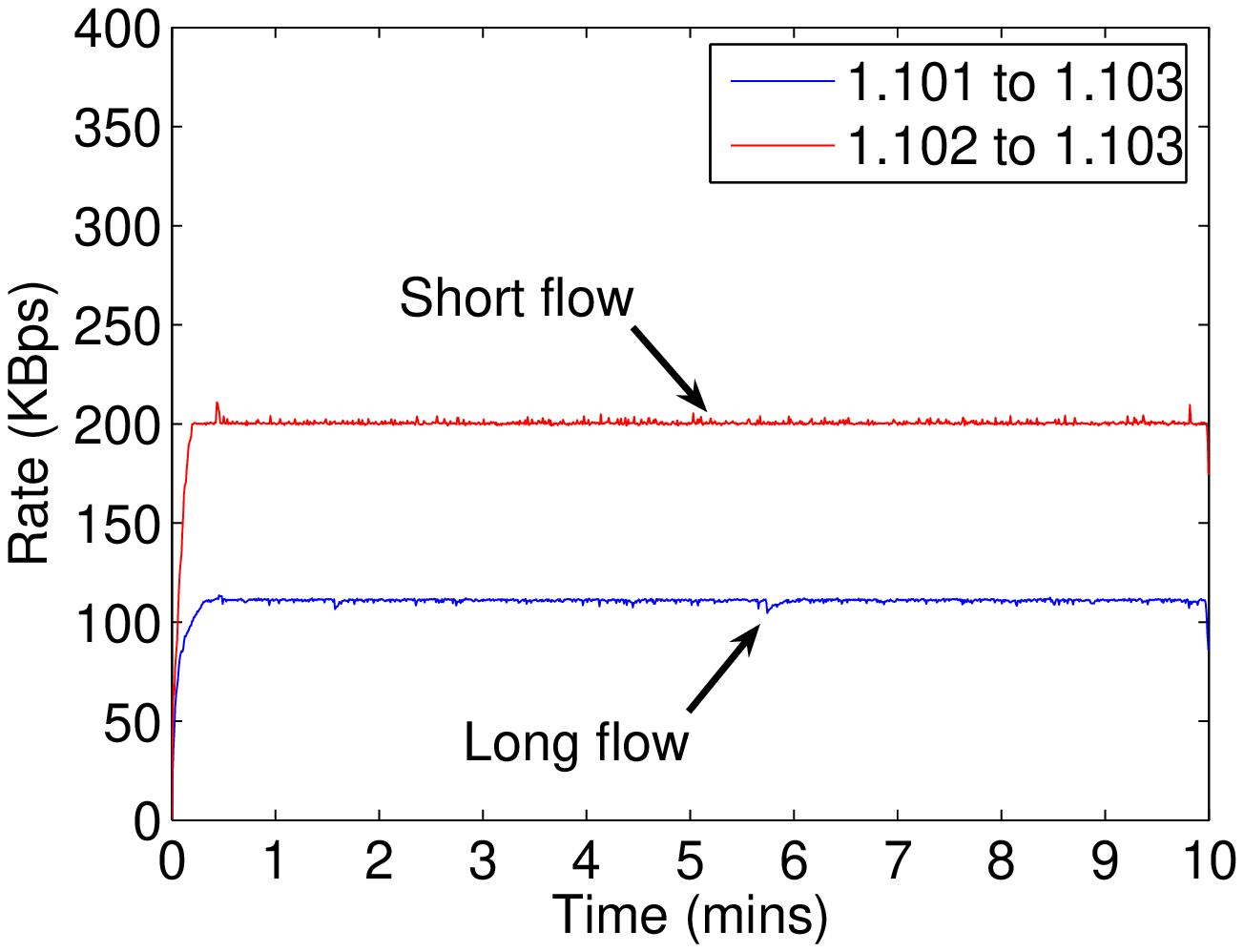} \label{fig:short_more_rate_trace}}
    \subfigure[$K_1=800,~K_2=200$. The long flow rate should be two times the short flow rate.]{\includegraphics[width=0.45\textwidth]{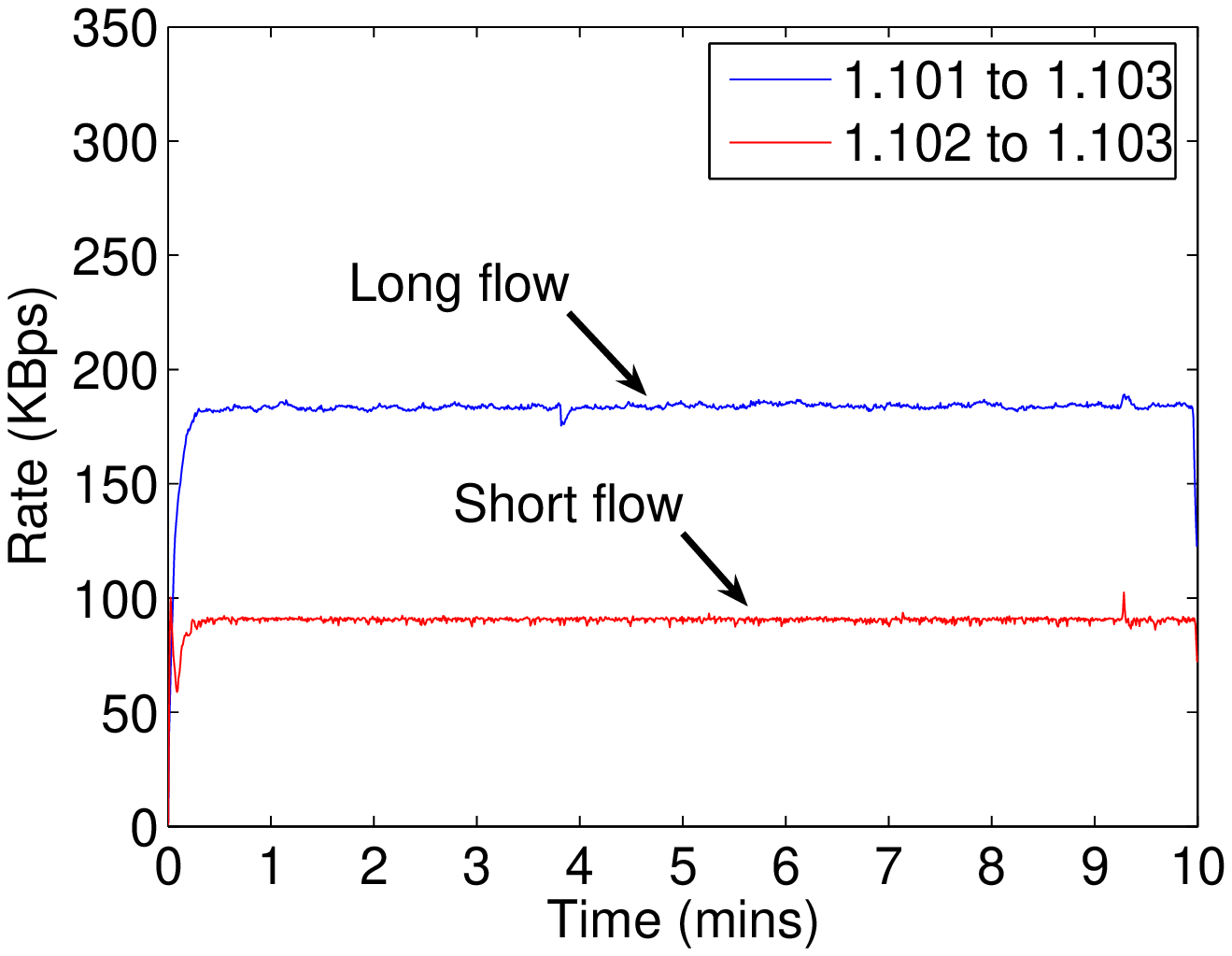} \label{fig:short_less_rate_trace}}\\
    \subfigure[$K_1=400,~K_2=200$. The long and short flows should get the same rate.]{\includegraphics[width=0.5\textwidth]{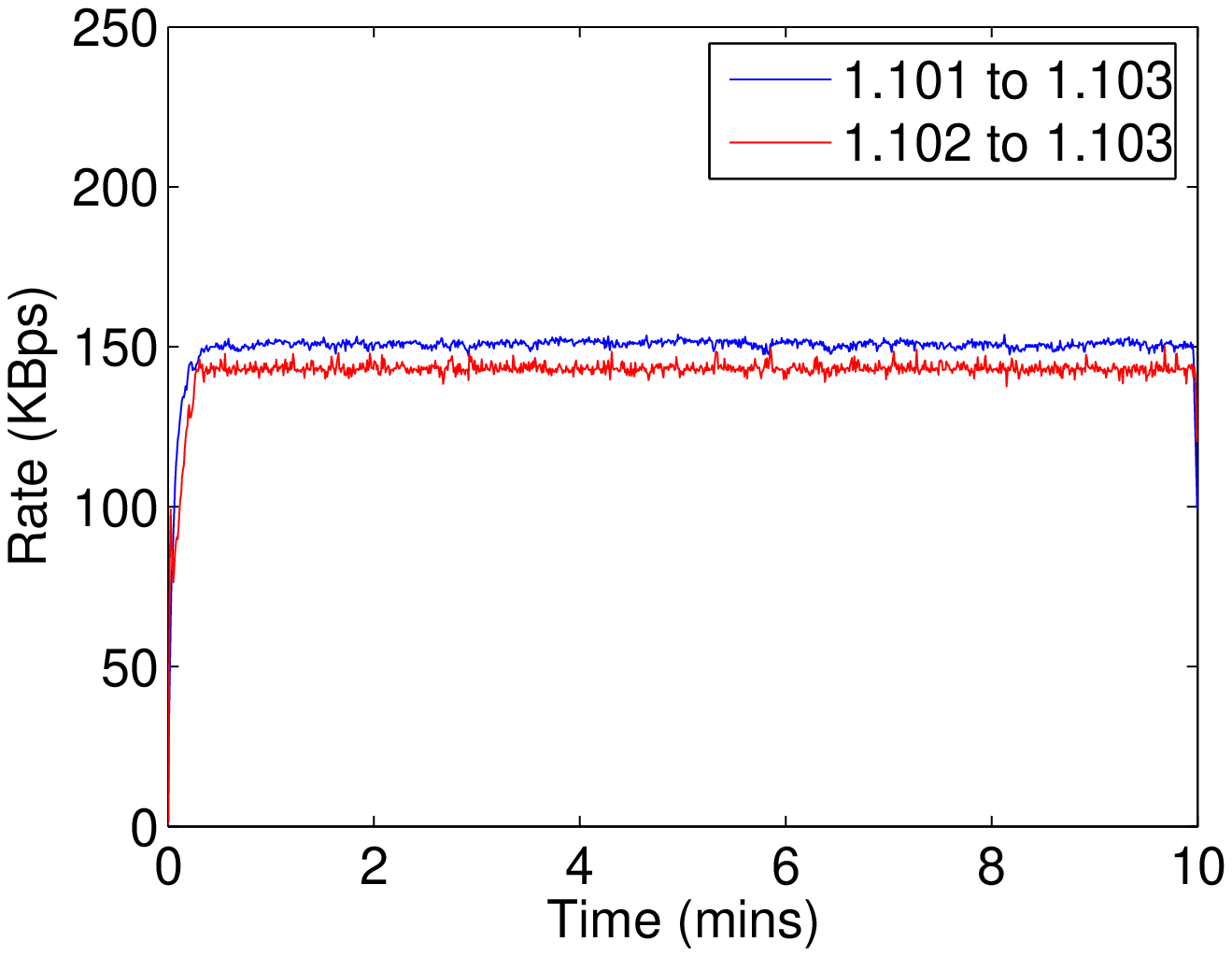} \label{fig:even_rate_trace}}
  \end{center}  
  \caption{Rate allocations in 1.101-1.102-1.103 line network.  We can control which flow gets more rate by controlling the utility function parameters $K_1$ and $K_2$.  As $K_1$ increases relative to $K_2$, the long flow gets more and more rate.}
\end{figure*}

On our MadWifi+Click platform, we implemented the two-scale BP with rate control algorithm for intermittently connected networks.  We emphasize that the focus of our work is \textit{not} an implementation of traditional back-pressure routing with the modified driver; rather, our aim is to decouple the two time scales (mobile-gateway and internal-internal) in such networks.  In order to do this, we first conducted an experiment with the traditional BP on a single time-scale line network.  The single time-scale line network consists of three nodes 1.101, 1.102, and 1.103 arranged in a line.  We have two flows: the long flow originates from 1.101 and is destined for 1.103.  The short flow originates from 1.102 and is destined for 1.103.  The utility functions for long and short flow are $K_1 \log (x_1)$ and $K_2 \log (x_2)$, respectively.  The rate allocations for various values of $K_1$ and $K_2$ are shown in Figures \ref{fig:short_less_rate_trace}, \ref{fig:short_more_rate_trace}, \ref{fig:even_rate_trace}.  We set $L=\{25,13,5\}$.  

The optimal rate allocation is given by the solution to the following optimization problem.  Let $f_1$ and $f_2$ be the fractions of time that the links 1.101-1.102 and 1.102-1.103 are active, respectively.  Due to the coupled wireless interference constraint, $f_1 + f_2 \leq 1$.  The optimal rate allocation can then be obtained by solving  
\begin{eqnarray}
  \text{maximize} && K_1 \log(x_1) + K_2 \log(x_2)\label{eq:line_net_opt} \\
  \text{subject to} && f_1 + f_2 \leq 1 \nonumber \\
  && x_1 \leq f_1 C \nonumber \\
  && x_1 + x_2 \leq f_2 C \nonumber 
\end{eqnarray}
assuming that both links 1.101-1.102 and 1.102-1.103 have the same link capacity $C$.  ($C$ is the one hop transmission rate \textit{between two nodes with no other transmissions in the range} and was measured to be around 465KBps.)  {When $K_1=K_2=200$, the short and long flow rates we experimentally obtained were 200KBps and 110KBps, respectively.  When $K_1=800$, $K_2=200$, the short and long rates obtained were 90KBps and 180KBps, respectively.  When $K_1=400$, $K_2=200$, the short and long rates were 145KBps and 150KBps, respectively.  The experimentally obtained rate allocations are approximately identical to the rate allocation obtained by solving the optimization problem (\ref{eq:line_net_opt})}

\begin{figure*}[t!]
  \centering
    \subfigure[$K_1=K_2=200$.  Intra-rate should be $2\times$ inter-rate.]{\includegraphics[width=0.45\textwidth]{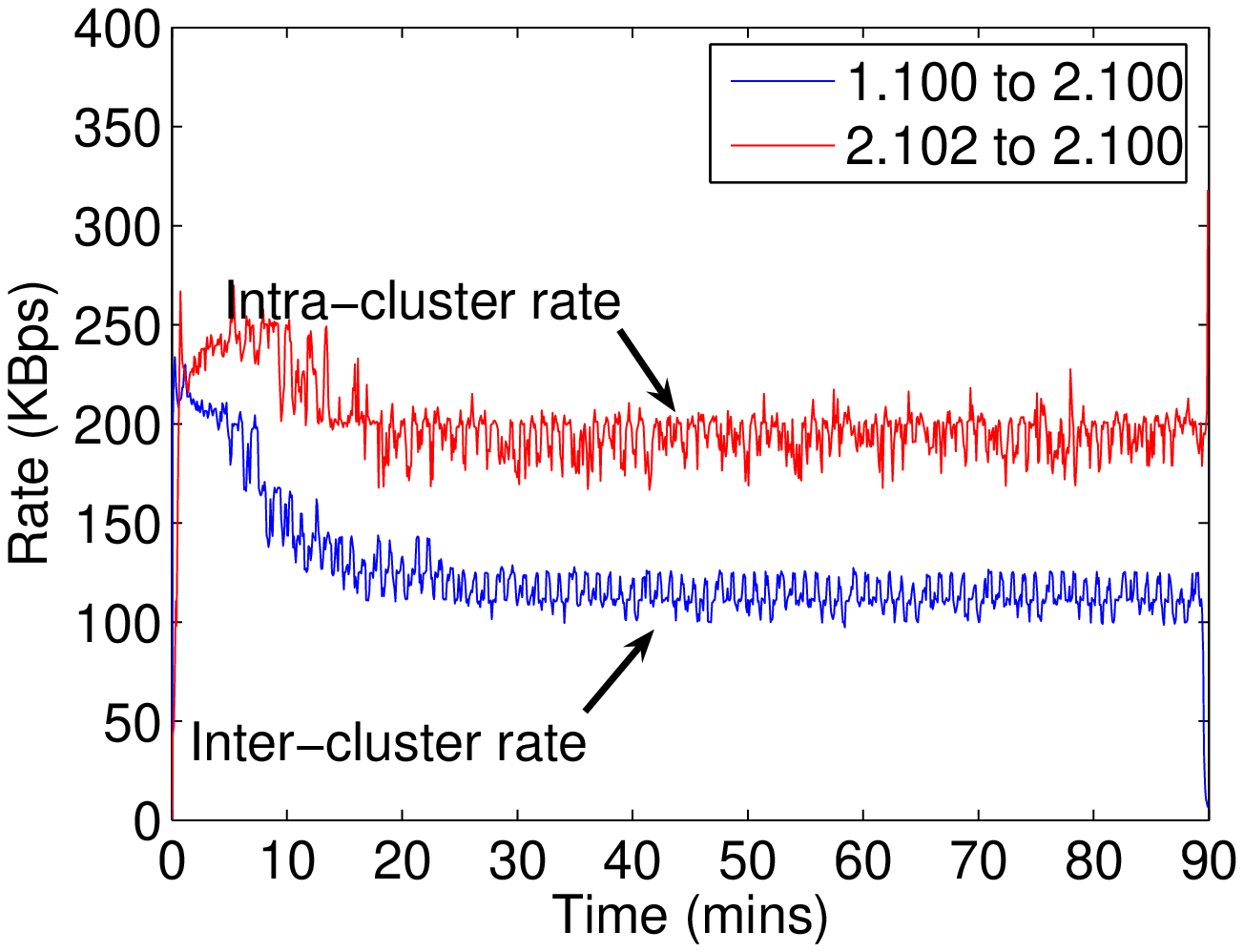} \label{fig:inter_less_rates}}
\subfigure[$K_1=800,~K_2 = 200$.  Inter-rate should be $2\times$ intra-rate.]{\includegraphics[width=0.45\textwidth]{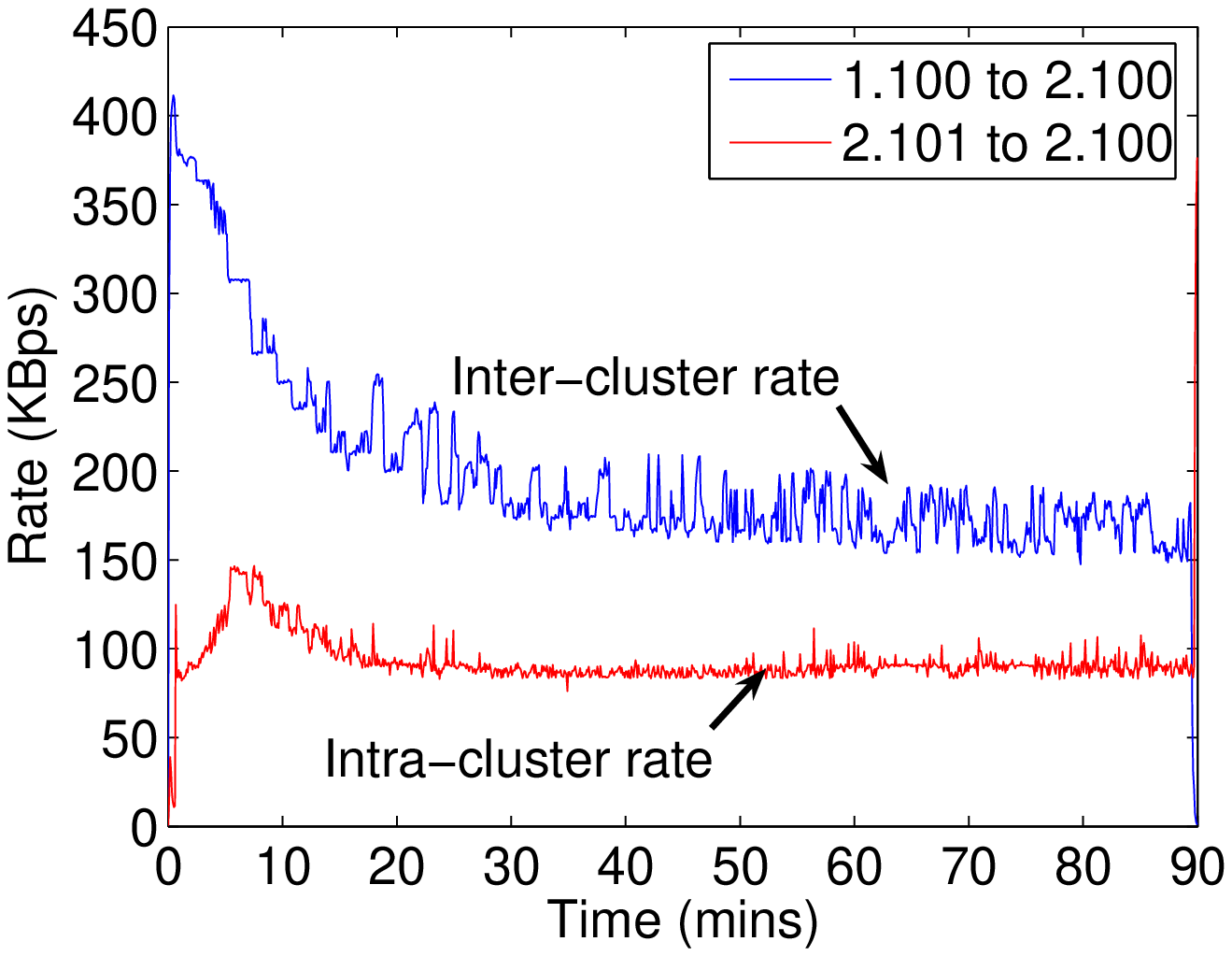} \label{fig:inter_more_rate}}\\
    \subfigure[$K_1=400,~K_2 = 200$.  Intra- and inter-rates should be equal.]{\includegraphics[width=0.45\textwidth]{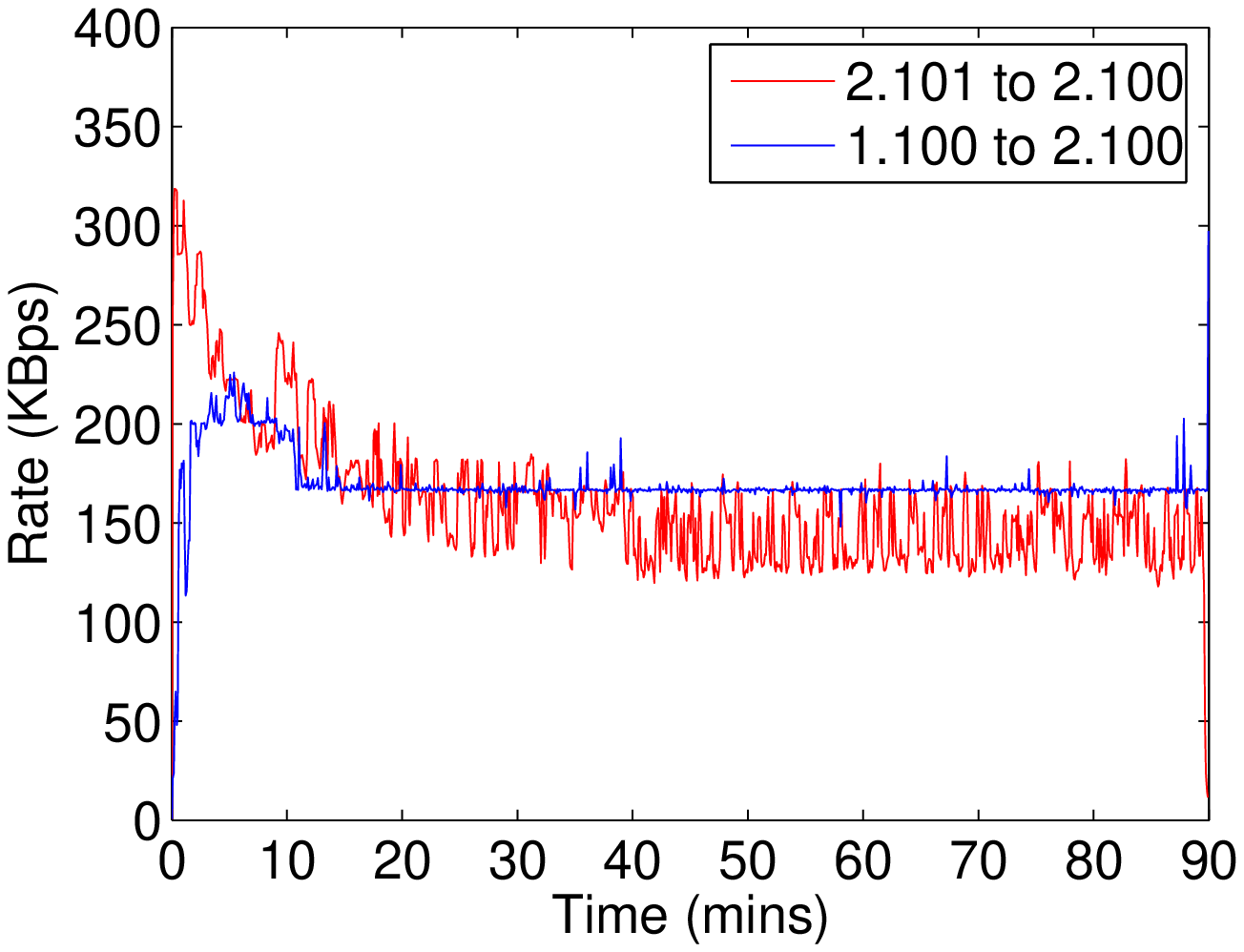} \label{fig:same_rate}}    
    \caption{Rate allocation in the network shown in Figure \ref{fig:test_network} (compare against optimal rate allocations in Figures \ref{fig:short_more_rate_trace}, \ref{fig:short_less_rate_trace}, and \ref{fig:even_rate_trace}).  The presence of the intermittent link is hidden both to the inter- and intra-cluster sources since they achieve the same rates in the network shown in Figure \ref{fig:test_network} as in the two-hop 1.101-1.102-1.103 line network.}
\end{figure*}


\subsection{Verification on a Line Network}
\subsubsection{Utility Maximizing Performance}
We experimentally verify our algorithm on a simple intermittently connected line  network shown in Figure
\ref{fig:test_network} with $N_c=2$.  We used a 100Mbps Cisco switch to emulate the mobile-gateway contacts.  Each gateway node is equipped with one wireless card and an Ethernet port.  The gateways use the wireless card to communicate with the internal nodes and the Ethernet port to communicate with the ``mobile.''

On each contact, up to 6000 packets can be transferred.  Each packet has a payload of 1KB, in addition to IP and Ethernet and the modified BP headers (roughly 6MB per contact).  The mobile contact node switched clusters every 10 seconds, so two consecutive contacts at a gateway are 20 seconds apart.  Thus, the average rate (averaged over many contacts) from the source cluster (the left cluster) to the destination cluster (the right cluster) is 300KBps.  We also chose $T=6000$ and $L=\{25,13,5\}$.

The purpose of the modified BP algorithm is to have the inter-cluster traffic source be totally unaware of the mobile-gateway contacts, and to disturb any intra-cluster traffic as little as possible.  We also want to obtain utility maximizing rate allocation, \textit{even though the clusters are physically separated}.

Let $x_1$ denote the inter-cluster rate, and $x_2$ denote the intra-cluster rate.  The utility functions for inter- and
intra-cluster traffics are $U_1(x_1) = K_1 \log(x_1)$ and $U_2(x_2) = K_2 \log(x_2)$, respectively.  We made sure the only bottleneck is the destination cluster.  (If the bottlenecks are either the source cluster or the mobile-gateway contacts, this could easily be learned by the source.)  Thus, the optimal rate allocation is the solution to the maximization problem (\ref{eq:line_net_opt}) (with $f_1$ and $f_2$ denoting the fraction of time the wireless channel 2.103-2.101 and 2.101-2.100
are active, respectively).  In summary:
\begin{itemize}
\item Recall that the traditional BP rate controller fails to give an optimal rate allocation (unless a {large} scaling is done, resulting in large queue sizes), and resulted in low inter-cluster rates as seen in Figure \ref{fig:bp_rate}.  However, using the modified BP, we get a high, sustained throughput for both the inter- and intra-cluster flows, and their rates are (shown Figures \ref{fig:inter_less_rates}, \ref{fig:same_rate} and \ref{fig:inter_more_rate}) close to the theoretically computed values.  The rates are also close to the ones {obtained in ``Routing and Rate Control'' portion of section \ref{sec:IMP}} (shown in Figures \ref{fig:short_more_rate_trace},  \ref{fig:short_less_rate_trace} and  \ref{fig:even_rate_trace}).  Thus, the modified BP successfully hides the presence of the intermittent links.
\item We also verify that large queues occur only at gateways; the queue sizes at gateways were $\approx 5\times 10^4$.  The queues at internal nodes were small, roughly $\approx 10$.  
\end{itemize}


\begin{figure}
  \centering
    \includegraphics[scale=0.5]{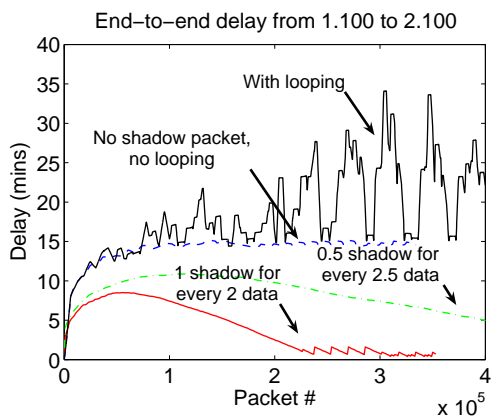} 
    \caption{We can decrease the end-to-end inter-cluster delay first by preventing packet looping (dashed) and second by using shadow packets.  As we use more and more shadow packets per data packet, the delay decreases faster.}\label{fig:delay_comb}
\end{figure}

\subsubsection{End-to-End Delay: Shadow Packets}
We compare the end-to-end delay that inter- and intra-cluster packets {experience}.  Since inter-cluster packets must pass through the gateways and mobiles with large queues, they incur large delays.  One factor causing the large inter-cluster packet delay is that some packets are ``looping'' between the gateways and the mobile.  (``Looping'' through large FIFO queues a few times can increase delays significantly.)  When we prevent this looping, we get inter-cluster delay shown in Figure \ref{fig:delay_comb}.   {(This looping is prevented only at gateways and mobiles.  Looping is prevented by never having the mobile transmit each packet back to the gateway from which it was received the packet.)}

Another factor that contributes to the large inter-cluster delays is that our utility-maximizing rate controller operates very close to the boundary of the capacity region.  This is known to require large queues and can thus cause long delays.  The authors in \cite{BuiSriSto09}, \cite{HuaNee09} deal with this problem by introducing the notion of shadow packets and queues.  Their essential idea here is to trade-off throughput for low delays.  Our implementation of shadow packets is as follows:

\noindent \underline{Shadow packet implementation:} For every $\kappa=3$ packets that the inter-cluster source injects into its queue according to Eq. (\ref{eq:rate_control}), it marks one red (or shadow).  The other two packets are marked blue {(data)}.  These shadow packets are dummy packets and do not contain any useful data (but still have 1KB payload).  The blue packets contain real data.  Thus, the real, useful rate is $0.66 x$.  The gateways and mobiles have two FIFO sub-queues for each inter-cluster destination.  The red and blue packets are separated into these two sub-queues.  The blue packets get transmission priority over the red packets, i.e. shadow packets are transmitted if and only if there are no blue packets that can be served.  The total size (blue queue size + red queue size) is used for back-pressure routing.

The end-to-end delay with shadow packets is shown in Figure \ref{fig:delay_comb}.  {The delay is measured only for the data packets.} (We also did another experiment where we send 1 shadow packet for every 5 data packets on average {(the green dash-dot curve in Figure \ref{fig:delay_comb})}.  {The two shadow packet experiments were done for 60 minutes.  When the rate of shadow packet generation is 0.2/per data packet, the experiment was not long enough for the delay to converge to the minimum, but from the figure, it is clear that the algorithm with shadow packets significantly has a significantly smaller delay than that without shadow packets.}) The delay curve first increases as we first need to build large queues at the gateways and mobiles.  But as the real packets have priority, only the dummy, shadow packets are left behind to hold the steady-state queue sizes required for back-pressure to work.  The inter-cluster delay decreased from roughly 15mins to roughly 1-2mins using shadow packets.  

\section{Experimental Results from Testbed}
We implemented the modified back-pressure algorithm on our 16-node {testbed}. The network we conducted our experiment on is shown in Figure \ref{fig:big_network}.  The nodes in each cluster were placed only a few feet apart. (The clusters were on different channels.)  Each node in a cluster uses packet filtering based on the source IP address; so for example, 1.101 can accept packets only from 1.102 and is only aware of 1.102's presence.  Thus, 1.101 will only transmit to 1.102.  We are aware that any transmission from, say, 1.101 causes interference on all other transmissions because the wireless range is large enough to cover the entire cluster.  However, a node can receive only one transmission at a time, and a failed transmission from, say, 1.103 (which will also transmit to 1.102 only) to 1.102 due to the interference caused by a simultaneous transmission from 1.101 would not have been received by 1.102 anyways even if the nodes were placed farther apart.  {(We are aware that the way we have closely laid out the nodes to conduct our experiment does not completely model the network depicted in Figure \ref{fig:big_network}.  For   example, as depicted in the figure the nodes 1.103 and 1.101 are supposedly placed far apart, but they are within each other's   transmission range.  Thus, there can be a collision between transmissions from 1.101 and 1.103 in the depicted network, but not   in our physical network (because of CSMA).  However, in practice the carrier sensing range of 802.11 is larger than the transmission range.  Hence, the network we are actually modeling is where the nodes 1.101 and 1.103 are out of each other's transmission range, but still within the carrier sense range.)}  Thus, placing the nodes close does not make our experimental results less valid.

Using a 100Mbps switch to emulate the mobile-gateway contacts, up to 6000 packets can be transmitted between a mobile and a gateway (so up to 12000 packets total) per contact.  We picked $T=6000$, $\kappa=\eta=3$, and used queue difference levels $L=\{25,13,5\}$.
After the contact is finished, the mobiles pick one of the other two gateways randomly, and initiate another contact 14 seconds later.  We are aware that 14 seconds is not long enough to model most mobility in the real world.  However, we picked 14 seconds to speed up our experiments so that we can have many contacts within a reasonable amount of time.  Finally, all flows have the same utility function of $200 \log(\cdot)$ in this study.

In summary, we have the following results from our experiment on the network in Figure \ref{fig:big_network}:

\begin{itemize}
\item We observed that even in this larger network, the intra-cluster queues remain very small (between 10-15 pkts).  Only the inter-cluster flows suffer large delays due to longer queues (between $10^4$ and $3\times 10^4$ pkts).

\item Furthermore, using our implementation of the shadow queues and packets (the idea was proposed by the authors in \cite{BuiSriSto09}; we have developed an implementation for intermittently connected networks), we can ``back-off'' from operating on the boundary of the throughput region (i.e., utility optimal), and improve the delay performance for the inter-cluster flows; the inter-cluster delay decreased from $\approx$ 20 mins (blue, solid trace in Figure \ref{fig:big_delay}) to $\approx$ 2 mins using our shadow packets (red, dashed traces in Figure \ref{fig:big_delay}).

\item To get a baseline on the approximate values we should expect, we used the fluid deterministic optimization (that ignores ACKs, collisions, retransmissions) to obtain the ``optimal'' rate allocations to be $x_1=106$, $x_2=83$, $x_3=191$, $x_4=146$ and
  $x_5=87$ (all KBps). Our experimental numbers are $x_1=84$, $x_2=68$, $x_3=120$, $x_4=107$ and $x_5=86$ (all KBps). The rates
  differ from the fluid approximation anywhere from a few percent to about 40\%.  One source of the discrepancy is that the theoretical framework as in Eq. (\ref{eq:line_net_opt}) assumes a fluid model with no contention loss and exponential back-offs{, and does not model the hop-by-hop ACKs we used}.  {In practice}
, there are many packet collisions that trigger the exponential back-offs that decrease the ``link capacity'' in the (idealized) fluid model. 
  
\end{itemize}
\begin{figure*}[t!]
  \centering
    \subfigure[Rates $x_1$ and $x_2$ in the 5.22GHz cluster]{\includegraphics[width=0.45\textwidth]{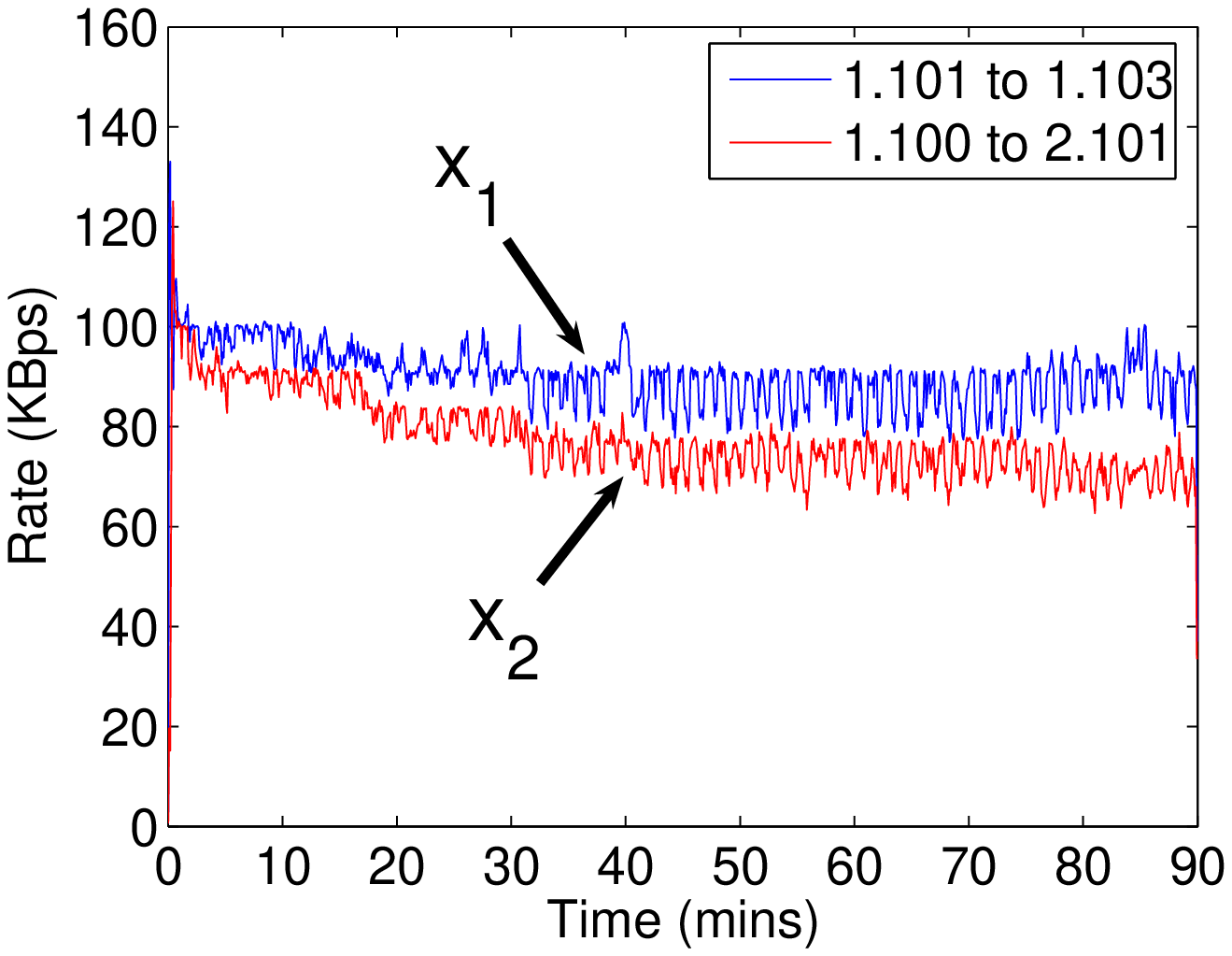} \label{fig:rate_trace1}}
    \subfigure[Rate $x_3$ in the 5.26GHz cluster]{\includegraphics[width=0.45\textwidth]{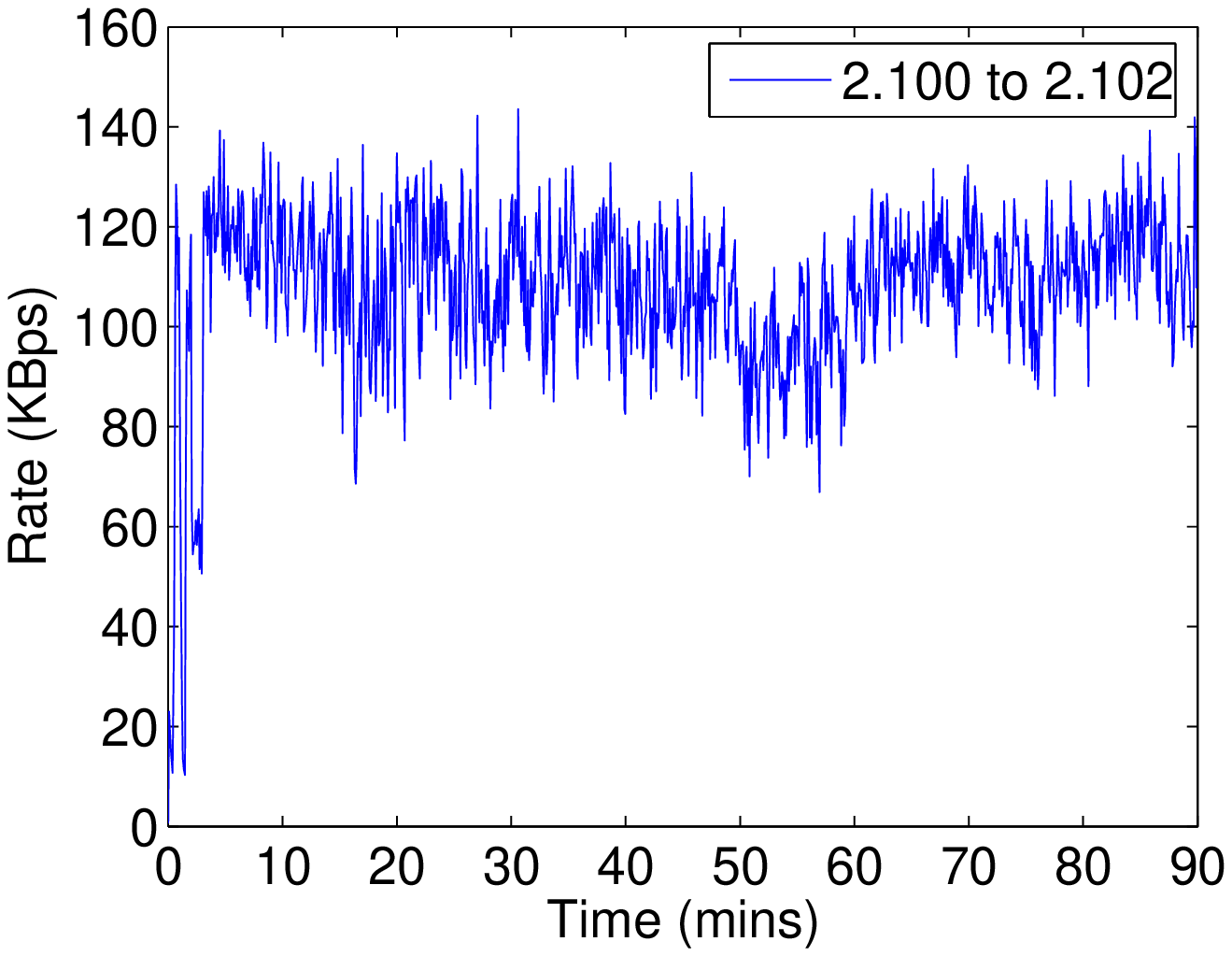} \label{fig:rate_trace2}}\\
    \subfigure[Rates for $x_4$ and $x_5$ in the 5.30GHz cluster]{\includegraphics[width=0.45\textwidth]{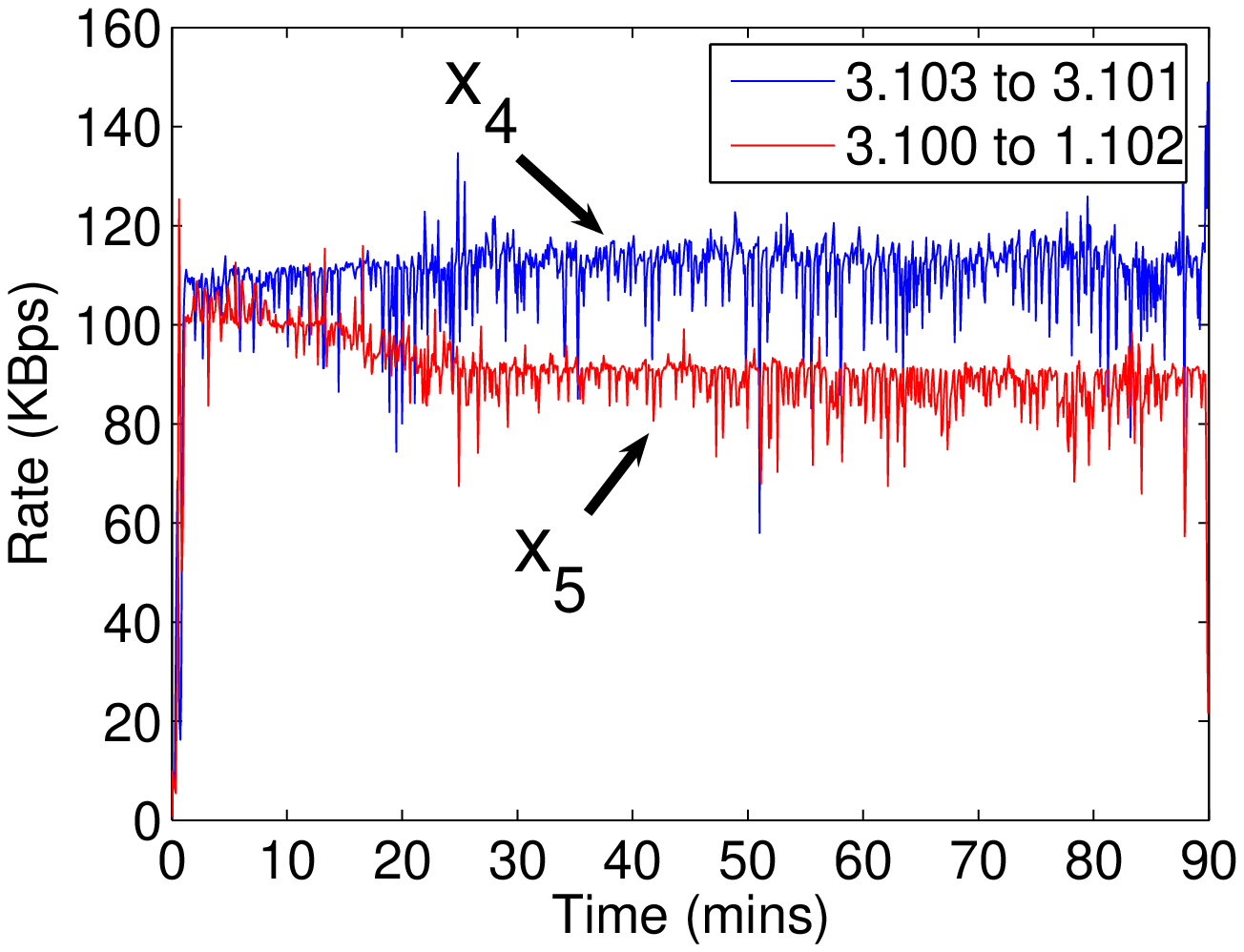} \label{fig:rate_trace3}}
    \caption{Rate allocation in the 16-node network.  See Figure \ref{fig:big_network} for the labels $x_1$, $x_2$, $x_3$, $x_4$ and $x_5$.  $x_2$ and $x_5$ are inter-cluster flows.}
\end{figure*}
\begin{figure}
  \centering
    \includegraphics[scale=0.6]{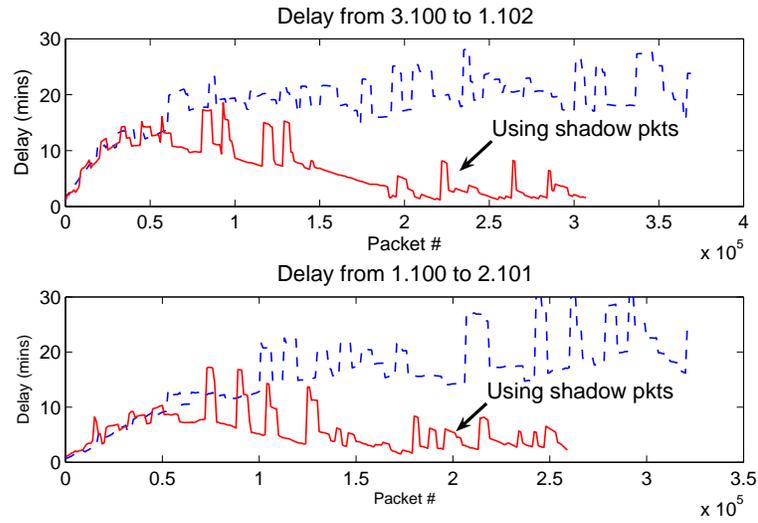}
    \caption{Inter-cluster delays in the network in Figure \ref{fig:big_network} (red, solid=using 1 shadow packet for every 2 data packets)}\label{fig:big_delay}
\end{figure}

\begin{figure}
  \centering
    \includegraphics[scale=0.7]{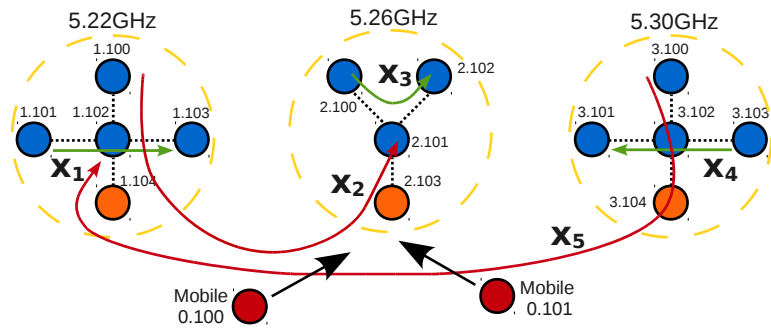}  
  \caption{Our experimental network with 16 nodes}\label{fig:big_network}
\end{figure}

\section{Appendix}
\subsection{More Comments on Overlay Network}\label{sec:overlay_comm}

It has been shown in \cite{NeeModLi05} that by using the packet transfer algorithms (\ref{eq:src_cond}), (\ref{eq:gg_cond}) and (\ref{eq:gd_cond}), the rate $\kappa_a^b[\tau]$ (pkts/time slot) at which node $a$ transfers packets out of $u_a^{b^*}$ to $q_a^b$ converges in each super time slot $\tau$.  (For eqns. (\ref{eq:src_cond}) and (\ref{eq:gd_cond}), $b^*$ is $b$; for eqn. (\ref{eq:gg_cond}), $b^*$ is $l_{a,b}[\tau]$.)  We assume that each super time slot $\tau$ is long enough for this convergence to take place.  (Hence, the rate at which $a$ transmits to $b$ converges to $\kappa_a^b[\tau]$.)   In addition, within each cluster $\C$
\begin{eqnarray*}
\sum_{g_s\in \HH_{\C}} \theta_{s}^{g_s}[\tau] \kappa_{s}^{g_s}[\tau] + \sum_{g_1,g_2\in \HH_{\C}} \theta_{g_1}^{g_2}[\tau] \kappa_{g_1}^{g_2}[\tau] +\sum_{g_d \in \HH_{\C}} \theta_{g_d}^{d} [\tau] \kappa_{g_d}^{d}[\tau]
\end{eqnarray*}
is maximized subject to the constraint
\begin{eqnarray*}
\textbf{x}_{\text{intra}(\C)}\bigcup \left\{\kappa_a^b[\tau]~:~a,b\in \C \right\}
\end{eqnarray*}
is supportable in $\C$.  

We will use $\kappa_{m}^{g}[\tau]$ (either 0 or $R/T$) to denote the rate (pkts/time slot) at which mobile $m$ transmits to gateway $t$ in super time slot $\tau$; likewise for $\kappa_{g}^{m}[\tau]$.

Let $\kappa_{(a,b)}^{c}[\tau]$ denote the rate at which packets from $u_a^c$ are transmitted from $a$ to $b$ in super time slot $\tau$.  For example, $\kappa_{(g_1,g_2)}^{g_d}[\tau]$ is the rate the gateway $g_1$ transfers to $g_2$ the packets destined for $g_d$ out of queue $u_{g_1}^{g_d}$.

\subsection{Notations and definitions needed for the proof of Theorem \ref{th:t_op}}
\subsubsection{$\Delta u_n^j[\tau] = u_n^j[\tau+1] - u_n^j[\tau ]$}\label{sec:del_u}

We group the inputs and outputs and the exogenous arrivals as follows to simplify our analysis of type II queues.  For each $u_n^l$ at node $n$, we have:
\begin{eqnarray*}
x_{in(n)}^{l} =  \left\{\begin{array}{ll}
\sum_{[n,d]\in \F}\sum_{g_d \in \HH_{\C(d)}} x_{n,l}^{g_d,d}  &\text{if } n\in \I_{\C(n)}, l\in \HH_{\C(n)}\\
\sum_{[s,l]\in \F} \sum_{g_s \in \HH_{\C(s)}} x_{s,g_s}^{n,l} & \text{if } n\in \HH_{\C(l)}, l \in \I_{\C(l)}\\
\sum_{[s,d]\in \F} x_{s,n}^{l,d} & \text{if } n\in \HH_{\C(n)},l\in \HH_{\C(l)}, \text{and }\C(n)\neq \C(l)  \\
\end{array}
\right. 
\end{eqnarray*}

\begin{eqnarray*}
\kappa_{in(n)}^{l} =  \left\{\begin{array}{ll}
\sum_{i\in \HH_{\C(n)} \bigcup \MM } \kappa_{(i,n)}^{j}  &\text{if } n\in \HH_{\C(n)} \bigcup \MM  \\
0 & \text{else} 
\end{array}
\right. 
\end{eqnarray*}
\
\begin{eqnarray*}
\kappa_{out(n)}^{l} =  \left\{\begin{array}{ll}
\kappa_{n}^{l} & \text{if } n\in \I_{\C(n)}, l\in\HH_{\C(n)}\\
\kappa_{n}^{l} & \text{if } n\in \HH_{\C(l)}, l\in \I_{\C(l)}\\
\sum_{i\in \HH_{\C(n)}\bigcup \MM } \kappa_{(n,i)}^{l} & \text{if } n\in \HH_{\C(n)}\bigcup \MM  
\end{array}
\right. 
\end{eqnarray*}

The conservation rule for type II queues is thus  $x_{in(n)}^j + \kappa_{in(n)}^{j} = \kappa_{out(n)}^j$.  For regulated type II queues, the queue dynamics are 
\begin{eqnarray*}
u_n^j[\tau+1] - u_n^j[\tau ] = T \left( y_n^j[\tau ] + \kappa_{in(n)}^{j}[\tau ] - \kappa_{out(n)}^{j}[\tau ] + z_n^j[\tau ]\right);
\end{eqnarray*}
for unregulated type II queues,
\begin{eqnarray*}
u_n^j[\tau+1] - u_n^j[\tau ] = T \left( x_n^j[\tau ] + \kappa_{in(n)}^{j}[\tau ] - \kappa_{out(n)}^{j}[\tau ] + z_n^j[\tau ]\right).
\end{eqnarray*}
$z_n^j[\tau]$ is the unused rate (pkts/time slot) in super time slot $\tau$ when there are not enough packets in the type II queue;
\begin{eqnarray*}
 z_n^j[\tau] = \max\{\kappa_{out(n)}^{j}[\tau ] -x_n^j[\tau ] - \kappa_{in(n)}^{j}[\tau ] - u_n^j[\tau],0 \}
\end{eqnarray*}
if $u_n^j$ is unregulated. 
\begin{eqnarray*}
 z_n^j[\tau] = \max\{\kappa_{out(n)}^{j}[\tau ] -y_n^j[\tau ] - \kappa_{in(n)}^{j}[\tau ] - u_n^j[\tau],0 \}
\end{eqnarray*}
if $u_n^j$ is regulated.  

Regulated or not, we have
\begin{eqnarray}
u_n^j[\tau+1] - u_n^j[\tau]& \leq& T \left( (1+\delta) x_n^j[\tau ] + \kappa_{in(n)}^{j}[\tau ] - \kappa_{out(n)}^{j}[\tau ] + z_n^j[\tau ]\right).~~~~~~~\label{eq:reg_u}
\end{eqnarray}
\subsubsection{$\Gamma$}\label{sec:gamma}

We define 
\begin{eqnarray*}
\mu_{(m(j),g(i,j))}&=&\frac{\pi_{m(j)}(i) R}{T},\\
\mu_{(g(i,j),m(j))}&=&\frac{\pi_{m(j)}(i) R}{T},
\end{eqnarray*} and
\begin{eqnarray*}
\vec{\mu}_{\MM} = \left\{\mu_{(m(j),g(i,j))}, \mu_{(g(i,j),m(j))} \right\}_{i,j}.
\end{eqnarray*}
Note that $\vec{\mu}_{\MM}$ is in the unit of packets/time slot.  We assume that the transmissions between mobiles and gateways do not cause interference to other transmissions.

Define the network graph $\GG = (\NN,\LL)$ where $\NN = \left(\bigcup_{\C_i} \NN_{C_i}\right) \bigcup \MM$ and
\begin{eqnarray*}
\LL = \left(\bigcup_{\C_i} \LL_{\C_i}\right) \bigcup \left\{(m(j),g(i,j)),(g(i,j),m(j))~:~i,j\right\}.
\end{eqnarray*}
Let
\begin{eqnarray*}
\Gamma = \left(\bigcup_{\C_i} \Gamma_{\C_i}\right)\bigcup \vec \mu_{\MM}.
\end{eqnarray*}

Let $\mu_{(n_1,n_2)}^{[s,d]}$ be the rate at which the flow $[s,d]$ is served over the link $(n_1,n_2)\in \LL$.  We say $\textbf{x}$ is supportable if there exists $\vec{\mu} \in \Gamma$ such that:
\begin{enumerate}
\item For any node $n_1 \in \NN$ and for all flows $[s,d]$,
\begin{eqnarray*}
x_{s}^{d} 1_{\{s=n_1\}} + \sum_{(n_2,n_1)\in \LL }\mu_{(n_2,n_1)}^{[s,d]} = \sum_{(n_1,n_3)\in \LL } \mu_{(n_1,n_3)}^{[s,d]}.
\end{eqnarray*}
\item $\left\{\sum_{[s,d]} \mu_{(n_1,n_2)}^{[s,d]} \right\} \in \Gamma$.
\end{enumerate}
We say $\textbf{x}_{\text{intra}(\C_i)}$ is supportable in $\C_i$ if the conditions 1) and 2) above hold with $\LL$ and $\Gamma$ replaced with $\LL_{\C_i}$ and $\Gamma_{\C_i}$, respectively.

\subsubsection{$\kappa_{\min}$, $\kappa_{\max}$, and $N_u$}\label{sec:k_min}
Let
\begin{eqnarray*}
\kappa_{\min} = \min\left\{ \min_{\C}\left\{ \max_{(a,b)\in \C~:~\textbf{x}_{\text{inter}(\C)}\bigcup \kappa_a^b \in \Gamma_{\C} } \kappa_a^b\right\},R/T\right\}
\end{eqnarray*}
and
\begin{eqnarray*}
\kappa_{\max} = \max \left\{ \max_{\C} \left\{\max_{(a,b)\in \C~:~\textbf{x}_{\text{inter}(\C)}\bigcup \kappa_a^b \in \Gamma_{\C}} \kappa_a^b\right\}, R/T \right\}.
\end{eqnarray*}
We assume that $\eta > \kappa_{\max}$.  We will see in lemma \ref{lemma:main_b} that there is at least one type II queue in each cluster that is guaranteed to transmit or receive at rate $\kappa_{\min}$ or higher.  $\kappa_{\max}$ is the maximum rate at which any type II queue in the network could change.  Lastly, let
\begin{eqnarray*}
N_u &=& N_s \left(N_g + \max_{\C} \{\max\{\text{\# of inter-cluster traffic srcs} \right.\\
&& \left. \text{in $\C$}, \text{\# of inter-cluster traffic dest. in $\C$}\}\}\right) + N_g
\end{eqnarray*}
and let $\vec \kappa = \{\kappa_a^b\}$.  Note that $N_u > $ \# of type II queues maintained at any node, and $N_u>$ \# of nodes with at least one type II queue.

\subsection{Proof of Theorem \ref{th:t_op}}
We will first bound the queue lengths when the routing algorithms (Eq. (\ref{eq:src_routing}), (\ref{eq:src_cond}), (\ref{eq:gg_cond}), and (\ref{eq:gd_cond})) are updated every $\tilde{T}$ super time slots; we will refer $\tilde{T}$ super time slots as a super-super time slot.  We assume that $\tilde{T}$ is large enough so that for any mobile $m$ and any gateway $g$ $m$ comes into contact with, $m$ makes at least $(1+\epsilon)^{-1}(\pi_m)_g \tilde{T}$ contacts with $g$ over $\tilde{T}$ super time slots.  Let $\tilde{\tau}$ denote $\tilde{\tau}$-th super-super time slot.  We will use this bound to obtain the upper bound when routing algorithms are updated every super time slot.

We prove stability using Lyapunov analysis.  The Lyapunov function we choose is a quadratic function of the type II queue length.  We will show that if any one of the type II queues reaches a certain threshold, the Lyapunov function will start to decrease.  This will show that the Lyapunov function is bounded, and that the queues are bounded as well.

We will use $\kappa_a^b[\tau]$ to denote the transmission rate from node $a$ to $b$ in the overlay network in super time slot $\tau$, and $\kappa_{(a,b)}^{c}[\tau]$ denote the rate at which packets from $u_a^c$ are transmitted from $a$ to $b$ in super time slot $\tau$. 

We define a Lyapunov function $V[\tilde{\tau}\tilde T] = \sum_{n} \sum_{j} (u_n^j[\tilde{\tau}\tilde T ])^2$.  Let
\begin{eqnarray*}
\Delta_{\tilde T} V[\tilde{\tau}\tilde T ]&=& V[(\tilde \tau +1)\tilde T] - V[\tilde \tau \tilde T]\\
&=&\sum_{n} \sum_{j} (u_n^j[(\tilde{\tau}+1) \tilde T ])^2 - (u_n^j[\tilde{\tau}\tilde T ])^2\\
&=& \sum_{n} \sum_{j}(u_n^j[(\tilde{\tau}+1)\tilde T ]- u_n^j[\tilde{\tau}T])\\
&& ~ ~ ~ ~ ~\times (u_n^j[(\tilde{\tau}+1)\tilde T ]+u_n^j[\tilde{\tau}\tilde T ])\\
&=& \sum_{n} \sum_{j} \Delta u_n^j[\tilde{\tau}\tilde T ] (u_n^j[(\tilde{\tau}+1)\tilde T ] + u_n^j[\tilde{\tau}\tilde T ])
\end{eqnarray*}
where
\begin{eqnarray}
\Delta u_n^j[\tilde{\tau}\tilde T ]& =& u_n^j[(\tilde{\tau}+1)\tilde T ] - u_n^j[\tilde{\tau}\tilde T] \nonumber \\
&=& \sum_{l=\tilde \tau \tilde T}^{(\tilde\tau+1)\tilde T-1 } u_n^j[ l +1] - u_n^j[ l] \nonumber \\
&\leq& T \sum_{l=\tilde \tau \tilde T}^{(\tilde\tau+1)\tilde T -1} \left( (1+\delta)x_n^j[l] + \kappa_{in(n)}^j[l] - \kappa_{out(n)}^j[l] + z_n^j[l] \right) \label{eq:del_u_ineq}
\end{eqnarray}
Eq. (\ref{eq:del_u_ineq}) follows from the following queue dynamics:
\begin{eqnarray}
u_n^j[\tau+1] - u_n^j[\tau]& \leq& T \left( (1+\delta) x_n^j[\tau ] + \kappa_{in(n)}^{j}[\tau ] - \kappa_{out(n)}^{j}[\tau ] + z_n^j[\tau ]\right)~~~~~~\label{eq:reg_u2}
\end{eqnarray}
which we derived in Eq. (\ref{eq:reg_u}).  Thus,
\begin{eqnarray*}
\Delta_{\tilde T} V[\tilde{\tau}\tilde T]&\leq& T \sum_{n} \sum_{j} \sum_{l=\tilde \tau \tilde T}^{(\tilde\tau+1)\tilde T -1} \left[(1+\delta) x_n^j[l] + \kappa_{in(n)}^{j}[l] \right. \\
&& \left. - \kappa_{out(n)}^{j}[l ] + z_n^j[l ]\right] (u_n^j[(\tilde{\tau}+1)\tilde T] + u_n^j[\tilde{\tau}\tilde T ])\\
&\leq& T \sum_{n} \sum_{j} \sum_{l}\left[(1+\delta) x_n^j[l ] + \kappa_{in(n)}^{j}[l ] \right. \\
&& \left. - \kappa_{out(n)}^{j}[l ] + z_n^j[l ]\right] (2u_n^j[\tilde{\tau}\tilde T] + \tilde T T \kappa_{\max})\\
&\leq&  T \sum_{n} \sum_{j} \sum_{l}2 u_n^j[\tilde{\tau}\tilde T]\left[(1+\delta) x_n^j[l ] \right. \\
&& \left. + \kappa_{in(n)}^{j}[l ] - \kappa_{out(n)}^{j}[l]\right]+C
\end{eqnarray*}
where $C=4 N_u^2 (\tilde T T \kappa_{\max})^2$ because if $u_n^j[\tilde \tau \tilde T] > \tilde T T \kappa_{\max}$, then $z_n^j[l]=0$ since there are enough packets; and $z_n^j[l] < \kappa_{\max}$ always.  ($\kappa_{\max}$ is defined in subsection \ref{sec:k_min}.  $\kappa_{\max}$ is the maximum rate (pkts/time slot) at which any type II queue can change.)

Let
\begin{eqnarray}
A_{\textbf{x}}[\tilde{\tau}\tilde T]& =& T\sum_{n} \sum_{j}\sum_{l=\tilde \tau \tilde T}^{(\tilde\tau+1)\tilde T-1} u_n^j[\tilde{\tau}\tilde T] x_{in(n)}^j[l]\nonumber\\
&=& T \tilde T \sum_n \sum_j u_n^j[\tilde \tau \tilde T] x_{in(n)}^j[\tilde \tau \tilde T]\label{eq:app_a_eq}
 \end{eqnarray}
since (\ref{eq:src_routing}) is assumed to be updated every $\tilde T$ super time slots; let
\begin{eqnarray}
B_{\kappa}[\tilde{\tau}\tilde T]& & = T\sum_{n} \sum_{j}\sum_{l=\tilde \tau \tilde T}^{(\tilde\tau+1)\tilde T - 1} u_n^j[\tilde{\tau}\tilde T](\kappa_{out(n)}^j[l] - \kappa_{in(n)}^j[l])\nonumber\\
&&  = T \sum_{j} \sum_{(n,m)}\sum_{l=\tilde\tau \tilde T}^{(\tilde \tau+1)\tilde T -1 } ( u_n^j[\tilde{\tau}\tilde T] - u_m^j[\tilde{\tau}\tilde T])\kappa_{(n,m)}^{j}[l].\label{eq:app_b_eq}
\end{eqnarray}
Note that since (\ref{eq:src_routing}) is updated every $\tilde T$ super time slots, 
\begin{eqnarray*}
 A_{\textbf{x}}[\tilde \tau \tilde T] = T \tilde T \sum_{[s,d]\in \F} \min_{g_s\in \HH_{\C(s)}, g_d\in \HH_{\C(d)}}\left(u_s^{g_s}[\tilde \tau \tT ]  + u_{g_s}^{g_d}[\tilde\tau \tT ] + u_{g_d}^{d}[\tilde \tau \tT ]\right) x_{s}^{d}.
\end{eqnarray*}

Then, 
\begin{eqnarray}
\Delta_{\tilde T} V[\tilde{\tau}\tilde T]\leq 2(1+\delta)A_{\textbf{x}}[\tilde{\tau}\tilde T] - 2 B_{\kappa}[\tilde{\tau}\tilde T] + C. \label{eq:ggxx}
\end{eqnarray}

Since $(1+\delta + \epsilon) \textbf{x}$ is supportable, there exists $\vec{\mu}\in \Gamma$ and $\vec \kappa$ such that $\vec{\mu}\in \Gamma$ and
\begin{eqnarray*}
0 = \sum_{n} \sum_{j} u_n^j[\tilde \tau\tilde T]\left( (1+\delta+\epsilon)x_{in(n)}^j + \kappa_{in(n)}^j - \kappa_{out(n)}^j\right).
\end{eqnarray*}
($\Gamma$ is defined in subsection \ref{sec:k_min}.  It is the convex hull of all possible schedules, including all possible mobile-gateway schedules.  $\vec{\kappa}$ is a vector of possible transmission rates in the overlay network.)  

Letting $\tilde{A}_{\textbf{x}}[\tilde{\tau}\tilde T ] = T\tilde T \sum_{n} \sum_{j} u_n^j[\tilde \tau\tilde T] x_{in(n)}^j$ and
\begin{eqnarray*}
\tilde B_{\kappa }[\tilde\tau\tilde T]& =& T \tilde T\sum_{n} \sum_{j} u_n^j[\tilde \tau\tilde T] (\kappa_{out(n)}^j - \kappa_{in(n)}^j) \\
&=& T \tilde T\sum_{j} \sum_{(n,m)} (u_n^j[\tilde\tau\tilde T] - u_m^j[\tilde \tau\tilde T]) \kappa_{(n,m)}^j,
\end{eqnarray*}
we have $(1+\delta+\epsilon) \tilde A_{\textbf{x}}[\tilde \tau\tilde T] = \tilde B_{\kappa} [\tilde \tau\tilde T]$ because $(1+\delta+\epsilon)\textbf{x}$ is supportable.

Hence,
\begin{eqnarray*}
\Delta_{\tilde T} V[\tilde{\tau}\tilde T] &\leq& 2(1+\delta)A_{\textbf{x}}[\tilde{\tau}\tilde T] - 2 B_{\kappa}[\tilde{\tau}\tilde T] + C \\
&&- 2(1+\delta+\epsilon)\tilde{A}_{\textbf{x}}[\tilde\tau\tilde T] + 2\tilde B_{\kappa}[\tilde \tau\tilde T]
\end{eqnarray*}

Since $A_{\textbf{x}}[\tilde\tau\tilde T] \leq \tilde A_{\textbf{x}}[\tilde\tau\tilde T]$ and $\tilde B_{\kappa}[\tilde\tau\tilde T] \leq B_{\kappa}[\tilde \tau\tilde T]$ (see Lemmas \ref{lemma:main_a}  and \ref{lemma:main_b} in the Appendix of this chapter),
\begin{eqnarray*}
\Delta_{\tilde T} V[\tilde \tau\tilde T] \leq -2\epsilon \tilde A_{\textbf{x}}[\tilde\tau\tilde T] + C. 
\end{eqnarray*}

If $\tilde A_{\textbf{x}}[\tilde \tau\tilde T] \geq C/(2\epsilon)$, then $\Delta_{\tilde T} V[\tilde \tau\tilde T] \leq -C$.  

If $\tilde A_{\textbf{x}}[\tilde \tau\tilde T] < C/(2\epsilon)$, then from Eq. (\ref{eq:ggxx}) and the fact that $A_{\textbf{x}}[\tilde \tau\tilde T]  \leq \tilde A_{\textbf{x}}[\tilde \tau\tilde T]$ we have
\begin{eqnarray*}
\Delta_{\tilde T} V[\tilde \tau\tilde T] \leq \tilde C - 2B_{\kappa}[\tilde \tau\tilde T] 
\end{eqnarray*}
 where $\tilde C = C + \frac{1+\delta}{\epsilon} C$.
If there exists $(m,n)$ and $j$ such that
\begin{eqnarray}
u_{m}^j[\tilde \tau\tilde T] - u_{n}^j[\tilde \tau\tilde T] \geq \frac{\tilde C}{T\tilde T \kappa_{\min}} \label{eq:th1_cond}
\end{eqnarray}
then $\Delta V_{\tilde T}[\tilde \tau\tilde T] \leq -\tilde C$ (plug Eq. (\ref{eq:th1_cond}) into Eq. (\ref{eq:app_b_eq})).  ($\kappa_{\min}$ is defined in subsection \ref{sec:k_min}.  It is the minimum rate at which at least one type II queue is guaranteed to change.)  

Such $(m,n)$ and $j$ that satisfy Eq. (\ref{eq:th1_cond}) exist if there exists $m$ such that $u_{m}^j[\tilde \tau\tilde T] \geq \frac{N_u \tilde C}{T \tilde T \kappa_{\min}}$
since $u_m^m[\tilde \tau\tilde T] = 0$ and there are $N_u$ nodes with type II queues ($N_u$ is defined in subsection \ref{sec:k_min}).

Thus, $V[\tilde \tau\tilde T] \leq \left( \bar K  \right)^2$ where
\begin{eqnarray*}
\bar K = N_u \left( \frac{N_u \tilde C}{T \tilde T \kappa_{\min}} + \kappa_{\max}T\tilde T\right)
\end{eqnarray*}
which implies that
\begin{eqnarray}\label{eq:u_bar_k}
u_n^j[\tilde \tau \tilde T] \leq \bar K
\end{eqnarray}
for all $n$ and $j$ when our algorithm is updated every $\tilde T$ super time slots.


 We now consider the real queue $\tilde u_n^j[\tilde \tau \tilde T]$ where $n$ and $j$ are gateways.  Let $M_n^j[\tilde \tau \tilde T]$ be the number of packets in the network that are yet to arrive at the regulated queue $u_n^j$ yet.  Assume there is $\tilde \tau_0$ such that for $\tilde \tau > \tilde \tau_0$, we have
 \begin{eqnarray*}
  M_n^j[\tilde \tau \tilde T] > N_u \bar K + T \tilde T \kappa_{\max}
 \end{eqnarray*}
 which implies $\tilde u_n^j[\tilde \tau \tilde T] > T\tilde T \kappa_{\max}$ due to Eq. (\ref{eq:u_bar_k}).  The packets that did not arrive at $u_n^j$ must be held some where, and because there are fewer than $N_u$ nodes with type II queues and each regulated type II queue is bounded as in Eq. (\ref{eq:u_bar_k}), we must have $\tilde u_n^j[\tilde \tau \tilde T] > T\tilde T \kappa_{\max}$.  Thus, we get
 \begin{eqnarray*}
 M_n^j[(\tilde \tau +1)\tilde T ]&=& \tilde T T \left( \sum_{[s,d]\in \F_{\text{inter}}} x_{s,n}^{j,d}[\tilde \tau T] \right)  + M_n^j[\tilde \tau \tilde T] - \tilde T T y_n^j[\tilde \tau T]\\
 &\leq&  M_n^j[\tilde \tau \tilde T]
 \end{eqnarray*}
 which implies that for $\tilde \tau > \tilde \tau_0$
 \begin{eqnarray*}
 M_n^j[\tilde \tau \tT] \leq \max \left\{ M_n^j [\tilde \tau_0 \tilde T], N_u \bar K + 2 T\tilde T \kappa_{\max}  \right\}.
 \end{eqnarray*}
 Thus, real queues $\tilde u_n^j[\tilde \tau \tilde T]$ are bounded as well.

 We now examine the real queue when $n$ is a gateway and $j$ is an inter-cluster traffic destination.  Let $\mathcal{B}$ denote the bound on $\tilde u_n^j[\tilde t \tT]$ where $n$ and $j$ are gateways.  Let $M_n^j[\tilde \tau \tilde T]$ be the number of packets in the network that are yet to arrive at the regulated queue $u_n^j$, $n \in \cup_{\C} \HH_{\C}$, $j\in \I_{\C(n)}$.  Assume there is $\tilde \tau_0$ such that for $\tilde \tau > \tilde \tau_0$, we have
 \begin{eqnarray*}
  M_n^j[\tilde \tau \tilde T] > N_u \bar K + N_g \mathcal{B}+ T \tilde T \kappa_{\max}
 \end{eqnarray*}
 which implies $\tilde u_n^j[\tilde \tau \tilde T] > T\tilde T \kappa_{\max}$ due to Eq. (\ref{eq:u_bar_k}) and since real, gateway-to-gateway queues are bounded by $\mathcal{B}$ and there are only $N_g$ source gateways where these packets can be.  Thus, we get
 \begin{eqnarray*}
 M_n^j[(\tilde \tau +1)\tilde T ]&=& \tilde T T \left(\sum_{s,g_s} x_{s,g_s}^{n,j}[\tilde \tau T]\right)  + M_n^j[\tilde \tau \tilde T] - \tilde T T y_n^j[\tilde \tau T]\\
 &\leq&  M_n^j[\tilde \tau \tilde T]
 \end{eqnarray*}
 which implies that for $\tilde \tau > \tilde \tau_0$
 \begin{eqnarray*}
 M_n^j[\tilde \tau \tT] \leq \max \left\{ M_n^j [\tilde \tau_0 \tilde T], N_u \bar K + N_g \mathcal{B} + 2 T\tilde T \kappa_{\max}  \right\}.
 \end{eqnarray*}
 Thus, real queues $\tilde u_n^j[\tilde \tau \tilde T]$ are bounded as well.


Now, consider when our algorithm is updated every super time slot (as is in Eq. (\ref{eq:src_routing}), (\ref{eq:src_cond}), (\ref{eq:gg_cond}) and (\ref{eq:gd_cond})). We have
\begin{eqnarray*}
\Delta_{\tilde T} V[\tilde \tau\tilde T]& =& \sum_n \sum_j \sum_{l=\tilde \tau \tilde T}^{(\tilde\tau+1)\tilde T-1}\left( u_n^j[l+1]\right)^2-\left(u_n^j[l] \right)^2\\
&=& \sum_n \sum_j \sum_{l}\left( u_n^j[l+1]-u_n^j[l]\right)\left(u_n^j[l+1]+u_n^j[l] \right)
\end{eqnarray*}
By Eq. (\ref{eq:reg_u}),
\begin{eqnarray*}
\Delta_{\tilde T} V[\tilde \tau\tilde T] &\leq& T\sum_n \sum_j \sum_{l=\tilde \tau \tilde T}^{(\tilde\tau+1)\tilde T-1}\left[ (1+\delta) x_n^j[l] + \kappa_{in(n)}^{j}[l] \right. \\
&& \left. - \kappa_{out(n)}^{j}[l] + z_n^j[l]\right](u_n^j[l+1]+u_n^j[l])\\
&\leq &T  \sum_n \sum_j \sum_{l}\left[ (1+\delta) x_n^j[l] + \kappa_{in(n)}^{j}[l] \right. \\
&& \left. - \kappa_{out(n)}^{j}[l] + z_n^j[l]\right](2u_n^j[l] + T\kappa_{\max})\\
&\leq&T  \sum_n \sum_j \sum_{l}2u_n^j[l]\left[ (1+\delta) x_n^j[l] \right.\\
&& \left. + \kappa_{in(n)}^{j}[l] - \kappa_{out(n)}^{j}[l] \right] + C'\\
&\leq& 2 (1+\delta) \hat{A}_{\textbf{x}}[\tilde \tau] - \hat{B}_{\kappa}[\tilde \tau] + C'
\end{eqnarray*}
where $C' = 4\tilde T N_u^2\left( T \kappa_{\max}\right)^2$ and
\begin{eqnarray*}
\hat A_{\textbf{x}}[\tilde\tau\tilde T] = T\sum_n \sum_j \sum_{l=\tilde \tau \tilde T}^{(\tilde\tau+1)\tilde T} u_n^j[l] x_n^j[l]
\end{eqnarray*}
and
\begin{eqnarray*}
\hat B_{\kappa}[\tilde\tau\tilde T] =T \sum_n \sum_j \sum_{l=\tilde \tau \tilde T}^{(\tilde \tau+1)\tilde T} u_n^j[l]\left( \kappa_{out(n)}^j[l] - \kappa_{in(n)}^j[l]\right).
\end{eqnarray*}
By Lemmas \ref{lem:a_aux} and \ref{lem:b_aux}, $\exists~C_A,C_B$ such that $\hat{A}_{\textbf{x}}[\tilde \tau \tilde T] \leq A_{\textbf{x}}[\tilde\tau \tilde T] + C_A$ and $\hat{B}_{\kappa}[\tilde \tau \tilde T] \geq B_{\kappa}[\tilde \tau \tilde T] - C_B$; thus, we have
\begin{eqnarray*}
\Delta_{\tilde T} V [\tilde \tau \tT] \leq 2(1+\delta) A_{\textbf{x}}[\tilde\tau \tilde T] - B_{\kappa}[\tilde \tau \tilde T] + C''
\end{eqnarray*}
where $C'' = C' + C_A + C_B$.  Then, following the same analysis as before, we get
\begin{eqnarray*}
u_n^j [\tilde \tau \tT] \leq N_u \left( \frac{N_u \tilde C''}{T\tilde T \kappa_{\min}} + \kappa_{\max} T \tilde T\right)
\end{eqnarray*}
using algorithms (\ref{eq:src_routing}), (\ref{eq:src_cond}), (\ref{eq:gg_cond}) and (\ref{eq:gd_cond}), where $\tilde C'' = C'' + (1+\delta)C''/\epsilon$.

Following similar reasoning in \cite{YinSriTow07}, we can bound the real queues.  

Note that the probability that the mobile does not exhibit stationary distribution in $\tT$ super time slots is exponentially decreasing in $\tT$.  Thus, we can obtain an expected bound on the type-II queues.  Furthermore, we can use theorem 1 of \cite{NeeModLi05} to bound the type-I queues since in eqs. (\ref{eq:src_cond}), (\ref{eq:gg_cond}) and (\ref{eq:gd_cond}), type-II queue sizes are used as a linear utility function.
\done

\subsection{$\tilde A_{\textbf{x}}[\tilde \tau \tilde T] \geq A_{\textbf{x}}[\tilde \tau \tilde T]$}
\begin{lemma}\label{lemma:main_a}
$\tilde A_{\textbf{x}}[\tilde \tau \tilde T] \geq A_{\textbf{x}}[\tilde \tau \tilde T]$.
\end{lemma}
\noindent{\textit{Proof: }}
\begin{eqnarray*}
\tilde A_{\textbf{x}}[\tilde \tau \tilde T]&=&  T\tilde T\sum_{[s,d]\in \F} \sum_{g_s\in \HH_{\C(s)}} \sum_{g_d \in \HH_{\C(d)}}\left(u_s^{g_s}[\tilde \tau \tT ] + u_{g_s}^{g_d}[\tilde\tau \tT ] + u_{g_d}^{d}[\tilde \tau \tT ]\right) x_{s,g_s}^{g_d,d}\\
&\geq& T \tilde T \sum_{[s,d]\in \F} \min_{g_s\in \HH_{\C(s)}, g_d\in \HH_{\C(d)}}\left(u_s^{g_s}[\tilde \tau \tT ]  + u_{g_s}^{g_d}[\tilde\tau \tT ] + u_{g_d}^{d}[\tilde \tau \tT ]\right) x_{s}^{d}\\
&=& A_{\textbf{x}}[\tilde\tau \tilde T]
\end{eqnarray*}
since $\sum_{g_s\in \HH_{\C(s)}} \sum_{g_d\in \HH_{\C(d)}} x_{s,g_s}^{g_d,d} = x_s^d$, and because of algorithm (\ref{eq:src_routing}) and the fact that the source routing is updated every $\tilde T$ super time slots.
\done\\
\subsection{$\tilde B_{\kappa}[\tilde \tau \tilde T] \leq B_{\kappa}[\tilde \tau \tilde T]$}
\begin{lemma}\label{lemma:main_b}
$\tilde B_{\kappa}[\tilde \tau \tilde T] \leq B_{\kappa}[\tilde \tau \tilde T]$.
\end{lemma}
\noindent{\textit{Proof: }}In each cluster $\C$, there are intra-cluster traffics mixed with source-to-gateway relay, gateway-to-gateway relay, and gateway-to-destination relay, all of which form parts of inter-cluster traffic.

We will use the theorems from \cite{NeeModLi05} to prove our lemma.

In the super-super time block $\tilde \tau$, let the utility of each intra-cluster traffic $[s,d]$ in $\C$ be $U_{[s,d]}(x_{[s,d]}) = \alpha x_{[s,d]}$, where $\alpha$ is some constant.

For each source-to-gateway relay that originates from $s\in \C$ ending at gateway $g_{s}\in \C$, let the utility function be
\begin{eqnarray*}
U_{[s,g_{s}]}(\kappa_{s}^{g_s}) = \theta_{s}^{g_{s}} [\tilde \tau \tT] \kappa_{s}^{g_{s}}.
\end{eqnarray*}

For each gateway-to-gateway relay between $g_1$ and $g_2$, let
\begin{eqnarray*}
U_{[g_1,g_2]}(\kappa_{g_1}^{g_2}) = \theta_{g_1}^{g_2}[\tilde \tau \tT ] \kappa_{g_1}^{g_2}.
\end{eqnarray*}

For each gateway-to-destination between $g_{d}$ and $d$, let
\begin{eqnarray*}
U_{[g_{d},d]}(\kappa_{g_{d}}^{d}) = \theta_{g_{d}}^{d} [\tilde \tau \tT] \kappa_{g_{d}}^{d}.
\end{eqnarray*}

We assume that $\alpha >> \theta_{s}^{g_{s}}[\tilde \tau \tT], \theta_{g_1}^{g_2}[\tilde \tau \tT], \theta_{g_{d}}^{d}[\tilde \tau \tT]$.

It is shown in Theorem 1 of \cite{NeeModLi05} that our transfer algorithms (\ref{eq:src_cond}), (\ref{eq:gg_cond}) and (\ref{eq:gd_cond}) solve the optimization problem
\begin{eqnarray*}
\text{Maximize: } &&  \sum U_{[s,d]}(x) + \sum U_{[s,g_d]}(\kappa_{s}^{g_s}) \\
&& + \sum U_{[g_1,g_2]}(\kappa_{g_1}^{g_2}) + \sum U_{[g_d,d]}(\kappa_{g_d}^d)\\
\text{Subject to: } && \{x_{[s,d]}: \F_{\text{intra}(\C)} \} \bigcup \{ \kappa_{a}^{b} \} \in \Gamma_{\C}\\
&& \{x_{[s,d]}\} \leq \textbf{x}_{\text{intra}(\C)}
\end{eqnarray*}
Since  $\alpha >> \theta_{s}^{g_{s}}[\tilde \tau \tT], \theta_{g_1}^{g_2}[\tilde \tau\tT], \theta_{g_{d}}^{d}[\tilde \tau \tT]$, we assume that the above optimization is solved with $\{x_{[s,d]}\} = \textbf{x}_{\text{intra}(\C)}$.  Then, the above optimization becomes
\begin{eqnarray*}
\text{Maximize: } &&  \sum U_{[s,g_d]}(\kappa_{s}^{g_s}) + \sum U_{[g_1,g_2]}(\kappa_{g_1}^{g_2}) + \sum U_{[g_d,d]}(\kappa_{g_d}^d)\\
\text{Subject to: } && \textbf{x}_{\text{intra}(\C)} \bigcup \{ \kappa_{a}^{b} \} \in \Gamma_{\C},
\end{eqnarray*}
which is solved by transfer algorithms (\ref{eq:src_cond}), (\ref{eq:gg_cond}) and (\ref{eq:gd_cond}) according to \cite{NeeModLi05}.

Let $\{ \kappa_{a}^{b} [\tilde \tau \tT]\}$ be the set of values that maximize the above optimization problem.  Then,
\begin{eqnarray}
\lefteqn{T\tilde T \sum_{j}\sum_{(n,m): n,m \notin \MM} \left(u_n^j[\tilde \tau\tT] - u_n^j[\tilde \tau\tT]\right)\kappa_{(n,m)}^j[\tilde \tau\tT]\geq } &&\nonumber \\
&&T \tilde T\sum_{j}\sum_{(n,m): n,m \notin \MM} \left(u_n^j[\tilde \tau\tT] - u_n^j[\tilde \tau\tT]\right)\kappa_{(n,m)}^j\label{eq:b_second}
\end{eqnarray}
for each cluster $\C$.

Let $1_{\{m_k =g_{i,k}\}}[l]=1$ if $m(k)$ and $g(i,k)$ are in contact in super time slot $l$.  We now consider the mobile-to-gateway contacts.
\begin{eqnarray}
\lefteqn{\sum_{l=0}^{\tilde T-1} \sum_{m_k\in \MM} \sum_{g_{i,k}}  1_{\{m_k =g_{i,k}\}}[\tilde\tau \tT + l]R}&&\nonumber\\
&& \left[\left(u_{m_k}^{j_{(m_k, g_{i,k})}[\tilde \tau\tT]}[\tilde \tau\tT] - u_{g(i,k)}^{j_{(m_k, g_{i,k})}[\tilde \tau\tT]}[\tilde \tau\tT] \right) \right.\nonumber\\
&&\left. + \left(u_{g(i,k)}^{j_{(g_{i,k},m_k)}[\tilde \tau\tT]}[\tilde \tau\tT] - u_{m(k)}^{j_{(g_{i,k},m_k)}[\tilde \tau\tT]}[\tilde \tau\tT] \right)\right]\label{eq:b_mg}\\
&=& T\tilde T\sum_{m_k\in \MM} \sum_{g_{i,k}} \kappa_{(m_k,g_{i,k})}[\tilde \tau\tT] \left(u_{m_k}^{j_{(m_k, g_{i,k})}[\tilde \tau\tT]}[\tilde \tau\tT] - u_{g_{i,k}}^{j_{(m_k, g_{i,k})}[\tilde \tau\tT]}[\tilde \tau\tT] \right) \nonumber\\
&& + \kappa_{(g_{i,k},m_k)}[\tilde \tau\tT] \left(u_{g_{i,k}}^{j_{(g_{i,k},m_k)}[\tilde \tau\tT]}[\tilde \tau\tT] - u_{m_k}^{j_{(g_{i,k},m_k)}[\tilde \tau\tT]}[\tilde \tau\tT] \right)\nonumber
\end{eqnarray}
since every super time slot $m_k$ and $g_{i,k}$ are in contact, $R$ packets are uploaded and downloaded and thus
\begin{eqnarray*}
 T\tilde T \kappa_{(m_k,g_{i,k})}[\tilde \tau\tT]& =& \sum_{l=0}^{\tilde T-1} 1_{\{m_k = g_{i,k}\}}[\tilde\tau \tT+ l] R, \\
 T \tilde T\kappa_{(g_{i,k},m_k)}[\tilde \tau\tT]& =& \sum_{l=0}^{\tilde T-1} 1_{\{m_k = g_{i,k}\}}[\tilde\tau \tT + l] R.
\end{eqnarray*}

Since we assumed that any mobile $m$ makes at least $(1+\gamma)^{-1}(\pi_m)_g \tilde{T}$ contacts with $g$ ($g$ is any gateway $m$ can come in contact with) over $\tilde{T}$ super time slots (see proof of theorem \ref{th:t_op}), we have
\begin{eqnarray*}
 \sum_{l=0}^{\tilde T-1} 1_{\{m_k = g_{i,k}\}}[\tilde\tau\tT + l] \geq (1+\gamma)^{-1}(\pi_{m_k})_{g_{i,k}} \tilde T
\end{eqnarray*}
and thus
\begin{eqnarray}
\text{Eq. }(\ref{eq:b_mg}) &\geq&  \sum_{m_k\in \MM} \sum_{g_{i,k}} (1+\gamma)^{-1}(\pi_{m_k})_{g_{i,k}} \tilde T R \nonumber \\
&&\left\{ \left(u_{m_k}^{j_{(m_k, g_{i,k})}[\tilde \tau\tT]}[\tilde \tau\tT] - u_{g_{i,k}}^{j_{(m_k, g_{i,k})}[\tilde \tau\tT]}[\tilde \tau\tT] \right)\right.\nonumber \\
&&\left. + \left(u_{g_{i,k}}^{j_{(g_{i,k},m_k)}[\tilde \tau\tT]}[\tilde \tau\tT] - u_{m_k}^{j_{(g_{i,k},m_k)}[\tilde \tau\tT]}[\tilde \tau\tT] \right)\right\}\label{eq:gamma10}
\end{eqnarray}
Since $(1+\gamma)(1+\delta+\epsilon)\textbf{x}$ is supportable, we have
\begin{eqnarray*}
T\tilde T \sum_{j} \kappa_{(m_k,g_{i,k})}^j &\leq& (1+\gamma)^{-1} (\pi_{m_k})_{g_{i,k}} \tilde T R \\
T \tilde T\sum_{j} \kappa_{(g_{i,k}, m_k)}^j &\leq& (1+\gamma)^{-1} (\pi_{m_k})_{g_{i,k}} \tilde T R
\end{eqnarray*}

In addition, because we have
\begin{eqnarray*}
u_{m_k}^{j_{(m_k, g_{i,k})}[\tilde \tau\tT]}[\tilde \tau\tT] - u_{g_{i,k}}^{j_{(m_k, g_{i,k})}[\tilde \tau\tT]}[\tilde \tau\tT]
 \geq u_{m_k}^j[\tilde \tau\tT] - u_{g_{i,k}}^j[\tilde \tau\tT]\\
u_{g_{i,k}}^{j_{(g_{i,k},m_k)}[\tilde \tau\tT]}[\tilde \tau\tT] - u_{m_k}^{j_{(g_{i,k}, m_k)}[\tilde \tau\tT]}[\tilde \tau\tT]
 \geq u_{g_{i,k}}^j[\tilde \tau\tT] - u_{m_k}^j[\tilde \tau\tT]
\end{eqnarray*}
for any $j$ by gateway$\leftrightarrow$mobile back-pressure algorithm equations (\ref{eq:gmb_1}) and (\ref{eq:gmb_2}), we have
\begin{eqnarray}
\text{RHS of Ineq.}~(\ref{eq:gamma10})&\geq& T\tilde T \sum_{m_k \in \MM} \sum_{g_{i,k}} \sum_j \kappa_{(m_k,g_{i,k})}^j\left( u_{m_k}^j[\tilde \tau\tT] - u_{g_{i,k}}^j[\tilde \tau\tT] \right)\nonumber \\
&& + \kappa_{(g_{i,k},m_k)}^j\left(  u_{g_{i,k}}^j[\tilde \tau\tT] - u_{m_k}^j[\tilde \tau\tT]\right).\label{eq:b_last}
\end{eqnarray}
LHS of Ineq. (\ref{eq:b_second}) + Eq. (\ref{eq:b_mg}) $=B_{\kappa}[\tilde \tau \tilde T]$, and RHS of Ineq. (\ref{eq:b_second}) + RHS of Ineq. (\ref{eq:b_last}) $=\tilde B_{\kappa}[\tilde \tau \tilde T]$.  Combining ineqs. (\ref{eq:b_second}), (\ref{eq:gamma10}) and (\ref{eq:b_last}), we have
\begin{eqnarray*}
B_{\kappa}[\tilde \tau \tilde T] \geq \tilde{B}_{\kappa}[\tilde \tau \tilde T].
\end{eqnarray*}
\done
\subsection{$\hat{A}_{\textbf{x}}[\tilde \tau \tilde T] \leq A_{\textbf{x}}[\tilde \tau \tilde T]+C_A$}
\begin{lemma}$\hat{A}_{\textbf{x}}[\tilde \tau \tilde T] \leq A_{\textbf{x}}[\tilde \tau \tilde T]+C_A$.\label{lem:a_aux}
\end{lemma}
\noindent{\textit{Proof: }}Let $\tau_l$ denote the $l$-th super time slot in super-super time slot $\tilde \tau$, where $l=0,1,\dots,\tilde T-1$.  For any $[s,d]\in \F_{\text{inter}}$,
since $\kappa_{\max}T $ is the maximum amount by which any type II queue can increase in a super time slot, we have
\begin{eqnarray*}
\lefteqn{\min_{g_s\in \HH_{\C(s)}, g_d\in \HH_{\C(d)}} \left(u_s^{g_s}[\tilde \tau \tilde T + \tau_{l+1}] + u_{g_s}^{g_d}[\tilde \tau \tilde T + \tau_{l+1}] + u_{g_d}^{d}[\tilde \tau \tilde T + \tau_{l+1}] \right)}&&\\
&&\leq \min_{g_s\in \HH_{\C(s)}, g_d\in \HH_{\C(d)}} \left(u_s^{g_s}[\tilde \tau \tilde T + \tau_l] + u_{g_s}^{g_d}[\tilde \tau \tilde T + \tau_l] + u_{g_d}^{d}[\tilde \tau \tilde T +\tau_l] \right)+3\kappa_{\max}T.
\end{eqnarray*}
Hence,
\begin{eqnarray*}
\lefteqn{\sum_{l=0}^{\tilde T-1}\min_{g_s\in \HH_{\C(s)},g_d\in \HH_{\C(d)}} \left(u_s^{g_s}[\tilde \tau \tilde T + \tau_l] + u_{g_s}^{g_d}[\tilde \tau \tilde T +\tau_l] + u_{g_d}^{d}[\tilde \tau \tilde T +\tau_l]\right)}&&\\
&& \leq \left[\min_{g_s\in \HH_{\C(s)},g_d\in \HH_{\C(d)}} \left(u_s^{g_d}[\tilde \tau \tilde T +\tau_0] + u_{g_s}^{g_d}[\tilde \tau \tilde T +\tau_0] + u_{g_d}^{d}[\tilde \tau \tilde T +\tau_0]\right)\right]\tilde T\\
&& + 3\kappa_{\max}T\sum_{l=0}^{\tilde T-1} l
\end{eqnarray*}
from which we get
\begin{eqnarray}
\lefteqn{ \sum_{l=1}^{\tilde T-1} \sum_{[s,d]\in \F_{\text{inter}}}\left[\min_{g_s, g_d} \left(u_s^{g_s}[\tilde \tau \tilde T +\tau_l] + u_{g_s}^{g_d}[\tilde \tau \tilde T +\tau_l] + u_{g_d}^{d}[\tilde \tau \tilde T +\tau_l]\right)\right]x_{s}^{d}T}&&\nonumber\\
&& \leq \sum_{[s,d]\in \F_{\text{inter}}}\left[\min_{g_s, g_d} \left(u_s^{g_s}[\tilde \tau \tilde T +\tau_0] + u_{g_s}^{g_d}[\tilde \tau \tilde T +\tau_0] + u_{g_d}^{d}[\tilde \tau \tilde T +\tau_0]\right)\right]\tilde T T  x_{s}^{d}\nonumber\\
&& + \frac{3(\kappa_{\max}T)^2(\tilde T-1)\tilde T}{2}.\label{eq:lem3eq}
\end{eqnarray}
Summing over $s$ and $d$, the LHS of Eq. (\ref{eq:lem3eq}) yields $\hat{A}_{\textbf{x}}[\tilde \tau \tilde T]$ and the RHS yields $A_{\textbf{x}}[\tilde \tau \tilde T] +C_A$ with $C_A = 1.5(N_u \kappa_{\max}T)^2(\tilde T-1)\tilde T$; hence, $\hat A_{\textbf{x}}[\tilde \tau \tilde T] \leq A_{\textbf{x}}[\tilde \tau \tilde T] + C_A$.
\done
\subsection{$\hat{B}_{\kappa}[\tilde \tau \tilde T] \geq B_{\kappa}[\tilde \tau \tilde T] - C_B$}
\begin{lemma} $\hat{B}_{\kappa}[\tilde \tau \tilde T] \geq B_{\kappa}[\tilde \tau \tilde T] - C_B$.\label{lem:b_aux}
\end{lemma}
\noindent{\textit{Proof: }}Let $\tau_l$ denote the $l$-th super time slot in super-super time slot $\tilde \tau$, where $l=0,1,\dots,\tilde T-1$.  For any $(m,n)\in \LL$,
let
\begin{eqnarray*}
 P_{(m,n)}^j[\tilde \tau \tilde T + \tau_l ] = u_m^j[\tilde \tau \tilde T + \tau_l] - u_n^j[\tilde \tau \tilde T + \tau_l].
\end{eqnarray*}

Since $\kappa_{\max}T$ is the maximum about by which any type II queue can increase/decrease in a super time slot, we have
\begin{eqnarray}
P_{(m,n)}^j[\tilde \tau \tilde T + \tau_{l+1}] \geq P_{(m,n)}^j[\tilde \tau \tilde T + \tau_l]-2\kappa_{\max}T\label{eq:ApB_eq2}
\end{eqnarray}
for any $(m,n)\in \LL$.  From Eq. (\ref{eq:ApB_eq2}), we get
\begin{eqnarray}
P_{(m,n)}^j[\tilde \tau \tilde T + \tau_{l+1}] \geq P_{(m,n)}^j[\tilde \tau \tilde T ] - 2lT \kappa_{\max}.\label{eq:ApB_eq3}
\end{eqnarray}
Let
\begin{eqnarray*}
\vec a = \argmax_{\vec{\mu}\in \Gamma} \sum_{(m,n)} \max_j \left( P_{(m,n)}^j[\tilde \tau \tilde T]\right) \mu_{(m,n)}
\end{eqnarray*}
and
\begin{eqnarray*}
\vec b[\tau_l] = \argmax_{\vec{\mu}\in \Gamma} \sum_{(m,n)} \max_j \left( P_{(m,n)}^j[\tilde \tau \tilde T + \tau_l]\right) \mu_{(m,n)}.
\end{eqnarray*}
From Eq. (\ref{eq:ApB_eq3}),
\begin{eqnarray}
\lefteqn{T\tilde T\sum_{(m,n)}\max_j \left( P_{(m,n)}^j[\tilde \tau \tilde T]\right) \left(\vec a \right)_{(m,n)} - 4  (\kappa_{\max}T \tilde T)^2 N_u^2 } && \nonumber \\
&&\leq T\sum_{l=0}^{\tilde T - 1}\sum_{(m,n)}\max_j \left( P_{(m,n)}^j[\tilde \tau \tilde T + \tau_l]\right) \left(\vec a \right)_{(m,n)}~~~~~~~~~\label{eq:ApB_eq4}\\
&& \leq T\sum_{l=0}^{\tilde T - 1}\sum_{(m,n)}\max_j \left( P_{(m,n)}^j[\tilde \tau \tilde T + \tau_l]\right) \left(\vec b [\tau_l]\right)_{(m,n)}.~~~~~~~~~\label{eq:ApB_eq5}
\end{eqnarray}
\begin{eqnarray*}
 B_{\kappa}[\tilde \tau\tilde T] = T\tilde T\sum_{(m,n)}\max_j \left( P_{(m,n)}^j[\tilde \tau \tilde T]\right) \left(\vec a \right)_{(m,n)} - 4  (\kappa_{\max}T \tilde T)^2 N_u^2 
\end{eqnarray*}
since algorithms (\ref{eq:src_cond}), (\ref{eq:gg_cond}), (\ref{eq:gd_cond}), (\ref{eq:gmb_1}),  and (\ref{eq:gmb_2}) are updated every $\tilde T$ super time slots.  
\begin{eqnarray*}
\hat B_{\kappa}[\tilde \tau \tilde T] = T\sum_{l=0}^{\tilde T - 1}\sum_{(m,n)}\max_j \left( P_{(m,n)}^j[\tilde \tau \tilde T + \tau_l]\right) \left(\vec b [\tau_l]\right)_{(m,n)}
\end{eqnarray*}
since those same algorithms are updated every super time slot. Thus, from Ineq. (\ref{eq:ApB_eq4}) and (\ref{eq:ApB_eq5}), our lemma holds with $C_B =4  (\kappa_{\max}T\tilde T)^2 N_u^2$.
\done

\chapter{Efficient Data Transport with Mobile Carriers}
\index{Efficient Data Transport with Mobile Carriers@\emph{Efficient Data Transport with Mobile Carrierss}}%

\section{Introduction}
Data delivery across ``disconnected clusters'' of nodes using mobile
nodes are of increasing interest. Applications include those in
Disruption Tolerant Networking \cite{Fal03}, battlefield networks, and
more generally in scenarios where there is a lack of
infrastructure. Mobile nodes potentially serve multiple
functions, e.g., surveillance and monitoring of the region, along with
supporting data delivery. Further, in many applications, there is
likely to be some flexibility in choosing the trajectories of these
mobile nodes (i.e., controllable mobility).

For concreteness, consider an exploration outpost in a remote corner
of the world. At such a location, it would be difficult to establish
infrastructure for traditional cellular or WiFi networks due to cost,
availability of power sources, etc. Relying on satellites can be
expensive and would only support low data rates. At such remote
locations, one can utilize a group of reconnaissance mobiles (such as
UAVs) to transport data from one part of the network to another. These
UAVs can be used to patrol the premise periodically in order to ensure
security, and they can be readily equipped with radio transceivers to
pick-up and drop-off data at different locations as they patrol, thus
serving a dual purpose.

Because these UAVs play a critical role in providing connectivity,
there has been a surge of interest in developing reliable and
efficient algorithms for these types of networks that use mobile data
carriers. However, due to the opportunistic
and intermittent nature of the mobile connections (the wireless
connections are formed and broken as the mobiles move about) and high
link delays, the traditional routing and rate control algorithms, such
as OSPF and TCP, used in the Internet suffer performance degradation
if used in highly intermittent and opportunistic environments.

One way to mitigate the problem of opportunistic and random
connectivity is through controlled mobility. By controlling the motion
of mobile data carriers, one can make the connections less
opportunistic and random, and more periodic or more predictable. There
is a vast amount of literature available on controlled mobility,  ranging from robotics to operations research \cite{BobDubGib85,Shi76}. Extensive
study has been done on problems such as minimizing the travel cost
subject to some constraints and finding an optimal routes for pick up
and drop off of goods \cite{RatRos83}.

In this dissertation, we focus on minimum cost dynamic routing with controlled
mobility.
Specifically, we study a network of stationary nodes that rely on
mobiles to transport data between them, and the data rates are not
known and may vary over time. The cost for data transport consists of two parts:

{\em First}, there is a per-packet per-route cost -- this reflects the
cost of transmitting a packet over a specific route. For instance,
such linear costs have been used in literature \cite{YinShaRed09} to
minimize hop count. This cost could be source dependent (e.g., a
hard-to-reach source might be penalized with a higher cost). In our
study, we allow any source and mobile dependent per-packet cost to
reflect this.

{\em Second}, there is a per-route cost -- a mobile
is allowed to periodically change trajectories, and the cost is a
function of the trajectory that is chosen. For instance, longer
trajectories (that potentially use more fuel) could be penalized with
a higher cost (in our model, we allow any positive cost per
trajectory).

In this dissertation, we design an algorithm that will \emph{1) guarantee
  throughput optimality} and \emph{2) minimize the sum cost over the
  entire network.} We do this by enabling the mobiles to control their
own routes of operation in response to the traffic demand. Without
controlled mobility, one would have to resort to fixing the routes a
priori, and this could lead to an unstable network, as we demonstrate
in the next section.
\begin{figure}[t!]
\begin{center}
\includegraphics[scale=0.5]{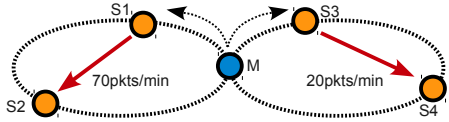}
\caption{A simple network with two modes: The mobile can choose to orbit the left route or the right route.  If it chooses the left (right) route, the mobile will come into contact with stationaries S1 and S2 (S3 and S4). On contact, the mobile can drop-off and pick-up data to and from the stationaries.  In this figure, S1 generates a stream of data for S2 (S3 for S4).  To serve the flow S1-S2, the mobile has to go into the left route, pick up data from S1 and drop them off to S2.  If the flow S1-S2 has higher rate than S3-S4, the mobile should go into left route more often than right route.}
\label{fig:simple_net}
\end{center}
\end{figure}

\begin{figure}[t!]
\begin{center}
\includegraphics[scale=0.5]{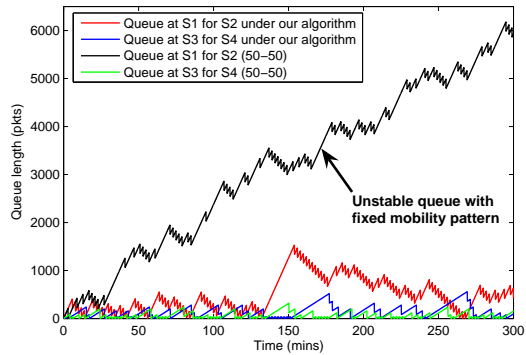}
\caption{If the route operations percentages are set to 50-50 a priori, the source rates we have in Figure \ref{fig:simple_net} can not be supported and will lead to unstable queues.  However, as long as the source rates are in the capacity region (which takes into account the patrol requirements), our algorithm will support them and stabilize the queues.}
\label{fig:simple_q}
\end{center}
\end{figure}

\section{Illustrative Example}
Consider the simple network shown in Figure \ref{fig:simple_net}. We
have one mobile that can choose from two ``routes.'' In the left
route, it will come into contact with stationary nodes S1 and S2 (in that order); in
the right route, with S3 and S4. On contact, the mobile can drop off
and pick up data to and from the stationary nodes. Each route requires
two minutes to finish, and after one route is finished, the mobile returns to the center
and can choose the next route. On each contact, 200 packets can
be picked up or dropped off by the mobile. In addition to transporting
data between nodes, the mobile also has a purpose of patrolling the
area, and must travel each route at least 15\% of the time.

S1 generates data destined to S2 at a rate of 70pkts/min, and S3
generates data destined to S4 at a rate of 20pkts/min. The mobile does
not know the data rates, and the rates can vary over time. If the
mobility pattern of the mobile is fixed, then the network may not be
able to support the two traffic flows. For example, if the mobile
patrols each route 50\% of the time, then the flow from S1 to S2
cannot be supported in the network and the queues are unstable, as
shown in Figure \ref{fig:simple_q}.

In this dissertation, we develop an algorithm that controls the mobility
pattern of the mobile dynamically so that it not only satisfies the
surveillance requirement \emph{but also} stabilizes the queues
whenever it is possible. In the simple example in Figure
\ref{fig:simple_net}, using our algorithm, the mobile may patrol the left
route 75\% of the time and the right route 25\% so that each route is
patrolled at least 15\% of the time (surveillance requirement), and
both traffic flows are supported and all queues are stable.

\section{Related Works}
The networks that utilize mobile carriers to transport data have been
studied extensively recently by \cite{Fal03,JaiFalPat04,WhiCon09,OttKutDwe06,HarAlm06,SpyTsoRag04,BurHooTorFalCerDurSco03,PenFleHas04,LinDorSch03,BurGalJenLev06,BalLevVen07,VahBec00,ThoNelBak10} and others.  The primary focus of \cite{SpyTsoRag04,LinDorSch03,ThoNelBak10,VahBec00} is to increase the data delivery
probability and reduce delivery latency through \emph{replication} in
the context of \emph{delay-tolerant networks} (DTN).  Replication is
useful in networks where mobile carriers move randomly because it
increases the opportunities to transfer data from mobile nodes to
static nodes and vice versa. In networks where the mobility patterns
of mobiles are fixed, replication is not necessary. However, the
drawback of fixed mobility patterns is that the network cannot
dynamically respond to changes in the traffic loads.  In addition,
as long as data is delivered to the destination, it is considered
sufficient, but in networks where the mobility pattern can be
controlled, we can not only guarantee data delivery, but also the most
efficient and optimal network resource utilization.

An extensive simulation study of a network where the mobile messengers
are used to transport data among clusters has been done in
\cite{HarAlm06}; there, the impact of different mobility patterns of
the messengers on delay and efficiency is examined.  In
\cite{CelMod09} and \cite{SugGup09}, algorithms that control the
mobility of a mobile data collector in a sensor network to reduce data
collection delay have been developed. Both papers explore the
trade-off between mobility and wireless transmissions energy. In
\cite{CiuCelMod10} a trajectory control algorithm such that the mobile
data collector dynamically switches its trajectory to be closer to
sensor nodes with more data to transmit in order to save TX power has
been developed. In addition, \cite{KanRahEst04,Schwager09} explore
sensor networks where nodes can reconfigure their positions
dynamically in order to enhance coverage and life span. The work in
\cite{LosNatCos10} is a study of a routing protocol based on
controlled mobility that minimizes the distance traveled by mobile
carriers to reach the destination. However, the mobile messengers in
\cite{HarAlm06,CelMod09,SugGup09,CiuCelMod10,KanRahEst04,LosNatCos10} \emph{cannot adapt their mobility patterns to traffic
  loads}, the result of which will be unstable queues as illustrated
in the previous section.  Experimental evaluation of a heuristic
mobility control algorithm that can respond to changing network
capacity and demand is presented in \cite{BurBroLev05}.  
  Optimization based approaches for mobile data collection have been
  studied in \cite{SomRamSri07}, where the authors study the problem
  of mobile sinks that need to collect data from various sources
  before their respective buffers overflow. The algorithm based on
  using the deadlines (time to fill buffers at various nodes) is shown
  to be NP-complete, and various heuristics are then explored to
  alleviate this.  Further, in \cite{HenBro08}, the authors study the
  problem of transporting data to a single collector from a collection
  of stationary data generators via reinforcement learning techniques.

The optimization framework based on back-pressure \cite{TasEph92} that
is used in our dissertation has been used extensively in
\cite{kellymaullotan2000,ErySri05,YinSriTow07,Sto05,NeelyThesis,ParShaCha10,YinShaRed09,NeeModLi05,Nee10} and many others for
developing efficient resource allocation schemes in wired and wireless
networks in the context of congestion control and back-pressure
routing and rate control. The networks studied in these papers consist
of static nodes, and the links are not intermittent, while in this
dissertation, we focus on intermittently connected networks, and develop an
algorithm that controls mobility to support network traffic flows
while guaranteeing some other objectives such as surveillance
requirement. The optimization algorithm that is most
  similar to ours is the one developed in \cite{Nee10}. The focus of
  \cite{Nee10} however is on an abstract problem of optimizing
  stochastic renewal systems; ours focuses on minimizing the cost of
  transporting data over mobile networks.

In this dissertation, we combine the optimization framework
  used for back-pressure routing with mobility control in order to
  develop a dynamic {\em throughput and cost optimal} mobility control
  algorithm that allows multiple mobiles to transport data among a
  collection of stationary nodes.  Our contributions include:
\begin{enumerate}
\item We formulate a cost minimization framework for the network where
  the mobile carrier adapts its mobility pattern to support traffic
  flows among stationary nodes while satisfying a secondary
  surveillance objective.  We present the min-cost mobility control
  algorithm that is throughput and cost optimal, and then develop a
  practical distributed algorithm.

\item We implement our practical distributed algorithm and present
  experimental results on our Pharos test bed \cite{stovall09}
  using the Click router \cite{click}, where we implement the radio
  and network aspects and emulate mobility.

\end{enumerate}

%
\section{Network Model}

The network consists of $L$ stationary nodes and one mobile
carrier\footnote{We assume one mobile carrier only to simplify the
  notations. This assumption however can be easily removed, see
  section \ref{sec:ext} for the multiple mobile formulation.}. The
stationary nodes do not move and can not communicate with each other
directly; they must rely on the mobile carrier to transport data among
them. We assume that stationary nodes generates data for other stationary
nodes. Let $d_l$ denote the destination node of the data stream
generated by stationary node $l$, and let $x_l^{d_l}$ (pkts/time slot)
be the corresponding average rate.  Let ${\bf{x}} =
\left\{x_{l}^{d_l}\right\}$. Define \begin{eqnarray*} 1_{\{l,l'\}} =
  \left\{\begin{array}{ll}
      1 & \text{ if } l^{\prime}=d_l  \\
      0 & \text{ else. }
\end{array}
\right.
\end{eqnarray*}

A stationary node can exchange data with the mobile carrier when the
two come into contact.  During each contact, the mobile carrier can
send $\eta_d$ packets to the stationary node, and receive $\eta_{p}$
packets from the stationary node.  We call the transmissions from a
stationary node to the mobile carrier a {\em pick up,} and the
transmissions from the mobile carrier to a stationary node a
\emph{drop off}.

We assume that there is a terminal $V$ in the network.  This terminal
does not generate data.  It is there to facilitate notation and
understanding of the definition of route, which we define next.

\begin{defn} \label{ass:route_ass} A {\bf{route}} of the mobile
  carrier starts and ends at the terminal.  The route is a (finite)
  set of tuples $(s,n_s)$ where $s$ is a stationary node and $n_s$ is the
  number of times the mobile visits $s$ on that route.  The route is
  further specified by the time required to patrol that route.\done
\end{defn} 

\begin{assumption}\label{ass:cost_ass}
  We assume that there are $J$ routes for the mobile carrier, which
  are indexed by $j$.  Stationary node $l$ is assessed a cost
  $a_{l,j}$ for every packet picked up by the mobile to be sent over
  route $j$ ($a_{l,j}$ is called {\em pick up cost}).  The mobile
  incurs a cost of $b_j$ per time slot when patrolling route
  $j$.\done\end{assumption}

An example of a route is $$R_1 = \{(V,2), (l_1,1), (l_2,4),
10mins\}.$$ On this route, the mobile starts at $V$, visits $l_1$ once
and $l_2$ four times before returning to $V$.  The time the mobile
takes to patrol $R_1$ is 10mins.  The route $R_1$ is different from
the route $R_2 = \{(V,2), (l_1,2), (l_2,4), 10mins\}$ because $l_1$ is
visited twice on $R_2$ but only once on $R_1$.  $R_1$ is also
different from the route $R_3=\{(V,2), (l_1,1), (l_2,4), 5mins\}$
because it takes less time to patrol $R_3$.  Note that the mobile
carrier must return to the terminal before switching onto another
route.  Though not shown, the terminal in the network in Figure
\ref{fig:simple_net} would be located where the left and right routes
meet (right under where the mobile is).

We can associate higher $b_j$ with the routes on which the mobile
moves faster since that would require more fuel.  We assume that
$a_{l,j} \leq a_{\max},~\forall l,j$ and $b_j \leq b_{\max}, ~\forall
j$.  We let $f_j$ denote the fraction of time the mobile carrier is on
route $j,$ and $T_j$ denote the time required to patrol route $j$ (in
units of time slots).  Assume that $T_{\min} \leq T_j \leq
T_{\max},~\forall j$.  If $N$ routes have been patrolled, and out of
the $N$ routes, the mobile patrolled route $j$ $N_j$ times, then $f_j
= \frac{N_j T_j}{\sum_{j'} N_{j'} T_{j'}}$.  Note that not all
stationary nodes may be included on one route, so the mobile may have
to switch from one route to another to transport packets from a source
to its destination.

Definition~\ref{ass:route_ass} specifies a route of a mobile via the
the collection of stationary nodes that a mobile visits, the number of
times that each of them is visited, and the total time taken to
physically traverse this route. Note that several physical paths
(i.e., the actual geographic paths) can share the same mobile route as
specified by this description (e.g., the difference between two
physical paths could be the order in which the stationary nodes are
visited, or that the actual trajectory could be different; however,
the path characteristics are summarized by Definition 1 could be the
same). In this case, multiple physical paths would be mapped to the
same route. The reason for our choice of these parameters to define a
route is that the list of stationary nodes along with the number of
times that they are visited describe the transfer capacity between the
mobile and stationary nodes, and this along with the time duration of
the mobile route is needed to describe the rate of data transfer
between the mobile and stationary nodes (rate = (number of
contacts)$\times$(packets transferred per contact) $/$ (time duration
of mobile route), see Table~\ref{table:notation_table}). Different
physical paths with the same route parameters in
Definition~\ref{ass:route_ass} lead to the same data transfer
constraints, hence we do not distinguish between them (as the rest of
the physical route properties are not relevant to our model for data
transfer), and our algorithm will treat them all as the same route.

Note that if for some reason, we need to distinguish between
physical paths that have the same route (e.g., with different costs),
it is easy to do so by one of two means: {\em (i)} simply change the
original route time durations (T, which is originally the same for the
two routes) to be $(T - \epsilon)$ and $(T + \epsilon)$, for some
arbitrarily small $\epsilon$; or {\em (ii)} augment the notation in
Definition~\ref{ass:route_ass} to have an additional route-index
parameter (that distinguishes between the two physical paths). All
that this will change is to add an extra route queue in the Min-Cost
Mobility Control Algorithm (see Section~\ref{sec:min_cost_mob}), and
the algorithm will proceed to load-balance between these two routes
based on the costs.

\noindent{{\bf{Surveillance Requirement: }}The mobile must periodically patrol route $j$ to guarantee that $f_j \geq p_j$ for some $p_j \geq 0$, $\forall~j$.\done


We let $\zeta_{l,j}$ denote the number of contacts that can be made between the mobile carrier and stationary node $l$ on route $j$.  Let $P_{l,j}$ and $D_{l,j}$ denote the rates that the mobile can receive from and transmit to node $l$ on route $j,$ i.e., $P_{l,j} = \frac{\eta_p \zeta_{l,j}}{T_j}$ and $D_{l,j} = \frac{\eta_d \zeta_{l,j}}{T_j}$.

Each stationary node $l$ maintains a queue $q_{l,j}$ for each route $j$, $j=1,...,J.$  Node $l$ deposits into queue $q_{l,j}$ the packets it wants the mobile to pick up on route $j$.  The mobile maintains a queue $Q_{l'}$ for each destination node $l'$, $l'=1,...,L$.
\begin{defn}
 We say the network is {\bf{stable}} if all queues are bounded. \done
\end{defn}
We let
\begin{eqnarray*}
 1_{\{P_{l,j} > 0 \}} = \left\{\begin{array}{ll}
1 & \text{ $P_{l,j} > 0$}  \\
0 & \text{ else. }
\end{array}
\right.
\end{eqnarray*}
Finally, we define
\begin{eqnarray*}
\eta_{\max} = \max \left\{\max_{l,j}\left\{ P_{l,j}T_j\right\}, \max_{l,j}\left\{ D_{l,j}T_j\right\}\right\}.
\end{eqnarray*}

\begin{table}[t!]
\begin{center}
\begin{tabular}{ | c | c | }
\hline
Notation & Description \\
\hline \hline
$x_l^{d_l}$ & rate stationary $l$ generates data for stationary $d_l$ \\
\hline
$y_{l,j}^{d_l}$ & rate mobile picks up data from $l$ on route $j$, destined for $d_l$\\
\hline
$p_j$ ($f_j$) & minimum (actual) fraction of time mobile spends on route $j$ \\
\hline
$T_j$ & time duration of mobile route $j$\\
\hline
$a_{l,j}$ & cost $l$ pays to have a packet picked up on route $j$\\
\hline
$b_j$ & cost per unit time mobile pays to spend on route $j$ \\
\hline
$q_{l,j}$ & queue $l$ maintains for packets to be picked on route $j$\\
\hline
$Q_{l}$ & queue mobile maintains for packets destined for $l$\\
\hline
$\delta_{l,j}$ & fraction of total resource on route $j$ stationary $l$ uses \\
\hline
$w_j$ & counter mobile uses to satisfy the surveillance req. (\ref{eq:pat_con})  \\
\hline
$K$ & tuning param. that controls optimality and queue sizes\\
\hline
$\kappa$ & tuning parameter that controls the size of the counter $w_j$ \\
\hline
$\eta_{P}$ & \# of packets picked up per contact\\
\hline
$\eta_{D}$ & \# of packets dropped off per contact\\
\hline
$\zeta_{l,j}$ & \# of contacts mobile makes with $l$ on route $j$\\
\hline
$P_{l,j}$ & data pick-up rate from $l$ on route $j$ ($\eta_{P}\zeta_{l,j}/T_j$)\\
\hline
$D_{l,j}$ & data drop-off rate to $l$ on route $j$ ($\eta_{D}\zeta_{l,j}/T_j$)\\
\hline
\end{tabular}
\caption{Description of variables used}\label{table:notation_table}
\end{center}
\end{table}

\section{Min-Cost Mobility Control}\label{sec:min_cost_mob}
In this section, we will introduce our min-cost mobility control
algorithm. The variables associated with our algorithm along with
their brief physical description are given in Table
\ref{table:notation_table}.  Define $y_{l,j}^{d_l}$ to be the
(average, long-term) rate at which data generated by stationary node
$l$ is picked up by the mobile carrier when patrolling route $j$.  The
objective of the min-cost mobility control is to find $\mathbf{f} =
\{f_j, j =1,...,J\}$
and $\mathbf{y}=\{y_{l,j}^{d_l}, l=1,...,N, j=1,...,J\}$ (pick up rate splitting) to {\em (i)} stabilize the
network, i.e., guarantee that the queue lengths are bounded, {\em
  (ii)} minimize the average cost $\sum_{l,j} y_{l,j}^{d_l} a_{l,j}
 + \sum_j b_j f_j$, and  {\em (iii)} satisfy the surveillance requirement.  These decisions are determined adaptively via a
dual-decomposition inspired mobility control algorithm.

\begin{defn} \label{lem:capacity}
We say the arrival rate $\mathbf{x}$ is {\bf{supportable}} if there exists $(\mathbf{f}, \mathbf{y})$ such that:
\begin{eqnarray}
\sum_{j} y_{l,j}^{d_l} &=& x_l^{d_l}\label{eq:sum_con}\\
\sum_{j} f_j &\leq& 1\label{eq:route_con}\\
f_j &\geq& p_j\in [0,1],~j=1,...,J \label{eq:pat_con}\\
y_{l,j}^{d_l} &=& \delta_{l,j} f_j P_{l,j},~\forall l,j\label{eq:pickup_con}\\
 \sum_{j,l}1_{\{l,l'\}} \delta_{l,j} f_{j} P_{l,j}&  \leq& \sum_{j} f_{j} D_{l',j},~\forall l'\label{eq:dropoff_con}\\
 0\leq &\delta_{l,j}&\leq 1.\label{eq:delta_con}
\end{eqnarray}
\done
\end{defn}
Let $\Lambda$ be the set of all supportable arrival rates.  Given
$\mathbf{x}\in \Lambda$, let $\Gamma_{\mathbf{x}}$ denote the set of
$(\mathbf{f},\mathbf{y})$ that satisfy the equations in definition
\ref{lem:capacity}. An arrival rate $\mathbf{x}$ is said to be
supportable if there exist $\mathbf{f}$ and $\mathbf{y}$ such that:
(i) The summation of $f_j$ (the fraction of time a mobile spends on
route $j$) is less than one (eq. (\ref{eq:route_con})) and the
surveillance requirement (eq. (\ref{eq:pat_con})) is satisfied; (ii)
The rate at which packets are transmitted from stationary node $l$ to
the mobile while it is patrolling route $j$ is upper bounded by the
pick up rate by the mobile on route $j$ (eq. (\ref{eq:pickup_con}));
(iii) The rate at which the data for destination node $l'$ is picked
up from all stationary nodes and on all routes is limited by the
combined rate at which data is dropped off to node $l'$ on all
routes (eq. (\ref{eq:dropoff_con})). The parameter $\delta_{l,j}$
controls the fraction of active contacts between node $l$ and the
mobile on route $j$; when the mobile and stationary node $l$ come into
contact, they have the option of using only part, if any, of the
transfer capacity of the contact; and (iv) The total rate the data is 
picked up from node $l$ on all routes should be equal to the rate at
which node $l$ generates data (eq. (\ref{eq:sum_con})).

Note that we cannot replace conditions (\ref{eq:pickup_con}) and (\ref{eq:dropoff_con}) with
\begin{eqnarray}
 y_{l,j}^{d_l} \leq f_j  P_{l,j}, \label{eq:w1} ~~~~~~~~~~~~~\\
\sum_{j,l} 1_{\{l,l'\}} f_j  P_{l,j} \leq \sum_{j} f_{j}  D_{l',j}. \label{eq:w2}
\end{eqnarray}
Consider two sources $s_1$ and $s_2$, with destinations $d_1$ and
$d_2$, respectively.  Assume that $s_1$ and $s_2$ can be visited by
the mobile only when the mobile is on route $R_0$, and $x_{s_1}^{d_1}
> x_{s_2}^{d_2}$.  Suppose the pick-up rates from $s_1$ and $s_2$ are
the same ($P_{s_1,R_0} = P_{s_2,R_0}$).  Then if we enforce constraint
(\ref{eq:w1}), then $x_{s_1}^{d_1} = y_{s_1,R_0}^{d_1} \leq f_{R_0}
P_{s_1,R_0}$ and $x_{s_2}^{d_2} = y_{s_2,R_0}^{d_2} \leq f_{R_0}
P_{s_2,R_0} = f_{R_0} P_{s_1,R_0}$.  Assume that $d_1$ and $d_2$ are
contacted by the mobile only when it is on routes $R_1$ and $R_2$,
respectively, and they are receiving data only from $s_1$ and $s_2$,
respectively.  In addition, suppose that the drop-off rates to $d_1$
and $d_2$ are the same ($D_{d_1,R_1} = D_{d_2, R_2}$).  If we force
constraint (\ref{eq:w2}), then $f_{R_0} P_{s_1,R_0} \leq
f_{R_1}D_{d_1, R_1}$ and $f_{R_0} P_{s_1,R_0} = f_{R_0} P_{s_2,R_0}
\leq f_{R_2}D_{d_2, R_2}$ which implies that the mobile has to spend
just as much time on $R_2$ as $R_1$, even though $x_{s_1}^{d_1} >
x_{s_2}^{d_2}$.  We resolve this issue by using $\delta_{l,j}$.

The objective of the minimum-cost mobility control is to solve
the following optimization problem:  Given $\mathbf{x} \in \Lambda$ and for any fixed $K > 0,$
\begin{eqnarray}
\text{minimize} \left\{K \sum_{l,j} y_{l,j}^{d_l} a_{l,j} + K\sum_j f_j b_j \right\} \label{eq:mob_opt}\\
\hbox{subject to }  (\mathbf{f}, \mathbf{y})\in \Gamma_{\mathbf{x}}.~~~~~~~~~~~~~~~~~\nonumber
\end{eqnarray}
Let $\left(\mathbf{\hat{f}}, \mathbf{\hat{y}}\right)$ denote an optimal solution to (\ref{eq:mob_opt}).



As we will see later, by suitably choosing the value of parameter $K >
0$ in the optimization problem, the solution generated by our algorithm will be sufficiently ``close'' to the optimal. Next, note that multiplying both sides of
constraint (\ref{eq:pat_con}) by any positive constant $\kappa > 0$
does not change the condition. Thus, the partial Lagrange dual of the
optimization problem (\ref{eq:mob_opt}) is the following\footnote{In the notations, $l'$ generally refers to the destination stationary, while $l$ generally refers to the source stationary.  If we want to be explicit, the destination of the flow originating from the source $l$ is denoted $d_l$.}:
\begin{eqnarray*}
\lefteqn{L(q_{l,j},Q_{l'}, w_j) =} && \\
&& \min_{y_{l,j}^{d_l}, f_j, \delta_{l,j}} \left\{ K \sum_{l,j} y_{l,j}^{d_l} a_{l,j} - \sum_{l,j} q_{l,j} \left(\delta_{l,j}f_jP_{l,j}  - y_{l,j}^{d_l}\right)  \right. \\
&& \left. - \sum_{l'} Q_{l'} \left(\sum_{j} f_j D_{l',j}  -  \sum_{j,l} 1_{\{l,l'\}} \delta_{l,j}f_{j} P_{l,j} \right) \right. \\
&& \left. + K\sum_j f_j  b_j  - \sum_{j} \kappa w_j \left( f_j  - p_j\right)\right\}
\end{eqnarray*}
subject to 1) $\sum_j y_{l.j}^{d_l} = x_l^{d_l}$, 2) $\sum_j  f_j \leq 1$, and 3) $\delta_{l,j}\in [0,1]$.  As we will see later, the parameter $\kappa$ is useful in our algorithm in order to match the time-scale of route selection with the time-scale of queue-length variation.

We now observe that we can decompose the Lagrange dual into two subproblems:
\begin{eqnarray}
\min_{y_{l,j}^{d_l}\geq 0} && K\sum_{j} y_{l,j}^{d_l} a_{l,j} + \sum_j q_{l,j} y_{l,j}^{d_l} \label{eq:sub_p1} \\
\text{s.t.} && \sum_{j}y_{l,j}^{d_l} = x_{l}^{d_l} \nonumber
\end{eqnarray}
for each $l=1,...,L$ and
\begin{eqnarray}
\max_{f_j\geq0, \delta_{l,j}} && \sum_{l,j}  q_{l,j} \delta_{l,j}f_j P_{l,j}  + \sum_j \kappa w_j \left(f_j  - p_j\right)  \label{eq:sub_p2} \\
&& + \sum_{l',j} Q_{l'} f_j \left( D_{l',j} - \sum_{l} 1_{\{l,l'\}}\delta_{l,j}P_{l,j}\right)    \nonumber \\
&& - K\sum_j f_j b_j \nonumber \\
\text{s.t.} && \sum_j f_j \leq 1 \text{ and } \delta_{l,j} \in [0,1]~\forall~l,j \nonumber ~~~~~~~~
\end{eqnarray}

Motivated by the dual decomposition, we propose the following min-cost
mobility control algorithm.  Here, the index $k$ denotes the $k$-th
route selection.  If the $k$-th route is $j$, the time duration
between $k$-th and $(k+1)$-th route selections is $T_j$.} 
In the min-cost mobility control algorithm, a stationary node deposits
its packets into a queue that solves the subproblem (\ref{eq:sub_p1}),
and a mobile station selects its route by solving subproblem
(\ref{eq:sub_p2}). In addition to packet queues, each mobile station
maintains a deficit counter for each route. The size of a deficit
counter indicates the number of times the mobile node needs to
further patrol the route to fulfill the surveillance requirement.

\vspace{0.1in}\hrule\vspace{0.1in}
\noindent{\bf Min-Cost Mobility Control Algorithm}
\begin{enumerate}
\item[(i)] Stationary node $l$ deposits $y_{l,j}^{d_l}(k)$ packets into queue $q_{l,j},$ where
\begin{eqnarray}\label{eq:x_cont}
y_{l,j}^{d_l}(k) = \left\{\begin{array}{ll}
x_{l}^{d_l} & \text{ if } j = j_l^*(k) \\
0 & \text{ else, }
\end{array}
\right.
\end{eqnarray}
and $j_l^*(k) = \argmin_{j}\left\{ K a_{l,j}+ q_{l,j}(k)\right\}$.
\item[(ii)] The $k$-th route $j^*(k)$ selected by the mobile carrier is such that
\begin{eqnarray}
j^*(k) \in \arg\max_j \left\{\sum_{l}  q_{l,j}(k) \delta_{l,j}(k) P_{l,j} -K  b_j \right. \nonumber \\
\left. + \sum_{l'} Q_{l'}(k) \left( D_{l',j} - \sum_{l} 1_{\{l,l'\}}\delta_{l,j}(k)P_{l,j}\right) \right. \nonumber \\
\left.  + \kappa w_j(k) (1-p_j) \right\} \label{eq:f_sol}~~~~~~~~~~~~~~~~~~~~~~~~~~~~~~~
\end{eqnarray}
where
\begin{align}
 \lefteqn {\delta_{l,j}(k) = } &&  \label{eq:delta_sol}\\
&& \left\{\begin{array}{ll}
1 & \text{ if } 1_{\{P_{l,j}>0\}}\left[  q_{l,j}(k) - \sum_{l'}1_{\{l,l'\}}Q_{l'}(k)\right] >0  \nonumber \\
0 & \text{ else. }
\end{array}
\right.
\end{align}
The mobile will pick up data from $l$ on route $j^*(k)$ if and only if $\delta_{l,j^*(k)}(k)=1$.  In addition,
\begin{align}\label{eq:f2_sol}
 f_{j}(k) = \left\{\begin{array}{ll}
1 & \text{ if } j = j^*(k) \\
0 & \text{ else, }
\end{array}
\right.
\end{align}
and
\begin{eqnarray}\label{eq:T_sol}
 T(k) = T_{j^*(k)}.
\end{eqnarray}

\item[(iii)] The queues are updated as follows:
\begin{eqnarray}
q_{l,j}(k+1) = \left[ q_{l,j}(k) + ~~~~~~~~~~~~~~~~\right. \nonumber \\
\left. T(k) \left(y_{l,j}^{d_l}(k) - \delta_{l,j}(k) P_{l,j}1_{\{f_j(k)=1\}}\right)\right]^+   \label{eq:stat_q_evo}
\end{eqnarray}
\begin{eqnarray}
Q_{l'}(k+1) = \left[Q_{l'}(k) + \sum_j 1_{\{f_j(k)=1\}} T(k) \right. \nonumber \\
\left. \times \left[\sum_{l}1_{\{l,l'\}}\delta_{l,j}(k) P_{l,j} - D_{l',j}\right]\right]^+   \label{eq:mob_q_evo}
\end{eqnarray}

\item[(iv)] The algorithm maintains a deficit counter $w_j(k)$ for route $j$ such that at step $k$, $w_j(k)$ is increases by $T(k) p_j$ and decreases by $T(k) f_j(k)$, i.e.,
\begin{align}\label{eq:w_up}
 w_j(k+1) = \left[w_j(k) + T(k)p_j - 1_{\{f_j(k)=1\}}T(k)\right]^+.
\end{align}
\end{enumerate}

\vspace{0.1in}\hrule\vspace{0.1in}


The next two theorems demonstrate the stability and optimality of our proposed algorithm, respectively.
\begin{theorem}\label{theo:stab}
Given $\mathbf{x}$ such that $(1+\epsilon)\mathbf{x}\in \Lambda$ for some $\epsilon>0$, under the iterative algorithm above, the network is stable.
\end{theorem}
\noindent{\emph{Proof: }}See Appendix A.\done
\begin{theorem}\label{theo:opt}
Given $\mathbf{x}$ such that $(1+\epsilon)\mathbf{x} \in \Lambda$ for some $\epsilon>0$, let $\left\{\hat f_j, \hat y_{l,j}^{d_l}\right\}$ be the solution to the optimization problem (\ref{eq:mob_opt}).  As $K\rightarrow \infty$, under our proposed algorithm,
\begin{eqnarray}
\sum_{l,j} a_{l,j} \hat y_{l,j}^{d_l} + \sum_j \hat f_j  b_j = \lim_{K\rightarrow \infty} \lim_{k'\rightarrow \infty} \left[ \frac{1}{\sum_{k\leq k'}T(k)}  \right. \nonumber\\
\left. \times \left( \sum_{k\leq k'} \left(\sum_{l,j}T(k)a_{l,j} y_{l,j}^{d_l}(k)
+\sum_j f_j(k) T(k) b_j \right) \right)\right].\label{eq:theo_opt}
\end{eqnarray}
\end{theorem}
\noindent{\emph{Proof: }}See Appendix B.\done

Note that $\lim_{k'\rightarrow \infty} \frac{1}{\sum_{k \leq k'} T(k)} \sum_{k\leq k'} T(k) y_{l,j}^{d_l}(k)$
in equation (\ref{eq:theo_opt}) is the time-average data rate that the mobile picks up from stationary node $l$ on route $j$, and
$\lim_{k'\rightarrow \infty} \frac{1}{\sum_{k \leq k'} T(k)} \sum_{k\leq k'} T(k) f_j(k)$
is the fraction of the time the mobile patrols route $j$.

\subsection{Impact of $K$ and $\kappa$}
We know from \cite{NeeModLi05} that $(\mathbf{f},\mathbf{y})$ obtained
by our algorithm is within a factor of $O(1/K)$ from $(
\mathbf{\hat{f}}, \mathbf{\hat{y}})$, and while
  $q_{l,j}$, $Q_{l'}$ and $\kappa w_j$ are $O(K \max \{ \eta_{\max},
  \kappa T_{\max} \})$, where $\max \{\eta_{\max}, \kappa T_{\max} \}$
  is the maximum amount by which $q_{l,j}$, $Q_{l'}$, and $\kappa
  w_{j}$ can increase between any two consecutive route selections.
  In order to keep $q_{l,j}$ and $Q_{l'}$ small and make the mobile go
  into ``surveillance'' route without having $w_j$ build up, we choose
  $\kappa$ such that $\eta_{\max} \approx \kappa T_{\max}$,
  i.e. $\kappa = \Theta \left( \eta_{max}/T_{\max}\right)$.

\subsection{Multiple Mobiles}\label{sec:ext}
So far, our model assumes only one mobile node to keep notation
simple.  However, one can easily extend the algorithm to a network
with multiple mobiles.  A straightforward way to extend our model is
to have the stationary nodes maintain a queue for each route
\emph{and} each mobile, i.e., stationary node $l$ maintains a queue
$q_{l,j,m}$ for route $j$ for mobile $m$.  Let $f_{m,j}$ denote the
fraction of time mobile $m$ operates on route $j$.  We then have the
constraint $\sum_{j\in M_m} f_{m,j} \leq 1$ for each mobile $m$, where
$M_m$ is the set of routes mobile $m$ can patrol. Each mobile also has
its own ``surveillance'' or secondary objective constraints on
$f_{m,j}$. Finally, each mobile solves the optimization problem
(\ref{eq:sub_p2}) independently, without any cooperation with other
mobiles.

It is easy to show that the results in this dissertation immediately carry
over to this more general setting (the proofs are analogous to those
presented here).  Lastly, we note that this more general formulation
supports multiple terminals, so that different mobiles can return to
different terminal after each patrol.

{\section{Practical Algorithm}\label{sec:algo}}
The min-cost mobility control algorithm we discussed in the previous section has two shortcomings.  One is that the source  nodes have to synchronize their queue selections with the mobile's selection.  The second issue is that the mobile has to know $q_{l,j}$'s to make the route selection.  However, we can take advantage of the secondary surveillance objective of the mobile.  While the mobile makes its surveillance round, it can collect the most up-to-date queue information while picking up and dropping off data, and use the information in selecting the next route.

Let $k$ denote the $k^{\text{th}}$ route the mobile operates on.  We can recast the update equations (\ref{eq:x_cont}), (\ref{eq:f_sol})  and (\ref{eq:delta_sol}) into the following practical, distributed decision controls.

{\bf{Stationary node: }}The first time the mobile contacts stationary node $l$, it communicates the values of $a_{l,j}$'s corresponding only to the routes on which it will visit stationary node $l$.  {For each new packet, the stationary node $l$ computes }
\begin{eqnarray}
j_l^* = \argmin_{j} K a_{l,j} + q_{l,j} \label{eq:stat_prac}
\end{eqnarray}
and deposits it into $q_{l,j_l^*}$.

{\bf{Mobile: }}Each time the mobile meets stationary node $l$, it collects all queue size information from node $l$ at the end of the contact.  At the end of the execution of a route, the mobile computes the next route $f_j, \delta_{l,j} \in \{0,1\}$ by maximizing
\begin{eqnarray}
\sum_{l\in j} {\bf q_{l,j}} \delta_{l,j} f_j P_{l,j} + \sum_{l' \in j}  Q_{l'} f_j \left( D_{l',j}  - \delta_{l,j}P_{l,j}\right)\nonumber \\
+ \sum_j \kappa w_j \left(f_j - p_j\right) - K\sum_j f_j b_j  ~~~~~~~\label{eq:mob_prac}
\end{eqnarray}
{where ${\bf q_{l,j}}$ denotes the most up-to-date information known to the mobile.}

\section{Experimental Results}

We implemented the practical version of our algorithm as described in
Section \ref{sec:algo} in Click \cite{click} on our testbed
\cite{stovall09} (see also \cite{rybhpash10} for more details).
The purpose of the experiment is to demonstrate that our practical
algorithm can achieve the optimal value and to show that we can come
arbitrarily close to the optimal value at the expense of longer queues
as $K\rightarrow \infty$. The algorithm presented in Section
\ref{sec:min_cost_mob} assumes perfect knowledge of the queue length
by the mobile, while the practical version does not.  To this end, we
build our experimental network with WiFi cards on the Proteus platform
\cite{stovall09}. We emulate the mobility by timed contacts,
i.e., the mobile node is, in our emulation, stationary, and would make
contact with one static node, then wait for some time before making
contact with another static node. These timed contacts are implemented
simply by turning on and off the appropriate wireless interfaces for
the various nodes.

\begin{figure}
\begin{center}
\includegraphics[scale=0.8]{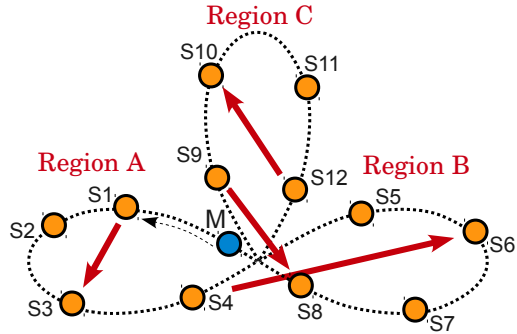}
\caption{A network with one mobile and 12 stationaries, 4 in each region.  The mobile has two modes per region: a fast and a slow route.  In fast route, the mobile goes through a region and comes back to the center in one minute; in slow route,  it takes two minutes.  All stationaries in the region are contacted in each route. (The terminal would be located where the three regions meet.)}
\label{fig:net1}
\end{center}
\end{figure}

\begin{figure}
\begin{center}
\includegraphics[scale=0.6]{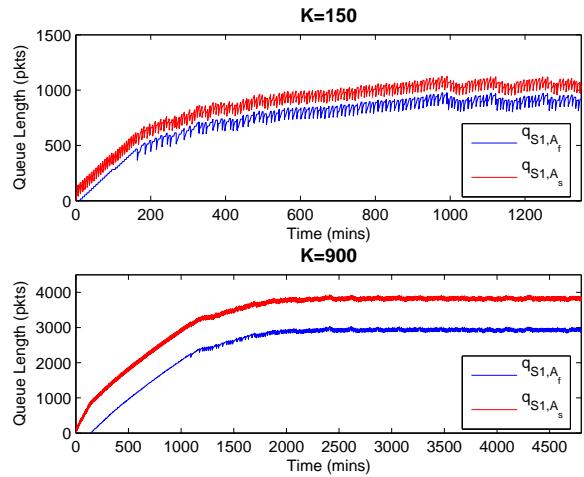}
\caption{Queue lengths observed at S1: Increasing $K$ results in more optimal rate allocations, but it comes at the price of longer queues and longer time for convergence. }
\label{fig:s1_q}
\end{center}
\end{figure}

\begin{figure}
\begin{center}
\includegraphics[scale=0.6]{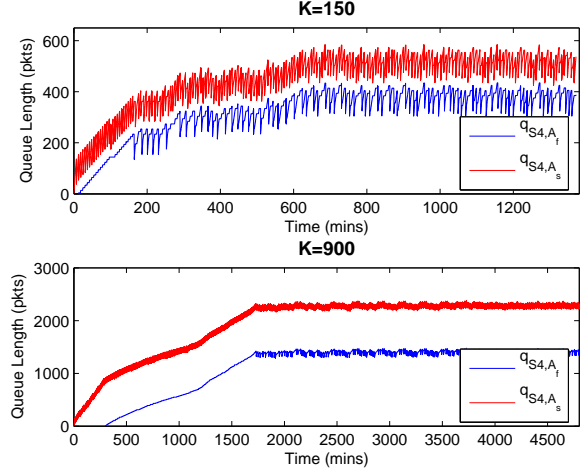}
\caption{Queue lengths observed at S4}
\label{fig:s2_q}
\end{center}
\end{figure}



\subsection{Experiment 1}
The network we used in this experiment is shown in Figure \ref{fig:simple_net}. We use the following routes:
\begin{eqnarray*}
 R_1 &=& \{(V,2), (S1,1), (S2,1), 1min \}  \\
 R_2 &=& \{(V,2), (S1,1), (S2,1), 2mins \} \\
 R_3 &=& \{(V,2), (S3,1), (S4,1), 1 min \} \\
 R_4 &=& \{(V,2), (S3,1), (S4,1), 2mins \}
\end{eqnarray*}
We used two flows, one from S1 to S2, at the rate of 40pkts/min and
another one from S2 to S3 at the rate of 30pkts/min.  The mobile and
the nodes can transmit 100pkts per contact.  The routes $R_2$ and
$R_4$ must be patrolled at least 10\% of the time, i.e., $p_2 = p_4 =
0.1$ and $p_1=p_3=0$. $a_{S1, R_1} = a_{S2, R_1} = K$ and $a_{S1,R_2}
= a_{S2,R_2}=0$; $b_{R_1}=b_{R_3}=K$ and $b_{R_2}=b_{R_4}=0$.  The
rate splitting over the modes $x_{S1,0}^{S2}$ and $x_{S1,1}^{S2}$ is
shown in Table \ref{table:net2_rates}. (The optimal values in Tables
\ref{table:net2_rates} and \ref{table:net3_rates} are obtained by
numerically solving the optimization problem (\ref{eq:mob_opt}) using
MATLAB.)


%
\begin{table}[t!]	
  \begin{center}
    \begin{tabular}{ | c | c | c | c | c | }
      \hline
	& \multicolumn{3}{|c|}{$K$} & \\
    \hline
        & 150 & 300 & 600 & Optimal \\
      \hline
      $y_{S1,R_1}^{S2}$ & 20 & 18.376 & 15.8 & 15 \\
      \hline
     $y_{S1,R_2}^{S2}$ & 20 & 21.624 & 24.2 & 25 \\
    \hline
    $y_{S2,R_1}^{S3}$ & 9.9 & 8.598 & 5.97 & 5 \\
      \hline
     $y_{S2,R_2}^{S3}$ & 20.1 & 21.402 & 24.03 & 25 \\
    \hline
    \end{tabular}
 \caption{Experiment 1: The mobile has to travel the left routes $R_1$ and $R_2$ to pick up data from S2 and travel the right routes $R_3$ and $R_4$ to drop off data to S3.  The mobile must patrol routes $R_2$ and $R_4$ at least 10\% of the time to satisfy the surveillance requirement.  The stationary nodes and the mobile will try to utilize the ``cheaper'' routes $R_2$ and $R_4$ (thus, more packets are picked up on those routes) before the more costly routes $R_1$ and $R_3$. (The units are pkts/min.)}\label{table:net2_rates}
  \end{center}
\end{table}

%


\subsection{Experiment 2}
In this experiment, we used the network shown in Figure \ref{fig:net1}.  The network is composed of three regions, A, B, and C.  In each region, the mobile has two routes, fast and slow.  We use $A_f$ and $A_s$ to denote the fast and slow routes in region A, respectively.  ($B_f$ and $B_s$ for region B and $C_f$ and $C_s$ for region C.)  On a fast route, the mobile goes through the region in one minute; on slow a route, the mobile takes two minutes.  Each stationary node in a region is contacted once on each route made through that region.

On each route, the mobile makes contacts starting from the lowest numbered stationary to the highest.  Each slow route is required to be patrolled at least 10\% of the time, i.e., $p_{A_s} = p_{B_s} = p_{C_s} = 0.1$.  On each contact, the mobile can pick up and drop off 100pkts (200pkts total).

S1 generates data for S3 at rate 23pkts/min; S4 generates data for S6 at 20pkts/min.  S9 generates data for S8 at rate 20pkts/min, and S12 generates data for S10 at rate 23pkts/min.  The packet pick up costs $a_{l,A_s}$, $a_{l,B_s}$, and $a_{l,C_s}$ for the slow route are 0 for all stationary nodes; the pick up costs $a_{l,A_f}$, $a_{l,B_f}$ and $a_{l,C_f}$ for the fast route are $K$.  The route costs are $b_{A_s} = b_{B_s} = b_{C_s} =0$ and $b_{A_f} = b_{B_f} = b_{C_f} = K$.  We let $K = 150$, $450$, and $900$.  $\kappa$ is set to 100.

We compute the optimal rate splitting by solving the optimization problem (\ref{eq:mob_opt}) (shown in the ``Optimal'' column) and compare against the observed rate splitting under different values of $K$.  As predicted by Theorem \ref{theo:opt}, the rate allocation approaches the optimal allocation as $K$ increases as shown in Table \ref{table:net3_rates}.  The price of being close to the optimal rates is long queues, as demonstrated in Figures \ref{fig:s1_q} and \ref{fig:s2_q} for $K=150$ and $900$.

\begin{table}[t!]	
  \begin{center}
    \begin{tabular}{ | c | c | c | c | c |}
      \hline
	& \multicolumn{3}{|c|}{$K$} & \\
    \hline
      $y_{l,j}^{d_l}$  & 150 & 450 & 900 & Optimal \\
      \hline\rule{0pt}{3ex}
      $y_{S1,A_f}^{S3}$ & 10.488 & 9.5 & 8.51 & 8.5\\
      \hline\rule{0pt}{3ex}
    $y_{S1,A_s}^{S3}$ & 12.512  & 13.5  & 14.49 & 14.5 \\
      \hline\rule{0pt}{3ex}
$y_{S4,A_f}^{S6}$ & 7.6 & 6.172 & 5.4 & 5.5  \\
      \hline\rule{0pt}{3ex}
$y_{S4,A_s}^{S6}$ & 12.4  & 13.828 & 14.6 & 14.5 \\
\hline\rule{0pt}{3ex}
$y_{S9,C_f}^{S8}$ & 7.64  &  6.112 & 5.378 & 5.5 \\
\hline\rule{0pt}{3ex}
$y_{S9,C_s}^{S8}$ & 12.36 & 13.888 & 14.622 & 14.5\\
\hline\rule{0pt}{3ex}
$y_{S12,C_f}^{S10}$ & 10.78 & 9.33 & 8.372 & 8.5\\
\hline\rule{0pt}{3ex}
$y_{S12,C_s}^{S10}$ & 12.22 & 13.67 & 14.628 & 14.5\\
      \hline
    \end{tabular}
 \caption{Experiment 2: As $K$ increases, the rate splitting approaches the optimal. (The units are pkts/min.)}\label{table:net3_rates}
  \end{center}
\end{table}

\section*{Appendix A: Proof of Theorem \ref{theo:stab}}\label{sec:theo_stab_proof}
Consider the Lyapunov function $V(k) = \sum_{l,j} \left( q_{l,j}(k) \right)^2 + \sum_{l'} \left( Q_{l'}(k) \right)^2  + \kappa \sum_j \left(w_j(k)\right)^2$.  Define $\Delta V(k) = V(k+1) - V(k)$.  Note that
\begin{eqnarray}
\Delta V(k) &\leq&  \sum_{l,j} \left( q_{l,j}(k+1) - q_{l,j}(k)\right)\left( 2 q_{l,j}(k) + \eta_{\max}\right) \label{eq:Vdown}  \\
&& + \sum_{l'} \left( Q_{l'}(k+1) - Q_{l'}(k) \right)\left(2 Q_{l'}(k) + \eta_{\max} \right) \nonumber\\
&& + \kappa \sum_j \left( w_j(k+1) - w_j(k) \right) \left( 2w_j(k) + T_{\max}\right) \nonumber
\end{eqnarray}
since $\left(q_{l,j}(k+1)\right)^2 - \left(q_{l,j}(k)\right)^2 =  \left(q_{l,j}(k+1) + q_{l,j}(k)\right)\left(q_{l,j}(k+1) - q_{l,j}(k)\right)$ and $\left(q_{l,j}(k+1) + q_{l,j}(k)\right) \leq 2q_{l,j}(k) + \eta_{\max}$.  (Likewise for $Q_{l'}(k)$ and $w_j(k)$.)  The maximum amount by which $q_{l,j}(k)$ and $Q_{l'}(k)$ can increase in one iteration is $\eta_{\max}$; the maximum amount by which $w_j(k)$ can increase in one iteration is $T_{\max}$.

We prove that there exists $q_{\max}$ such that if $q_{l,j}(k)$ or $Q_{l'}(k) > q_{\max}$ for some $l$, $j$, $l'$, and $k$, then
\begin{eqnarray}\label{eq:v_down_cond}
  \Delta V(k)  < -\alpha
\end{eqnarray}
where $\alpha>0$.  Equation (\ref{eq:v_down_cond}) would prove the evolution of $\left\{q_{l,j}(k), Q_{l'}(k)\right\}$ is upper bounded.

Define:
\begin{eqnarray}
 \tilde q_{l,j}(k) &=& \max \left\{ T(k) \delta_{l,j}(k) f_{j}(k)P_{l,j} - q_{l,j}(k), 0 \right\}\label{eq:t_q_max}\\
\tilde Q_{l'}(k) &=& \max \left\{ T(k) \sum_j f_{j}(k)D_{l',j} - Q_{l'}(k), 0 \right\}\label{eq:t_Q_max}\\
\tilde w_j(k) &=& \max \left\{ T(k) - w_j(k), 0 \right\}.\label{eq:t_w_max}
\end{eqnarray}
Note that if $q_{l,j}(k) \geq \eta_{\max}$ and $Q_{l'}(k) \geq \eta_{\max}$, then $\tilde q_{l,j}(k)=0$ and $\tilde Q_{l'}(k)=0$, respectively, and if $w_j(k) \geq T_{\max}$, then $\tilde w_j(k)=0$.

Using equations (\ref{eq:stat_q_evo}), (\ref{eq:mob_q_evo}), (\ref{eq:w_up}) (\ref{eq:t_q_max}), (\ref{eq:t_Q_max}), and (\ref{eq:t_w_max}), the Lyapunov down drift equation (\ref{eq:Vdown}) can be bounded as
\begin{eqnarray*}
 \Delta V(k) &\leq & 2 \sum_{l,j}   \left[T(k) \left(y_{l,j}^{d_l}(k) - \delta_{l,j}(k) f_j(k)P_{l,j}\right) + \tilde q_{l,j}(k) \right] q_{l,j}(k) \\
&&  + 2 \sum_{l'} \left[ f_j(k)T(k)\left( \sum_{l}1_{\{l,l'\}} \delta_{l,j}(k)P_{l,j} - D_{l',j}\right)  + \tilde Q_{l'}(k) \right] Q_{l'}(k) \\
&&+2 \kappa \sum_j  \left(T(k)p_j - T(k)f_j(k) + \tilde w_j(k)\right) w_j(k) \\
&& + 2\sum_{l,j}\eta_{\max}^2 + 2\sum_{l'}\eta_{\max}^2 + 2\kappa \sum_j T_{\max}^2
\end{eqnarray*}
because 1) $q_{l,j}(k+1) - q_{l,j}(k) \leq 2\eta_{\max}$, 2) $Q_{l'}(k+1) - Q_{l'}(k)\leq 2\eta_{\max}$, 3) $w_{j}(k+1) - w_{j}(k)\leq 2T_{\max}$, and 4)
\begin{eqnarray*}
\lefteqn{q_{l,j}(k+1) - q_{l,j}(k)  \leq} && \\
&&  T(k) \left(y_{l,j}^{d_l}(k) - \delta_{l,j}(k)f_j(k)P_{l,j}\right) + \tilde q_{l,j}(k),
\end{eqnarray*}
\begin{eqnarray*}
\lefteqn{Q_{l'}(k+1) - Q_{l'}(k)\leq}&&\\
&& f_j(k)T(k) \left(\sum_{l} 1_{\{l,l'\}}\delta_{l,j}(k) P_{l,j} - D_{l'.j} \right)+ \tilde Q_{l'}(k),
\end{eqnarray*}
and $w_j(k+1) - w_j(k) \leq T(k)\left(p_j - f_j(k)\right) + \tilde w_j(k)$.

Because of equations (\ref{eq:t_q_max}), (\ref{eq:t_Q_max}) and (\ref{eq:t_w_max}), we have
\begin{eqnarray}
 \tilde q_{l,j}(k)q_{l,j}(k) &\leq& \eta_{\max}^2 \label{eq:tqq_max}\\
\tilde Q_{l'}(k)Q_{l'}(k) &\leq& \eta_{\max}^2 \label{eq:tQQ_max}\\
 \tilde w_j(k) w_j(k) &\leq& T_{\max}^2.\label{eq:tww_max}
\end{eqnarray}

Let
\begin{eqnarray}\label{eq:A_def}
 A(k) &=& \sum_{l,j}  y_{l,j}^{d_l}(k) q_{l,j}(k)
\end{eqnarray}
and
\begin{eqnarray}
B(k) &=& \sum_{l,j} \delta_{l,j}(k) f_j(k)P_{l,j} q_{l,j}(k)\nonumber \\
&&+ \sum_{l',j} f_j(k) Q_{l'}(k) \left(D_{l',j} - \sum_{l}1_{\{l,l'\}}\delta_{l,j}(k)P_{l,j}\right).\label{eq:B_def}
\end{eqnarray}
Let
\begin{eqnarray}\label{eq:C_def}
 C & = & 2\sum_{l,j}\eta_{\max}^2 + 2\sum_{d_l}\eta_{\max}^2 + 2\kappa\sum_j T_{\max}^2.
\end{eqnarray}
Then,
\begin{eqnarray}
 \Delta V(k)  &\leq& 2A(k)T(k) - 2B(k)T(k) + 2C \nonumber \\
&&- 2\kappa \sum_j T(k) (f_j(k)-p_j)w_j(k)\label{eq:delta_v}
\end{eqnarray}
due to equations (\ref{eq:tqq_max}), (\ref{eq:tQQ_max}), (\ref{eq:tww_max}), (\ref{eq:A_def}), (\ref{eq:B_def}), and (\ref{eq:C_def}).

Since $(1+\epsilon){\bf{x}}\in \Lambda$, there exist $\{ \tilde
y_{l,j}^{d_l}\}$, $\{\tilde f_j \}$, and $\{\tilde \delta_{l,j}\}$ in
$\Gamma_{(1+\epsilon)\mathbf{x}}$ that satisfy the
  definition \ref{lem:capacity}.
Let
\begin{eqnarray}
 \tilde B(k) &=& \sum_{l,j} \tilde \delta_{l,j} \tilde f_j P_{l,j} q_{l,j}(k)\nonumber \\
&& + \sum_{l',j}   \tilde f_jQ_{l'}(k) \left(D_{l',j} - \sum_l 1_{\{l,l'\}}\tilde \delta_{l,j}P_{l,j}\right).~~~\label{eq:tB_def}
\end{eqnarray}
Then,
\begin{eqnarray*}
\Delta V(k)  &\leq& 2T(k) \left[ A(k) -  \left(B(k) -\tilde B(k) + \tilde B(k)\right)  \right. \\
&& \left. - \kappa \sum_j (f_j(k)-p_j)w_j(k)\right] + 2C.
\end{eqnarray*}
To the RHS of the above inequality, we add and subtract the following two terms 1) $2K\sum_j f_j(k) b_j T(k)$ 2) $2K\sum_j \tilde f_j b_j T(k)$ and add $2 \kappa T(k) \sum_j \left( \tilde f_j  -p_j\right)w_j(k)$ ($\geq 0$ since $\tilde f_j$ satisfies inequality (\ref{eq:pat_con})).  Thus, we get
\begin{eqnarray}
 \Delta V(k)  &\leq& 2T(k) \left[ A(k)  - \left( B(k) - \tilde B(k) + \tilde B(k) \right.\right. \nonumber \\
&& \left.\left. - K\sum_j f_j(k) b_j  +  K \sum_j \tilde f_j b_j \right)  - K\sum_j f_j(k) b_j  \right. \nonumber \\
&&    \left.  +  K\sum_j \tilde f_j b_j - \kappa \sum_j  (f_j(k)-p_j)w_j(k) \right. \nonumber \\
&& \left. +  \kappa \sum_j \left( \tilde f_j    -p_j\right)w_j(k) \right] +2C  \nonumber\\
&\leq & 2A(k) T(k) - 2 \tilde B(k)T(k)  + 2C   \nonumber \\
&& - 2K\sum_j f_j(k) b_j T(k)  +  2K\sum_j \tilde f_j b_j T(k). \label{eq:ll_ineq}
\end{eqnarray}
Inequality (\ref{eq:ll_ineq}) holds because
\begin{eqnarray*}
B(k) - K \sum_j f_j(k) b_j  +\kappa \sum_j  (f_j(k)-p_j)w_j(k) \\
\geq \tilde B(k) - K \sum_j \tilde f_j b_j   + \kappa \sum_j \left( \tilde f_j -p_j\right)w_j(k)
\end{eqnarray*}
 by algorithms (\ref{eq:f_sol}) and  (\ref{eq:delta_sol}).  Hence,
\begin{eqnarray}
\Delta V(k)  & \leq&   2A(k)T(k) - 2  \tilde B(k)T(k)  + 2C  - 2KT(k) \sum_j b_j \left(  f_j(k)   -   \tilde f_j \right)  \nonumber \\
&\leq& 2 T(k) \left[ \sum_{l,j}  y_{l,j}^{d_l}(k) q_{l,j}(k)   - \sum_{l,fj}   \tilde \delta_{l,j} \tilde f_{j} P_{l,j} q_{l,j}(k)   \right. \nonumber \\
&& \left. - K\sum_j b_j \left(  f_j(k)  -   \tilde f_j  \right)\right]+2C\label{eq:DV2} \\
&\leq& 2 T(k) \left[ \sum_{l,j}  y_{l,j}^{d_l}(k) q_{l,j}(k)    - 2\sum_{l,j}  \tilde y_{l,j}^{d_l} q_{l,j}(k)   \right. \nonumber \\
&&  \left. - 2K \sum_j b_j \left(  f_j(k)   -   \tilde f_j  \right)\right] + 2C.\label{eq:DV3}
\end{eqnarray}
Equation (\ref{eq:DV2}) follows because $\{\tilde f_{j}\}$ and $\{ \tilde \delta_{l,j}\}$ satisfy (\ref{eq:dropoff_con}).  Equation (\ref{eq:DV3}) follows because $\{\tilde y_{l,j}^{d_l}\}$ satisfies (\ref{eq:pickup_con}).
Adding and subtracting 
\begin{eqnarray*}
2K\sum_{l,j}a_{l,j}T(k) \left(\tilde y_{l,j}^{d_l}+y_{l,j}^{d_l}(k)\right) 
\end{eqnarray*}
to RHS of equation (\ref{eq:DV3}), we have
\begin{eqnarray}
\Delta V(k)  &\leq& 2 \sum_{l,j} T(k) y_{l,j}^{d_l}(k) \left(Ka_{l,j} + q_{l,j}(k)\right)  +2C   \nonumber \\
&& - 2\sum_{l,j} T(k)  \tilde y_{l,j}^{d_l} \left( Ka_{l,j} + q_{l,j}(k) \right)\nonumber\\
&& - 2K\sum_{l,j}a_{l,j}T(k) \left( y_{l,j}^{d_l}(k) - \tilde y_{l,j}^{d_l} \right) \nonumber \\
&& - 2KT(k) \sum_j b_j \left(  f_j(k)   -   \tilde f_j  \right). \label{eq:delV}
\end{eqnarray}
Note that up until this point, we have only used the fact that
$(1+\epsilon)\textbf{x} \in \Lambda$ (and various substitutions) to
arrive at the upper bound (\ref{eq:delV}).  Because of algorithm
(\ref{eq:x_cont}), we have
\begin{eqnarray}
\sum_{l,j}T(k)  y_{l,j}^{d_l}(k) \left( K a_{l,j}+ q_{l,j}(k) \right) =\nonumber \\
\sum_{l}x_{l}^{d_l} T(k) \min_{j} \left\{Ka_{l,j} + q_{l,j}(k)  \right\}.\label{eq:delV2}
\end{eqnarray}

Since $\sum_{j} \tilde y_{l,j}^{d_l} = (1+\epsilon)x_{l}^{d_l}$,
\begin{eqnarray}
 (1+\epsilon)\sum_{l}x_{l}^{d_l} T(k) \min_{j} \left\{Ka_{l,j} + q_{l,j}(k) \right\}\nonumber \\
\leq \sum_{l,j} T(k) \tilde y_{l,j}^{d_l} q_{l,j}(k) + K\sum_{l,j}a_{l,j}T(k) \tilde y_{l,j}^{d_l}.\label{eq:delV3}
\end{eqnarray}

Thus,
\begin{eqnarray}
\Delta V(k)  & \leq& 2C - 2K\sum_{l,j}T(k) a_{l,j}\left( y_{l,j}^{d_l}(k) -\tilde y_{l,j}^{d_l}\right)  \nonumber \\
&& - 2\epsilon \sum_{l}x_{l}^{d_l}T(k)\min_{j}\left\{ q_{l,j}(k) + K a_{l,j}\right\} \nonumber\\
&& - 2KT(k) \sum_j b_j \left(  f_j(k)   -   \tilde f_j  \right).\nonumber
\end{eqnarray}
Since $a_{l,j}\leq a_{\max}$, $b_j \leq b_{\max}$, $T_{\min} \leq T_j \leq T_{\max}$, and $T(k) y_{l,j}^{d_l}(k), T(k) \tilde y_{l,j}^{d_l} \leq \eta_{\max}$, we have
\begin{eqnarray*}
 q_{\max}' &=& 2C + 2K\sum_{l,j}\eta_{\max} + 2K\sum_{j'}b_{\max} T_{\max} \\
&\geq&2C - 2K\sum_{l,j}T(k) a_{l,j}\left( y_{l,j}^{d_l}(k) -\tilde y_{l,j}^{d_l}\right)  \\
&& - 2KT(k) \sum_j b_j \left(  f_j(k)   -   \tilde f_j  \right),
\end{eqnarray*}
and
\begin{eqnarray*}
 \Delta V(k)  & \leq& q_{\max}'  - 2\epsilon \sum_{l}x_{l}^{d_l}T(k)\min_{j}\left\{ q_{l,j}(k) + K a_{l,j}\right\}.
\end{eqnarray*}
Let
\begin{eqnarray*}
 \nu = \min \left\{ x_{l}^{d_l}~ s.t.~ x_{l}^{d_l} >0 \right\},
\end{eqnarray*}
i.e., $\nu$ is the smallest source rate greater than 0 among all
sources in the network .  If there is $l$ such that
$q_{l,j}(k)>q_{\max} = q_{\max}'/(2\epsilon T_{\min}\nu)$ for all $j$,
then equation (\ref{eq:v_down_cond}) holds.  Thus,
\begin{eqnarray}\label{eq:q_bound}
 q_{l,j}(k) \leq q_{\max} + \eta_{\max},~\forall l,j,k.
\end{eqnarray}

The control decision in equation (\ref{eq:delta_sol}) combined with
the bound (\ref{eq:q_bound}) prevents $Q_{l'}(k)$ from being $>
q_{\max} + \eta_{\max}$ for any $l',k$.  Thus, $Q_{l'}(k) \leq
q_{\max} + \eta_{\max}$ $\forall l',k$.

Because of equation (\ref{eq:q_bound}), $A(k)T(k) \leq
\sum_{l,j}\eta_{\max}q_{\max}$. By simply adding $2K\sum_{j}
f_{j}(k)b_j T(k) > 0$ to eq. (\ref{eq:delta_v}), we have
\begin{eqnarray*}
 \Delta V(k) &\leq& 2A(k)T(k) -2B(k)T(k) +2C \\
&& -2\kappa \sum_j T(k) (f_j(k) - p_j)w_j(k) + 2K\sum_j f_j(k) b_j T(k)
\end{eqnarray*}

If there is $w_j(k)$ such that 
\begin{eqnarray*}
w_j(k) >
\frac{\sum_{l,j}\eta_{\max}q_{\max} + C + 2\sum_j
  Kb_{\max}T_{\max}}{\kappa\left(1-p_j\right)T_{\min}}, 
\end{eqnarray*}
then
\begin{eqnarray*}
B(k)T(k) + \kappa \sum_j T(k) \left(f_j(k) - p_j\right) w_j(k) - K \sum_j f_j(k) b_j T(k) \\
\geq \sum_{l,j}\eta_{\max}q_{\max} + C +
K\sum_j b_{\max} T_{\max} ~~~~~~~~~~~~~~~~~~~~~~~~~~
\end{eqnarray*}
by algorithm (\ref{eq:f_sol}), which
implies that $\Delta V(k) < 0$.  Thus, if 
\begin{eqnarray*}
w_j(k) >
\frac{\sum_{l,j}\eta_{\max}q_{\max} + C + K\sum_j
  b_{\max}T_{\max}}{\kappa\left(1-p_j\right)T_{\min}} 
\end{eqnarray*}
for some $j,k$,
then $\Delta V(k)$ satisfies equation (\ref{eq:v_down_cond}).
Therefore, 
\begin{eqnarray*}
w_j(k) \leq \frac{\sum_{l,j}T_{\max}q_{\max} + C + K\sum_j
  b_{\max}T_{\max}}{\kappa\left(1-p_j\right)T_{\min}} ~~~ \forall~j,k.
\end{eqnarray*}

\section*{Appendix B: Proof of Theorem
  \ref{theo:opt}}\label{sec:theo_opt_proof}

Let $C$ be as defined in eq. (\ref{eq:C_def}).  Then, by
eq. (\ref{eq:delV}), we have
\begin{eqnarray*}
 \Delta V(k)  & \leq& 2C - 2K\sum_{l,j}T(k) a_{l,j}\left( y_{l,j}^{d_l}(k) - \hat y_{l,j}^{d_l}\right) \\
&&- 2KT(k) \sum_j b_j \left(  f_j(k)   -   \hat f_j  \right) ,
\end{eqnarray*}
because of eqs. (\ref{eq:delV2}) and (\ref{eq:delV3}) and
  since $\sum_j \hat y_{l,j}^{d_]} = x_l^{d_l}$.  This implies that
\begin{eqnarray*}
\lefteqn{ \lim_{k'\rightarrow\infty} \frac{1}{\sum_{k \leq k'} T(k)}\sum_{k\leq k'} \Delta V(k)} && \\
  && = \lim_{k'\rightarrow\infty} \frac{k'}{\sum_{k \leq k'} T(k)} \frac{1}{k'}\sum_{k\leq k'} \Delta V(k) \\
&& \leq 2C - 2K\lim_{k'\rightarrow\infty} \frac{\sum_{k\leq k'} \sum_{l,j}T(k)a_{l,j}\left( y_{l,j}^{d_l}(k) - \hat y_{l,j}^{d_l}\right)}{\sum_{k\leq k'}T(k)}  \\
&& ~~~ - 2K\lim_{k'\rightarrow \infty} \frac{1}{\sum_{k\leq k'}T(k)}   \sum_{k\leq k'} \sum_j b_j f_j(k) T(k) + 2K \sum_j \hat f_j b_j
\end{eqnarray*}

Since $\left\{q_{l,j}(k), Q_{l'}(t), w_j(k) \right\}$ are bounded by Theorem \ref{theo:stab}, we have
 \begin{eqnarray*}
  \lim_{k'\rightarrow\infty} \frac{1}{k'}\sum_{k\leq k'} \Delta V(k)   = 0,
 \end{eqnarray*}
which implies that
\begin{eqnarray}
 \lim_{k'\rightarrow \infty} \frac{ \sum_{k\leq k'} \left(\sum_{l,j}T(k)a_{l,j} y_{l,j}^{d_l}(k) +\sum_j f_j(k) T(k) b_j \right)}{\sum_{k\leq k'}T(k)} \nonumber \\
\leq \sum_{l,j} a_{l,j} \hat y_{l,j}^{d_l} + \sum_j \hat f_j  b_j + C/K.\label{eq:theo_opt_ub}~~~~~~~~~~~
\end{eqnarray}
In addition, we have
\begin{eqnarray}
 \sum_{l,j}T(k) a_{l,j} y_{l,j}^{d_l}(k) + \sum_j f_j(k) b_j T(k) \nonumber \\
\geq \sum_{l,j} T(k) a_{l,j} \hat y_{l,j}^{d_l} + T(k) \sum_j \hat f_j b_j \label{eq:theo_opt_lb}
\end{eqnarray}
for each $k$. Eq. (\ref{eq:theo_opt_lb}) holds because $\{ \hat f_{j},
\hat y_{l,j}^{d_l} \}$ is an optimal solution to
eq. (\ref{eq:mob_opt}).Eqs. (\ref{eq:theo_opt_lb}) and
(\ref{eq:theo_opt_ub}) show that $\sum_{l,j} a_{l,j} \hat
y_{l,j}^{d_l} + \sum_j \hat f_j b_j$ is $\leq$ and $\geq$ to RHS of
eq. (\ref{eq:theo_opt}), respectively.\done

\chapter{Conclusion}
Currently exsiting communications algorithms depend on the dynamics of the network being either so fast that any fluctuations can be averaged out or slow enough to be tracked.  However, if these algorithms are deployed on mobile communication networks, they fail to operate or are highly inefficient since the time-scale separation assumption on which these algorithms are built does not hold in mobile networks.  The routing and rate control algorithms we have presented in this dissertation solve this problem by not relying on tracking or averaging out the network dynamics but on exploiting local queue information and the network's ability to dynamically adjust itself, all the while maintaining high efficiency and not sacrificing throughput.  

In Chapter 2, we proposed modifications to the TCP controller to adapt it to the hybrid downlink networks.  The throughputs obtained via our modifications were shown to be proportional to $\mathbb{E}\left[P\right]$ in the multi-path/multi-homing scenario, without the source tracking the channel quality information.

In Chapter 3, we have presented back-pressure rate-control/routing algorithms adapted for intermittently connected networks.  Our proposed algorithms solved the time-scale coupling of the traditional back-pressure algorithm; namely, intermittent connectivity feeds back the wrong congestion signal to the inter-cluster source, making it believe that the connection is a low-rate link, or in order to have high inter-cluster rate, one has to maintain large queues at internal nodes that are no where near the intermittent links.  We have verified that our algorithms work on a simple line network, and on a larger 16-node network.

Lastly, in Chapter 4, we studied a network that uses a mobile carrier to transport data between stationary nodes.  In our work, the mobile can change its movement dynamically to respond to the data traffic in the network.  We have developed a cost minimization framework for such a network and developed a joint mobile-stationary algorithm that minimizes the sum cost, and demonstrated our algorithm on a wireless testbed.

There are many potential research topics related to our work here on mobile communication networks.  We highlight a few of them here.
\begin{itemize}
 \item Generic mobile network: The mobile transport network we have studied in Chapter 4 had the assumption that the mobiles return to one decision point to choose the next route to be on.  Though useful in many scenarios, the network model is not generic enough to allow all types of mobile transport network imaginable.  A possible future topic could be to extend the network model and design a distributed cost minimizing algorithm for the extended model.
 \item Delay reduction: In Chapter 3, we have designed a delay reduction algorithm for wireless networks using shadow packets.  One disadvantage of shadow packets is that it is not energy efficient.  In some networks and applications, minimum delay is more important than maximum throughput.  Designing an energy efficient delay reduction algorithm using shadow packets would be another possible topic.  
 \item Communication security: Improving security in wireless communication is an active area of research.  Advancements in full-duplex radio technologgy may enable secure wireless communication in cellular downlink networks we have studied in Chapter 2 or combined with mobile transport networks of Chapters 3 and 4 for military applications.
\end{itemize}

%
%
%
%
%

\bibliographystyle{plain}  
\bibliography{dissref}        
\index{Bibliography@\emph{Bibliography}}



\begin{vita}
Jung Ho Ryu obtained his B.S. degree in electrical engineering and mathematics in June, 2003 from Northwestern University in Evanston, IL.  He enrolled at The University of Texas at Austin to pursue a doctoral degree in the department of electrical and computer engineering in 2004, and obtained his M.S. degree in electrical engineering in August, 2006.  

His current research interests include wireless and mobile network architectures and design, full-duplex radio system, communication security, routing and rate control, multiple-input and multiple-output communication systems, PHY-layer research, and distributed radio channel access algorithms.  
\end{vita}

\end{document}